\documentclass[preprint2,twocolappendix]{aastex63}

\usepackage{amsmath}

\usepackage{longtable}
\usepackage{verbatim}

\newcommand{\kepler}{\textit{Kepler}}
\newcommand{\gaia}{\textit{Gaia}}
\newcommand{\psamp}{$\mathcal{P}$}
\newcommand{\ssamp}{$\mathcal{S}$}
\newcommand{\csamp}{$\mathcal{C}$}
\newcommand{\earthrad}{$R_\oplus$}
\newcommand{\teff}{T_\mathrm{eff} }
\newcommand{\fstar}{\mathcal{F}_\star}

\newcommand{\vect}[1]{\boldsymbol{#1}}

\accepted{to AJ}

\graphicspath{{./}{figures/}}

\shorttitle{Planet Occurrence Rates with 10 Unique Chemical Elements}
\shortauthors{Wilson et al.}

\begin{document}

\title{The Influence of 10 Unique Chemical Elements in Shaping the Distribution of \kepler\ Planets}

\correspondingauthor{Robert F. Wilson}
\email{robert.f.wilson@nasa.gov}

\newcommand{\uva}{Department of Astronomy, University of Virginia, Charlottesville, VA 22904-4325, USA}

\author[0000-0002-4235-6369]{Robert F. Wilson}
\affiliation{\uva}
\affiliation{NASA Goddard Space Flight Center, 8800 Greenbelt Road, Greenbelt, MD 20771, USA}

\author[0000-0003-4835-0619]{Caleb I. Ca\~nas}
\altaffiliation{NASA Earth and Space Science Fellow}
\affiliation{Department of Astronomy \& Astrophysics, The Pennsylvania State University, 525 Davey Laboratory, University Park, PA 16802, USA}
\affiliation{Center for Exoplanets and Habitable Worlds, The Pennsylvania State University, 525 Davey Laboratory, University Park, PA 16802, USA}

\author{Steven R. Majewski}
\affiliation{\uva}

\author[0000-0001-6476-0576]{Katia Cunha}
\affiliation{Observat\'orio Nacional, Rua General Jos\'e Cristino, 77, Rio de Janeiro, RJ 20921-400, Brazil}
\affiliation{Steward Observatory, University of Arizona, 933 North Cherry Avenue, Tucson, AZ 85721-0065, USA}

\author[0000-0002-0134-2024]{Verne V. Smith}
\affiliation{NSF's NOIRLab, 950 North Cherry Avenue, Tucson, AZ 85719, USA}

\author[0000-0003-4384-7220]{Chad F. Bender}
\affiliation{Steward Observatory, University of Arizona, 933 North Cherry Avenue, Tucson, AZ 85721-0065, USA}

\author[0000-0001-9596-7983]{Suvrath Mahadevan}
\affiliation{Department of Astronomy \& Astrophysics, The Pennsylvania State University, 525 Davey Laboratory, University Park, PA 16802, USA}
\affiliation{Center for Exoplanets and Habitable Worlds, The Pennsylvania State University, 525 Davey Laboratory, University Park, PA 16802, USA}

\author[0000-0003-0556-027X]{Scott W. Fleming}
\affiliation{Space Telescope Science Institute, 3700 San Martin Dr., Baltimore, MD 21218, USA}

\author{Johanna Teske}
\affiliation{Carnegie Earth and Planets Laboratory, 5241 Broad Branch Road, NW, Washington, DC 20015}

\author[0000-0002-9089-0136]{Luan Ghezzi}
\affiliation{Universidade Federal do Rio de Janeiro, Observat\'orio do Valongo, Ladeira do Pedro Ant\^onio, 43, Rio de Janeiro, RJ 20080-090, Brazil}

\author[0000-0002-4912-8609]{Henrik J\"onsson}
\affil{Materials Science and Applied Mathematics, Malm\"o University, SE-205 06 Malm\"o, Sweden}

\author[0000-0002-1691-8217]{Rachael~L.~Beaton}
\altaffiliation{Carnegie-Princeton Fellow; Much of this work was completed while this author was a NASA Hubble Fellow at Princeton University.}
\affiliation{Department of Astrophysical Sciences, Princeton University, 4 Ivy Lane, Princeton, NJ~08544}
\affiliation{The Observatories of the Carnegie Institution for Science, 813 Santa Barbara St., Pasadena, CA~91101}

\author[0000-0001-5388-0994]{Sten Hasselquist}
\altaffiliation{NSF Astronomy and Astrophysics Postdoctoral Fellow}
\affiliation{Department of Physics \& Astronomy, University of Utah, Salt Lake City, UT, 84112, USA}

\author[0000-0002-3481-9052]{Keivan Stassun}
\affil{Department of Physics and Astronomy, Vanderbilt University, VU Station 1807, Nashville, TN 37235, USA}

\author[0000-0003-4752-4365]{Christian Nitschelm}
\affil{Centro de Astronom{\'i}a (CITEVA), Universidad de Antofagasta, Avenida Angamos 601, Antofagasta 1270300, Chile}

\author[0000-0002-1693-2721]{D. A. Garc\'{i}a-Hern\'{a}ndez}
\affiliation{Instituto de Astrof\'{i}sica de Canarias (IAC), E-38205 La Laguna, Tenerife, Spain}
\affiliation{Universidad de La Laguna (ULL), Departamento de Astrof\'{i}sica, E-38206 La Laguna, Tenerife, Spain}

\author{Christian R. Hayes}
\affiliation{Department of Astronomy, Box 351580, University of Washington, Seattle, WA 98195}

\author[0000-0002-4818-7885]{Jamie Tayar}
\altaffiliation{Hubble Fellow}
\affiliation{Institute for Astronomy, University of Hawai‘i at Mānoa, 2680 Woodlawn Drive, Honolulu, HI 96822, USA}
\affiliation{Department of Astronomy, University of Florida, Bryant Space Science Center, Stadium Road, Gainesville, FL 32611, USA}


\begin{abstract}
The chemical abundances of planet-hosting stars offer a glimpse into the composition of planet-forming environments. To further understand this connection, we make the first ever measurement of the correlation between planet occurrence and chemical abundances for ten different  elements (C, Mg, Al, Si, S, K, Ca, Mn, Fe, and Ni). Leveraging data from the Apache Point Observatory Galactic Evolution Experiment (APOGEE) and Gaia to derive precise stellar parameters ($\sigma_{R_\star}\approx2.3\%$, $\sigma_{M_\star}\approx4.5\%$) for a sample of 1,018 Kepler Objects of Interest, we construct a sample of well-vetted Kepler planets with precisely measured radii ($\sigma_{R_p}\approx 3.4\%$). After controlling for biases in the Kepler detection pipeline and the selection function of the APOGEE survey, we characterize the relationship between planet occurrence and chemical abundance as the number density of nuclei of each element in a star's photosphere raised to a power, $\beta$. $\beta$ varies by planet type, but is consistent within our uncertainties across all ten elements.  For hot planets ($P=$1-10 days), an enhancement in any element of 0.1 dex corresponds to an increased occurrence of $\approx$20\% for Super-Earths ($R_p=$1-1.9$\,R_\oplus$) and $\approx$60\% for Sub-Neptunes ($R_p=$1.9-4$\,R_\oplus$). Trends are weaker for warm ($P=$10-100 days) planets of all sizes and for all elements, with the potential exception of Sub-Saturns ($R_p=$ 4-8$\,R_\oplus$).  Finally, we conclude this work with a caution to interpreting trends between planet occurrence and stellar age due to degeneracies caused by Galactic chemical evolution and make predictions for planet occurrence rates in nearby open clusters to facilitate demographics studies of young planetary systems.
\end{abstract}


\section{Introduction} \label{sec:intro}

A clear host-star chemical influence on associated planets was recognized in early spectroscopic surveys primarily aimed at discovering planets through radial velocity (RV) variations, which found that stars hosting giant planets tend to have enhanced metallicities\footnote{In this study, we use metallicities and iron abundance interchangeably, where iron abundances are paramaterized by the number density of iron nuclei in a star’s photosphere relative to the amount of hydrogen normalized to some zero-point, typically the Solar abundance: [Fe/H], where $[X/Y] \equiv \log(N_X/N_Y) - \log(N_X /N_Y)_0$.}
\citep{gonzalez1997,heiter&luck2003, santos2004}. 
More detailed population studies of RV-detected planets confirmed this trend 
between host star [Fe/H] and the frequency at which giant planets are found \citep{santos2004,fischer2005}, a trend that appears to decrease in significance with lower planet mass and/or radius \citep{sousa2008,ghezzi2010,schlaufman&laughlin2011,buchhave2012,wang&fischer2015,ghezzi2018}. 
This correlation is typically interpreted as evidence for the core accretion model of planet formation \citep[e.g.,][]{rice&armitage2003,ida&lin2004,alibert2011,mordasini2012,maldonado2019}, where host star metallicity is a proxy for the solid surface density of the protoplanetary disk; higher metallicities translate to more planet-forming material, which facilitates quick planetary core growth up to a critical mass of $\sim$10~$\rm{M}_\oplus$, in turn allowing more time to accrete gaseous envelopes before gas dissipation in the protoplanetary disk.

The Planet-Metallicity Correlation (PMC) partly motivated large spectroscopic surveys of candidate and confirmed \kepler\ planet-hosting stars \citep[e.g.,][]{bruntt2012, buchhave2012, buchhave2014, everett2013, dong2014, fleming2015, brewer2016, johnson2017}. 
Within this population of close-in, transiting planets, more intricate relationships between stellar metallicity, planet radius, and orbital period have come to light.
It is generally found that planets with larger radii have hosts with super-solar metallicity \citep{buchhave2014,schlaufman2015,wang&fischer2015}. This correlation appears strongest for large planets ($R_P$~$\gtrsim$~4~$R_\oplus$), and nearly disappears for the smallest planets ($R_p$~$\lesssim$~1.7~$R_\oplus$). 
While the PMC is weaker for small planets in general, that is not the case for small planets in short period ($P\lesssim$~10~days) orbits.  
The presence of such planets is positively correlated with metallicity, suggesting that an abundance of solids facilitates the growth and/or migration of small, close-in planets \citep{mulders2016,wilson2018,petigura2018,narang2018,ghezzi2021}. Thus, the amount of available solids in the protoplanetary disk seems to be a key variable in setting the planet mass, radius, \emph{and} period distributions. 
While these works in particular demonstrated the intricate relationships between host-star chemistry and the formation/evolution of planetary systems, they also demonstrated the precision and resources needed to unveil such relationships.

While correlations of planetary architecture to bulk metallicity are well-established,
some results indicate that these trends may be integrating over more detailed chemical relationships. 
For example, \cite{adibekyan2012a} found that an increase in the abundance of certain $\alpha$-elements, such as Mg and Ti, increases the likelihood of planet occurrence.
This work supported that of \cite{brugamyer2011}, who found that, beyond the PMC, planet detection rates are positively correlated with enhanced Si abundances, but not with enhanced O abundances. 
\cite{brugamyer2011}~inferred from this that core accretion is driven by grain nucleation rather than icy mantle growth, and that $\alpha$-elements may drive the formation of planetesimals more efficiently than other elements. 
These investigations show the potential for detailed, multi-element stellar abundance studies to advance models of planet formation.

Measuring variations in the planet occurrence rate with the enhancement or depletion of specific elements could put credible constraints on theories of planet formation. 
For example, if the occurrence of short period planets are positively correlated with a volatile element, an element likely to be in gaseous form at close orbital separations \citep{lodders2003}, one may infer that the core of such planets formed at greater orbital distance where those elements were contained in solid form (i.e., exterior to the respective molecule's ice line) before migrating interior to the respective molecules ice line \citep{oberg2011,marboeuf2014}. 
However, these inferences can be complicated by effects such as cosmic ray ionisation and pebble migration \citep[e.g.,][]{eistrup2018}.

Another interpretation for a trend in planet occurrence between different elements may be due to the density of the planetary core. If it is assumed that the mineralogical makeup of planetesimals dictates the planet's interior structure, and planetesimals' mineralogical makeup may be inferred from stellar abundances \citep{dorn2017a, dorn2017b, hinkel2018}, then one expectation would be that the abundance of elements that result in a denser core would be more likely to prevent atmospheric evaporation. Such a trend may be observable as a strong, positive correlation between the occurrence of planets with a H/He envelope and the enhancement of elemental ratios that result in more dense cores.
In these ways, measuring the correlation between planet occurrence rate and the enhancement of differing chemical elements may provide a means for testing theories ranging from planet migration to exogeology.

However, the data collection needed to study the relationships between planetary properties and the detailed chemical makeup of their host stars properly is particularly resource-intensive, as it requires high resolution, high signal-to-noise spectra of not only hundreds of planet-hosting stars, but also a significant fraction of the stars searched for planets (typically on the order of 10$^{4-5}$ stars for \kepler). Because of this, an occurrence rate study with detailed chemical abundances has not been performed for the \kepler\ field, where much of our knowledge of small planets has originated.

The Apache Point Galactic Evolution Experiment \citep[APOGEE;][]{majewski2017} provides a unique opportunity to perform such a study. APOGEE began in the third phase of the Sloan Digital Sky Survey \citep[SDSS-III][]{eisenstain2011}, and is now in its second phase, APOGEE-2, as a part of SDSS-IV \citep{blanton2017}. The APOGEE survey collects spectra with a multiplexed, high-resolution ($R\sim22,500$), near-infrared ($\lambda \sim 1.5-1.7\, \mu\mathrm{m}$) fiber-fed spectrograph \citep{wilson2012,wilson2019} mounted on the Sloan 2.5-meter telescope \citep{gunn2006} at Apache Point Observatory. The primary goal of APOGEE is to study the Milky Way through the RVs and chemical abundances of nearly 750,000 stars across multiple stellar populations and Galactic regions. 
Additional science programs are also included in the survey, with one such program monitoring stars with candidate planets from \kepler\ (\kepler\ Objects of Interest; KOIs) to search for false positives through RV variations \citep{fleming2015,zasowski2017}. This effort, the APOGEE-KOI Goal Program, has observed 1177 \kepler\ stars, with a median of 17 (mean: 17.7) epochs, as of the sixteenth Sloan data release \citep[DR16;][]{ahumada2020,jonsson2020}. Because of the large number of epochs, the combined, RV-aligned spectra are of high $S/N$ (median: 155, mean: 217), enabling precise derivations of stellar atmospheric parameters and chemical abundances.

In this paper, we utilize the data from the APOGEE-KOI program to explore the role of ten different chemical species (C, Mg, Al, Si, S, K, Ca, Mn, Fe, and Ni) in sculpting the population of \kepler\ planets. In \S\ref{sec:dataandmethods} we describe our data, the derivation of stellar parameters for the KOIs in this study, and the resulting precision in planet radii for our sample. In \S\ref{sec:samples} we describe the sample selection for measuring occurrence rates. 
In \S\ref{sec:allresults} we describe the chemical abundance trends present in the selected sample, and the results of our occurrence rate analyses. Finally, we end this paper with a discussion and reiterate our conclusions in \S\ref{sec:discussion} and \S\ref{sec:conclusion}, respectively. 

\section{Data and Methods}\label{sec:dataandmethods}

\subsection{The APOGEE-KOI Goal Program}

\begin{figure}
\centering
\includegraphics[width=0.45\textwidth]{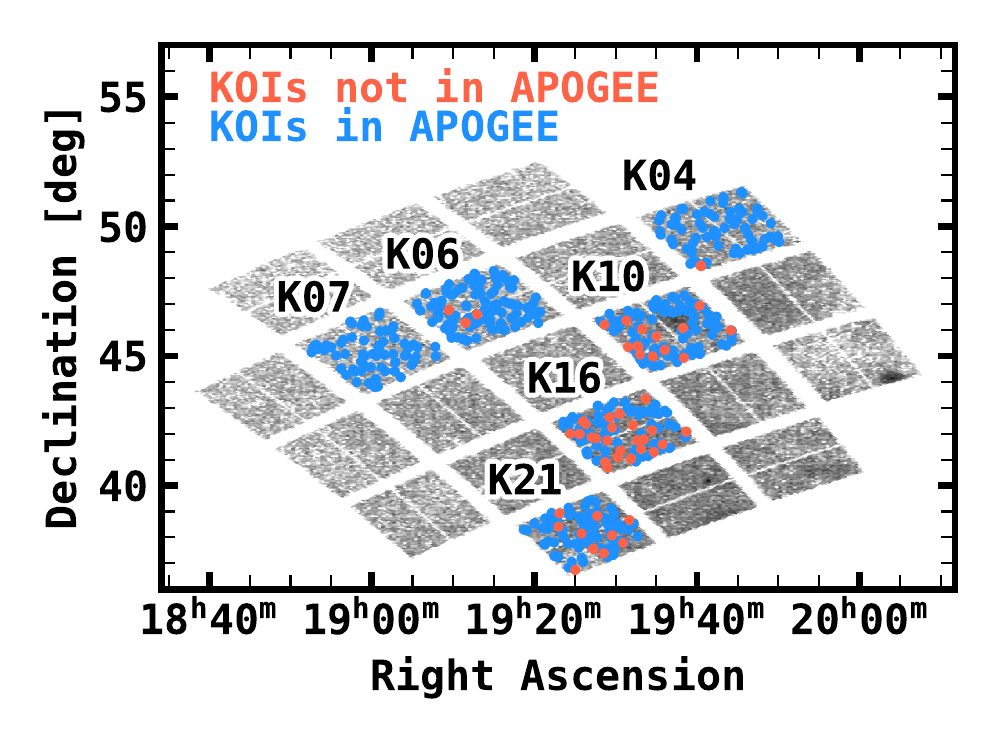}
\caption{The right ascension and declination of stars in the APOGEE-KOI sample. The grayscale points show the density of stars in the \kepler\ stellar properties table at a particular sky coordinate, while the points show the DR24 KOIs observed (blue), and not observed (red) by the APOGEE-KOI program in a temperature range with reliable abundance-ratio measurements (see Figure \ref{fig:calibrate}). The name of each field is listed to the top left of the field.}
\label{fig:sample}
\end{figure}

 The APOGEE-KOI Goal Program targets were chosen with the intention of observing all possible ``Confirmed" or ``Candidate" KOIs with $H<14$ on six different \kepler\ tiles, one of which was observed as a pathfinder program in SDSS-III. One \kepler\ tile is roughly the size of the APOGEE footprint, thus allowing for a near one to one match between an APOGEE field and \kepler\ tile. Some KOIs were excluded from the sample on the basis of nonphysical impact parameters and putative planet radii consistent with stellar values. In total the DR16 APOGEE catalog contains observations for 1299 stars (totaling 1461 unique planet candidates without a ``False Positive" disposition) in the \kepler\ Q1-Q17 DR24 KOI catalog \citep{mullally2015}. Of the 1299 stars, 1177 are part of the APOGEE-KOI radial velocity survey and 122 stars were observed throughout the \kepler\ field as parts of other APOGEE programs \citep[see e.g.,][]{zasowski2013,zasowski2017}. 
 In APOGEE DR16, six fields have been observed in total, labeled as K04, K06, K07, K10, K16, and K21 (see Figure \ref{fig:sample}). Each field was selected on the basis of maximizing the number of available KOIs at the time of target selection. 
 For three of the fields (K04, K06, and K07), KOIs were selected from the Q1-Q17 DR24 KOI catalog, while the other three fields (K10, K16, and K21) were queried from the NexSci Exoplanet Archive\footnote{https://exoplanetarchive.ipac.caltech.edu/} immediately prior to the
design of each field: 2014 March for K10, K21 and 2013 August for K16. These publicly available catalogs were dynamic, and therefore do not have a static or well-studied selection function. As a result, there are a number of KOIs that were discovered after sources were chosen for inclusion in the APOGEE-KOI program (these planet candidates are displayed as red dots in Figure \ref{fig:sample}). 
In \S\ref{subsec:completeness}, we account for biases that may arise from the exclusion of these planets in our analysis.

\subsection{Stellar and Planetary Parameters}

For each KOI observed in APOGEE, we re-derive fundamental stellar properties (e.g., $M_\star, R_\star$) and planet radii. The primary motivation for re-deriving stellar properties in our sample is to improve the precision of the planet radii by incorporating precise spectroscopic parameters derived from the high S/N, high resolution APOGEE spectra. This approach has the additional benefit of maintaining a uniform analysis in deriving properties for the planets in our sample so as not to add additional bias. While we only make use of the stellar radii in our analysis, we provide additional stellar properties for the sake of comparison and any future investigations.

\subsubsection{Spectroscopic Parameters and Abundances: $T_{\mathrm{eff}}$, $\log g$, [Fe/H], [$X$/Fe]} \label{sec:xfetrends}

The spectroscopic parameters in this work are adopted from APOGEE DR16 \citep{ahumada2020,jonsson2020}. All of the spectra from APOGEE are processed through automated data reduction pipelines \citep{nidever2015,holtzman2018}. The spectroscopic parameters used for stars in the APOGEE-KOI program are derived from the Automated Stellar Parameters and Chemical Abundances Pipeline \citep[ASPCAP;][]{aspcap}. In DR16, ASPCAP consists of two components: a \texttt{fortran90} optimization code \citep[\texttt{FERRE} \footnote{Available at https://github.com/callendeprieto/ferre};][]{allendeprieto2006} and an IDL wrapper used for book-keeping and preparing the input APOGEE spectra. \texttt{FERRE} performs a $\chi^2$ minimization across an interpolated library of synthetic stellar atmosphere models \citep[e.g.,][]{zamora2015}, to find a best fit set of input parameters (effective temperature, $T_\mathrm{eff}$; bulk solar-scaled metallicity, [M/H]; surface gravity, $\log g$; microturbulent velocity, $\xi_t$; and C, N, and $\alpha$ abundances).

Once these best-fitting fundamental atmospheric parameters are found, ASPCAP fits individual spectral windows from a carefully curated linelist \citep{shetrone2015,smith2021}  optimized for each chemical element.
In APOGEE DR16 both ``raw" and calibrated spectroscopic parameters and abundance measurements are provided.
$T_\mathrm{eff}$ is calibrated to reproduce the photometric values of \citet{gonzalezhernandez2009}, $\log g$ in the case of dwarfs is calibrated using a combination of asteroseismic values and fits to isochrones. 
Calibrated abundances are zero-point shifted so that stars with solar [M/H] in the solar neighborhood have a mean [X/M]=0 \citep{jonsson2020}.
Unless otherwise stated, we use the calibrated parameters in this study. ASPCAP values of [X/Fe] are reported, which we change to [X/H] via the following equation, [X/H]~$\equiv $~[X/Fe] + [Fe/H].

Abundance ratios for the ten chemical species in this study are defined in the same way as for [Fe/H], i.e., $\mathrm{[X/Fe]} \equiv \log(N_X/N_{Fe}) - \log(N_X/N_{Fe})_0$. However, the chosen zero-point varies by chemical species and is not necessarily the corresponding Solar abundance \citep{jonsson2020}. The APOGEE data products report two different values for carbon abundance ratios, one measured from atomic lines (\texttt{CI\_FE} in the APOGEE DR16 data model) and one measured from molecular CO lines (\texttt{C\_FE} in the APOGEE DR16 data model). For this work, we use the carbon abundance ratio as measured from atomic carbon lines, unless otherwise stated.

When deriving fundamental stellar properties (\S\ref{sec:fittracks}), we use the errors reported by ASPCAP for $\teff$, as comparisons in the literature have shown scatter consistent with these uncertainties \citep[e.g.,][]{wilson2018}. However, the errors reported by ASPCAP are sometimes underestimated for $\log g$ and [Fe/H]. Therefore, when using these parameters to fit to evolutionary tracks in \S\ref{sec:fittracks}, we inflate the uncertainties on $\log g$ and [Fe/H]. We do this by multiplying all reported errors by a given value to define the median uncertainty. For [Fe/H], we inflate the errors so that the median uncertainty is 0.03 dex, a factor of 1.5$\times$ the median uncertainty determined from repeat observations of high $S/N$ spectra \citep{jonsson2020}. We choose to inflate these errors because the typical uncertainty measured in \cite{jonsson2020} was determined using a combined sample of giant and dwarf spectra, and ASPCAP generally measures more precise abundances for giant stars than for dwarf stars. The ASPCAP calibrated $\log g$ are systematically underestimated in FG dwarfs, forcing the fits to the evolutionary tracks to adopt models with systematically lower temperatures than the initial input measurements. 
To adjust for this, we inflated the ASPCAP $\log g$ uncertainties until the input and output temperatures showed no trend. In all, we inflated the $\log g$ uncertainties to have a median error of 0.15 dex, $\sim$1.8$\times$ larger than the ASPCAP reported uncertainties.

To reduce the influence of any systematic trends present in the ASPCAP abundances, we check for correlations with [X/Fe] and $\teff$. 
To test this, we select a sample of dwarf stars observed by APOGEE with high $S/N$ spectra. We start with the DR16 catalog, and remove all stars with log g $<$ 3.5, a distance, d $>$ 1 kpc, as measured from the geometric parallax in \gaia~DR2 \citep{gaiadr2,bailer-jones2018}. 
In addition to these selection cuts designed to remove stars that are not broadly representative of our sample, we also apply a number of cuts designed to remove poor quality data. We remove stars with a spectrum $S/N <$~100, and stars with any of the following ASPCAP or Star Flags set\footnote{for a description of these flags, see https://www.sdss. org/dr16/algorithms/bitmasks/}: TEFF\_BAD, LOGG\_BAD, METALS\_BAD, ALPHAFE\_BAD, STAR\_BAD, and VERY\_CLOSE\_NEIGHBOR.

\begin{figure*}
    \centering
    \includegraphics{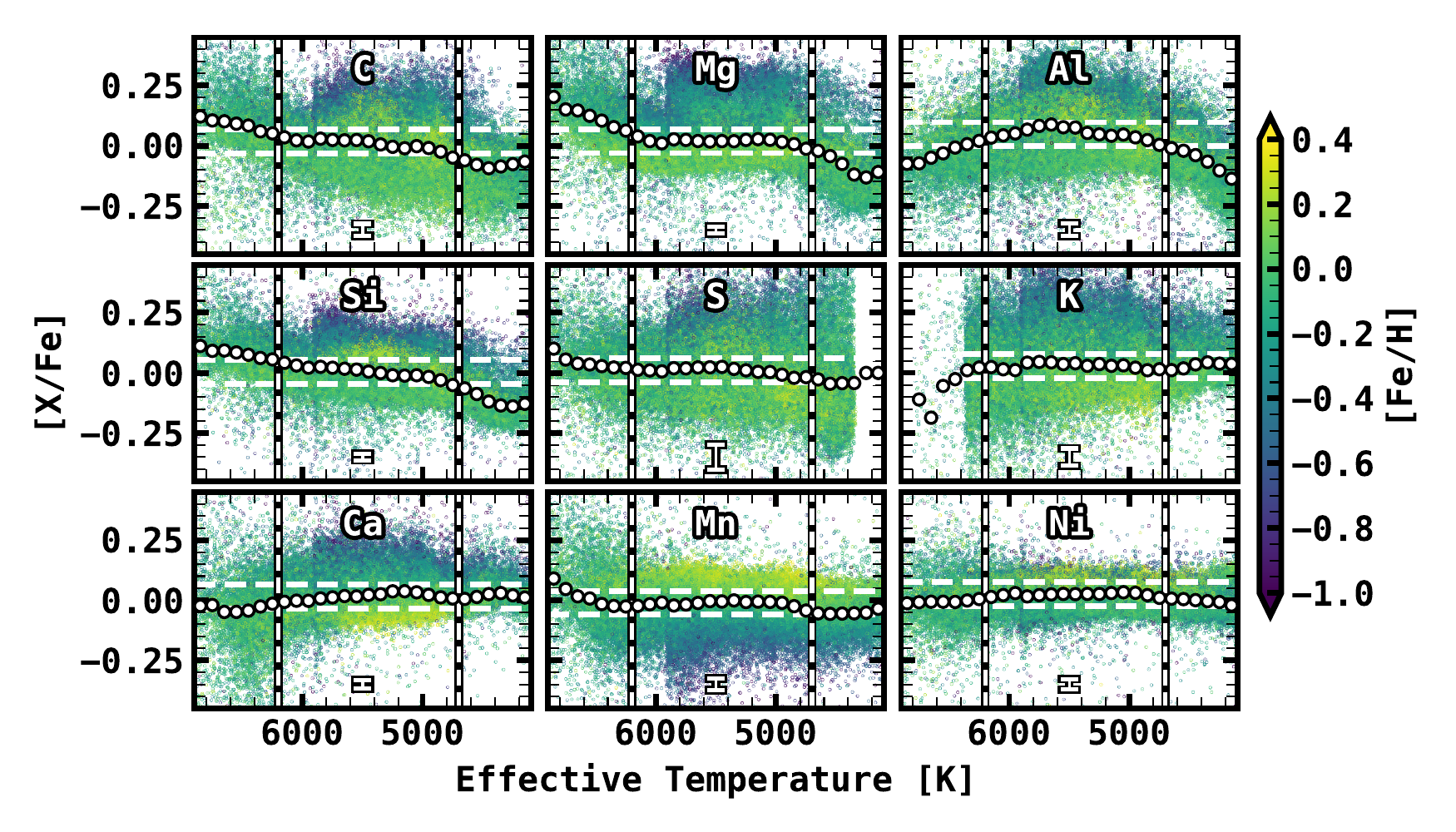}
    \caption{The trends between abundance ratio and $\teff$ for dwarfs in the Solar neighborhood observed by APOGEE, for each element considered in this study. The color of the points in each figure corresponds to the metallicity of the star. The white points show the median abundance ratio in $\teff$ bins of 100K, the dashed vertical lines show our adopted $\teff$~range for this study ($4800\,\mathrm{K}<\teff<6200\,\mathrm{K}$), and the horizontal lines denote the abundance range of $\pm0.05$~dex of the median abundance within the adopted temperature range. The median uncertainty for each abundance is shown as an error bar in the bottom of each panel.}
    \label{fig:calibrate}
\end{figure*}

With this sample of dwarf stars in APOGEE, we assume that there should be no trend in abundance-ratio with effective temperature. If a trend exists, it is more likely to indicate a systematic error in ASPCAP than an astrophysical source.
Our goal is to identify a range of effective temperatures where the APOGEE abundance-ratio measurements are reliable and will not bias our inferences of the planet population. 
In general, we find two prominent features in the ASPCAP-derived abundance ratios at high and low $\teff$~range for ASPCAP that we consider to be systematic in nature and wish to avoid in our analysis (see Figure \ref{fig:calibrate}). At $\teff \lesssim 4700$~K there is a ``hook" feature on the order of up to 0.1 dex, where the ASPCAP-derived abundances decrease dramatically then rise again, present for C, Mg, Si, and Al abundance ratios. We find this same feature in dwarfs in M67, which should all have the same abundance-ratios, leading us to conclude it is systematic in nature. 
On the hotter end, we find an increase in the abundance ratio at $\teff \gtrsim 6200$~K for most of the elements in our sample, which we believe is also a systematic trend. Thus, for this study we only use stars in the temperature range  $4700\,\mathrm{K} < \teff < 6200\,\mathrm{K}$~for our occurrence rate analyses.

Despite our best efforts, there are still a number of elements that display noticeable trends with $\teff$ and abundance ratio (see Figure \ref{fig:calibrate}). Most elements all have a trend with a magnitude (estimated as the range of the median abundance ratios in $\teff$~bins of width 100~K) that is $\leq$0.05 dex, less than a factor of 2-3 of the typical 1$\sigma$ uncertainties. In these cases, any trends with $\teff$~should be negligible. C, Al, and Si, however, all have trends with a magnitude between 0.08-0.1~dex, significantly greater than ($\gtrsim$3-5$\sigma$) their typical uncertainties. Such a trend may introduce a bias in our analysis, as effective temperature is strongly correlated with radius for stars on the main sequence and therefore the \kepler\ plant detection efficiency \citep{pepper2003}. 
We explore this possibility in the Appendix (\S\ref{sec:xfebiases}), but come to the conclusion that biases arising from these systematic trends in ASPCAP are not significant enough to impact our analysis.

\subsubsection{Non-Spectroscopic Parameters:  $\pi$, $Ks$, $E(B-V)$}

For this study, we adopt the parallax, $\pi$, from \gaia~DR2 \citep{gaiadr2}. We apply the global parallax systematic offset as derived by \cite{zinn2019}, adding $\delta_{\pi} = 52.8\pm2.4\, \mu\mathrm{as}$ to the reported $\pi$ from \gaia\ DR2, and adding the uncertainty on the zero-point offset in quadrature with the reported $\sigma_{\pi}$. In conjunction with $\pi$, the stellar apparent magnitude sets a strict semi-empirical constraint on the stellar luminosity.  To minimize the impact of dust extinction in our analysis we adopt the $Ks$-band magnitude from 2MASS \citep{skrutskie2006}, as it is the longest wavelength ($\lambda \sim 2.2 \,\mu \mathrm{m}$) photometric band uniformly available for our sample.

To account for extinction from dust, we employ the 3D dust map from \cite{green2019} which we access using the python package \texttt{dustmaps} \citep{dustmaps}. 
We add the uncertainty from the \cite{green2019} three-dimensional dust map in quadrature with $\sigma_{E(B-V)} = 0.001$~mag to account for the typical uncertainties in the color excess ratios measured in \cite{wang2019} from which we adopt our reddening law.

\subsubsection{Fit to Stellar Evolutionary Tracks}\label{sec:fittracks}

To infer fundamental stellar parameters (e.g., $R_\star$, $M_\star$) for the stars in our sample we apply the python package \texttt{isofit}\footnote{Available at https://github.com/robertfwilson/isofit}. For the sake of brevity, we detail the methodology employed by the \texttt{isofit} package in the appendix (\S\ref{sec:isofit}). In short, \texttt{isofit} compares observations to a grid of MESA Isochrones and Stellar Tracks (MIST) models \citep{dotter2016,choi2016} with masses ranging from 0.1 to 8.0 $M_\odot$, metallicities ranging from $-$2 to 0.5 dex, and evolutionary states ranging from the Zero-Age Main Sequence to the beginning of the White Dwarf Cooling track. After finding an initial best model, a Markov Chain Monte Carlo (MCMC) analysis is applied to estimate the credible ranges for each parameter.

For each host star in our initial planet candidate sample, we run \texttt{isofit} with the following observable quantities and associated uncertainties: $\pi$, $Ks$, $E(B-V)$, $\teff$, $\log g$, and [Fe/H]. We instantiate the MCMC sampling using 30 walkers, with 350 steps and 200 burn-in steps. While modest, we find that this returns posterior distributions in stellar mass and radius that are consistent with the distributions returned after convergence\footnote{This is true for stars on the main sequence, and for parameters that are well constrained, such as stellar radius and luminosity. These settings do not typically return an adequate posterior distribution for other parameters, such as age, or in parameter spaces where degeneracies are likely, such as near the base of the Red Giant Branch.},
and these settings significantly reduce our computational load. We report the stellar parameters as the median for each parameter in the posterior distribution and the upper and lower limits as the 84th and 16th percentile of the posterior, respectively. In all, we derive fundamental stellar parameters for 1,018 stars (281 stars did not have reliable ASPCAP solutions). The stellar parameters derived from \texttt{isofit} are given in Table \ref{tab:stellarprops}.

\begin{centering}
\begin{table*}
    \caption{Derived Properties for 1,018 KOIs in APOGEE. (This table is available in its entirety in machine-readable form) } 
    \begin{tabular}{lll} \hline \hline
         Column & Column Label & Column Description \\ \hline
         1 &  KIC &  \kepler\ Input Catalog Identification Number \\
 2 &  APOGEE\_ID &   The APOGEE Star Identification
 \\
  3 &  Teff &   effective temperature of the star in K
 \\
 4 &  Teff\_e &  16th percentile of derived posterior in Teff  
 \\
 5 &  Teff\_E &  84th percentile of derived posterior in Teff  
 \\
 6 &  logg &  logarithm of the surface gravity of the star in cm/s$^2$   
 \\
 7 &  logg\_e &  16th percentile of derived posterior in logg  
 \\
 8 &  logg\_E &  84th percentile of derived posterior in logg  
 \\
 9 &  feh &  metallicity  of the star, [Fe/H] 
 \\
 10 &  feh\_e &  16th percentile of derived posterior in feh  
 \\
 11 &  feh\_E &  84th percentile of derived posterior in feh  
 \\
 12 &  mass &   mass of the star  in $M_\odot$ 
 \\
 13 &  mass\_e &  16th percentile of derived posterior in mass  
 \\
 14 &  mass\_E &  84th percentile of derived posterior in mass  
 \\
 15 &  radius &  radius of the star in $R_\odot$  
 \\
 16 &  radius\_e &  16th percentile of derived posterior in radius  
 \\
 17 &  radius\_E &  84th percentile of derived posterior in radius  
 \\
 18 &  logL &  logarithm of the bolometric luminosity of the star in $L_\odot$  
 \\
 19 &  logL\_e &  16th percentile of derived posterior in logL  
 \\
 20 &  logL\_E &  84th percentile of derived posterior in logL  
 \\
 21 &  density &  density of the star in $\rho_\odot$  
 \\
 22 &  density\_e &  16th percentile of derived posterior in density  
 \\
 23 &  density\_E &  84th percentile of derived posterior in density  
 \\
 27 &  distance &  distance of the star in pc  
 \\
 28 &  distance\_e &  16th percentile of derived posterior in distance  
 \\
 29 &  distance\_E &  84th percentile of derived posterior in distance  
 \\
 30 &  ebv &  the reddening of the star in units of $E(B-V)$  
 \\
 31 &  ebv\_e &  16th percentile of derived posterior in ebv  
 \\
 32 &  ebv\_E &  84th percentile of derived posterior in ebv  
 \\ \hline 
    \end{tabular}
    \label{tab:stellarprops}
\end{table*}
\end{centering}

\subsubsection{Accuracy and Precision of Stellar Properties}

We assume the larger of the absolute value between the median and upper or lower limits to be a reliable metric for the precision of the stellar parameters inferred in our sample.
These uncertainties are displayed in Figure \ref{fig:comparisons}.
For stellar radius, we find a mean uncertainty of $\sigma_{R_\star} = 2.7$\% and median uncertainty of $\sigma_{R_\star} =  2.3\%$. This error is largely limited by the uncertainty in $\teff$ and $Ks$. It is more difficult to say what sets the minimum uncertainty in $M_\star$, given that there are several inputs that are correlated. In all, we find the median uncertainty $4.5\%$ and mean uncertainty to be $4.7\%$. However, we caution that for some stars our reported uncertainty in $M_\star$ is likely underestimated. Grid effects may prevent the walkers from exploring the full range of parameter space in $M_\star$, especially for stars with $\sigma_{M_\star} \lesssim 3\%$.  
We also note once again for emphasis that the reported uncertainties in stellar mass do not take model uncertainties into account, and are entirely model-dependent. While $\log g$ does offer a semi-empirical mass constraint when combined with the inferred radius, which only depends on the bolometric correction as a model-dependent constraint, it is not as limiting in our case where we inflate the $\log g$ uncertainties to have a median of 0.15 dex. 
To this end, comparing the masses derived with different sets of model grids are likely to reveal larger uncertainties in the inferred mass, but such an exercise is outside the scope of this work.

\begin{figure}
    \centering
    \includegraphics{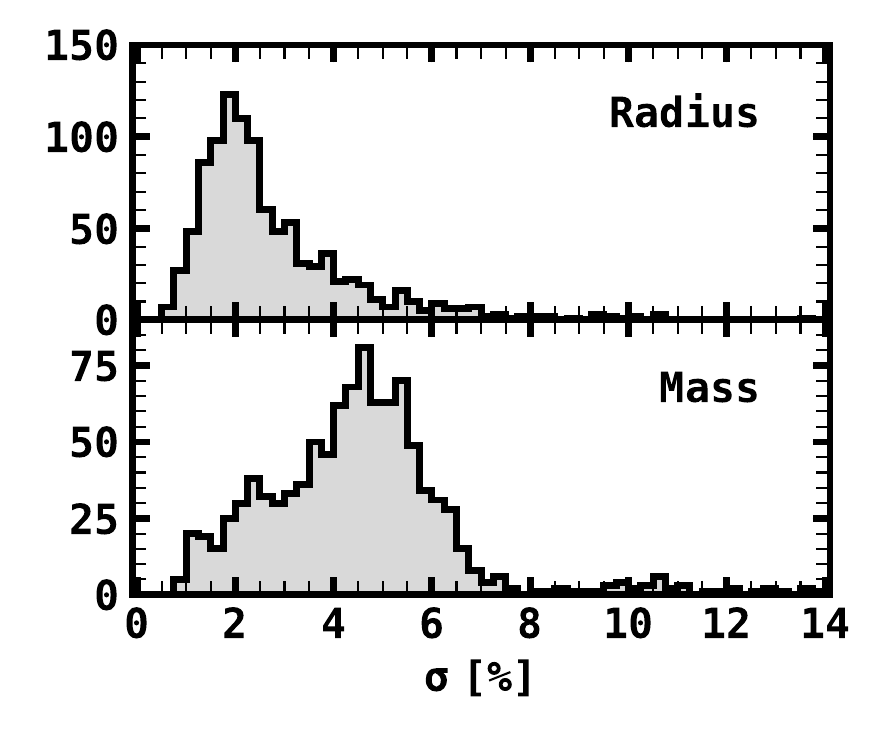}
    \caption{The relative errors of the stellar radius (top) and mass (bottom) in the APOGEE-KOI sample derived with \texttt{isofit}. The mean and median stellar radius uncertainties are 2.7\% and 2.3\%, respectively. The mean and median uncertainties on the stellar mass are 4.5\% and 4.7\%, respectively. }
    \label{fig:errorhist}
\end{figure}

To judge the accuracy of the stellar parameters in our sample, we compare the results from \texttt{isofit} to the parameters derived in \cite{berger2020}, which has a measured mass and radius for each star in our sample. \cite{berger2020} derived masses and radii for $\sim$186,000 stars in the \kepler\ field by comparing photometric effective temperatures, \gaia\ parallaxes, and 2MASS $Ks$-band magnitudes to a custom set of MIST model grids, and spectroscopic [Fe/H] where applicable. For stars with no spectroscopic [Fe/H], the authors assumed a thin disk metallicity prior.
These comparisons are highlighted in Figure \ref{fig:comparisons}.

We find overall agreement consistent with our reported uncertainties. The mean difference in radii, calculated as $(R_\star - R_\mathrm{B20})/R_\star$, gives a mean and scatter of $-0.68 \pm 3.44 \%$, where $R_\mathrm{B20}$~is the radii inferred by \cite{berger2020}. This is well within the combined uncertainties defined in our sample and in \cite{berger2020}. However, there are some systematic differences. While there is generally excellent agreement in $R_\star$, the radii in the APOGEE sample are systematically lower by as much as $\sim$5\%~for lower-mass stars ($\lesssim 0.7 M_\odot$). This may be caused by the use of slightly different model grids. 
Most stellar model grids are inconsistent with empirical constraints when deriving parameters for late M-type dwarfs. While we do not make any corrections in our model grid to account for this,  \cite{berger2020} adjust their model grids for stars with $M_\star \lesssim 0.75 M_\odot$ by adopting empirical relations from \cite{mann2015,mann2019}. 
However, because our analysis is with FGK dwarfs, and our radii still largely agree with those from \cite{berger2020} within our combined uncertainties and the limiting systematic uncertainties of $\sim$2\% \citep{mann2019,zinn2019b}, there is no strong motivation to make adjustments for this range of parameter space.

\begin{figure*}
    \centering
    \includegraphics[width=0.48\textwidth]{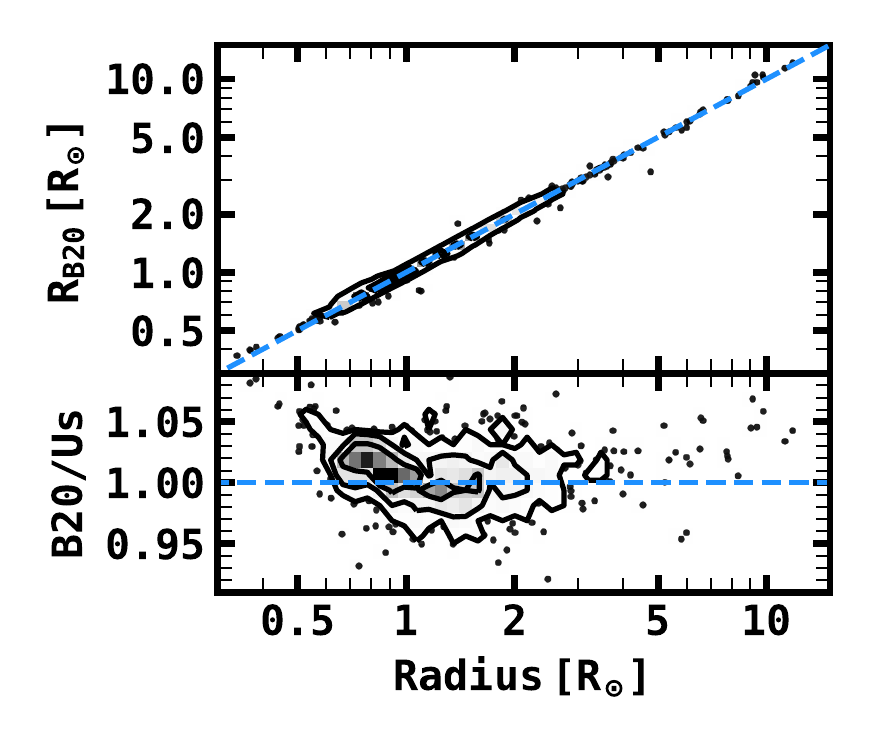}
    \includegraphics[width=0.48\textwidth]{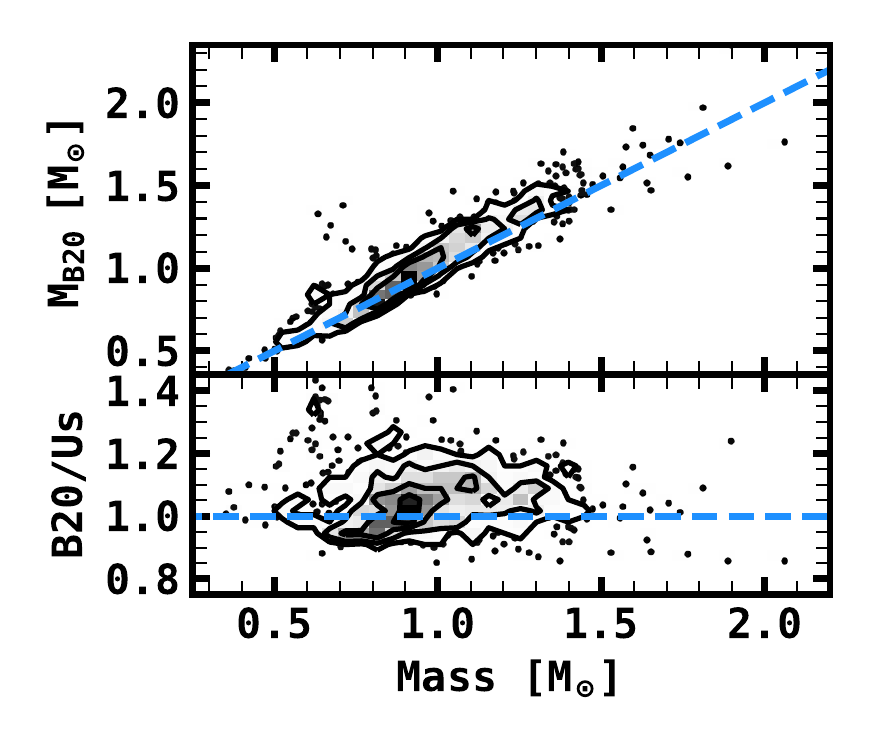}
    \caption{Comparison of the fundamental stellar properties derived in this work versus the stellar properties derived by \cite{berger2020} for the same stars (B20). The dashed blue lines in each case represent the one-to-one agreement between the two samples. \textit{Left:} Comparisons of the stellar radii derived in this work. Overall there is excellent agreement, with scatter in the ratio of radii of 3.4\%, and an average offset of $<1\%$. \textit{Right:} Comparisons of the stellar masses derived in this work and in \cite{berger2020}. They agree overall within the scatter, but have an offset of $\approx 6\%$, in that the APOGEE sample has a lower mass on average.
    }
    \label{fig:comparisons}
\end{figure*}

Performing the same comparison for $M_\star$, we find the mean and scatter of $(M_{\star} - M_\mathrm{B20})/M_\star = -0.061 \pm 0.081$, where $M_{B20}$ is the mass derived in \cite{berger2020}. While there is a somewhat significant offset, it is still within the reported scatter for the comparison. However, this offset is larger than our reported uncertainties ($\sim$4-5\%) in $M_\star$, but as mentioned above, $\sigma_{M_\star}$~is likely underestimated for a fraction of stars in our sample. 
This offset is most likely due to a difference in the $\teff$~of the two samples. We find that the effective temperatures between our sample and those of \cite{berger2020} have~$\teff - T_\mathrm{eff,\,B20} = -78 \pm 193$~K. This lower temperature explains the differences in the inferred stellar mass.  However, this difference is mostly for stars with effective temperatures near 5000-6000~K. The difference in effective temperature is minimal for stars with $\teff \lesssim 5000$~K.

\begin{figure*}
    \centering
   \includegraphics[width=0.48\textwidth]{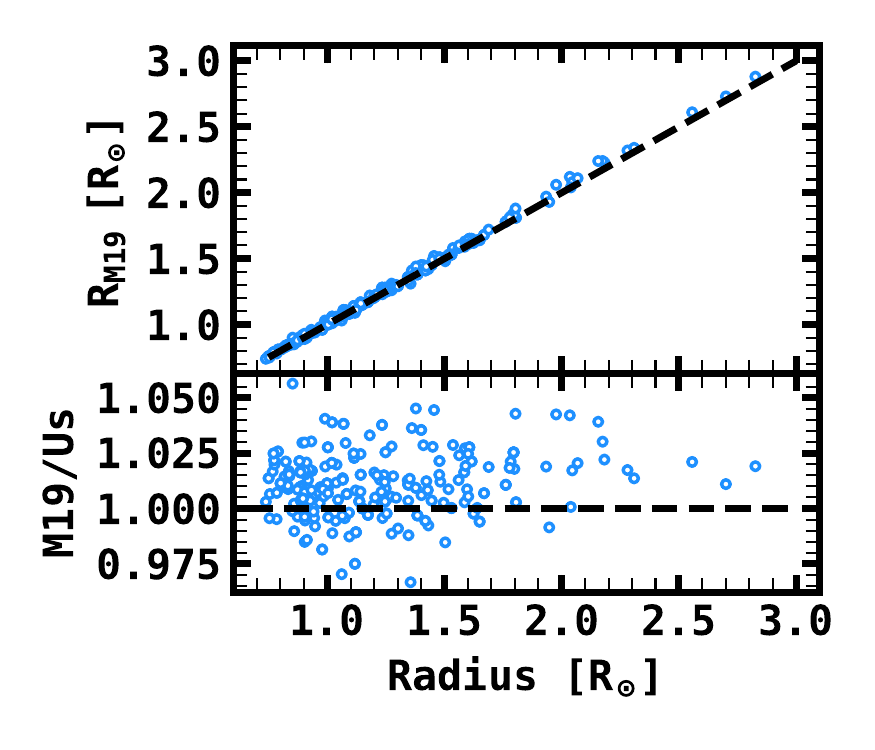}
    \includegraphics[width=0.48\textwidth]{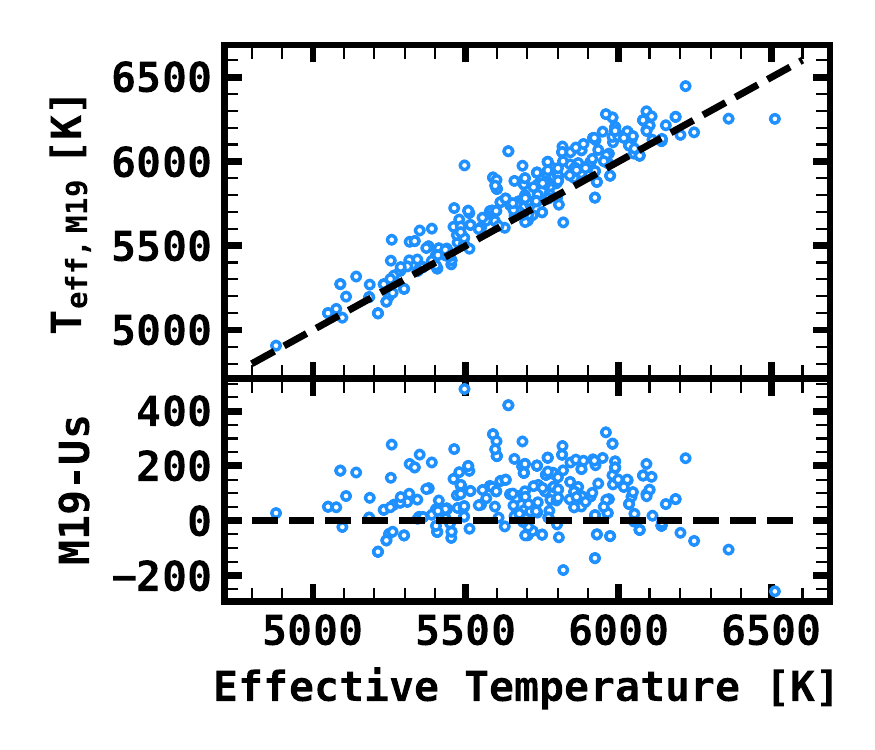}
    \caption{Comparison of the fundamental stellar properties derived in this work versus the stellar properties derived by \cite{martinez2019} for the same stars (M19). In each panel, the dashed black line denotes agreement. \textit{Left:} Comparison of the stellar radii. We find relative agreement, with an average offset and scatter of 1.1$\pm$1.3\% in the ratio of the radii. \textit{Right:} Comparison of the effective temperatures derived by ASPCAP and the effective temperatures from M19. There is a mean offset and scatter of $108\pm171$~K between the two samples. The systematically lower $\teff$~in ASPCAP is the likely reason for the systematic offset in stellar radii. }
    \label{fig:m19comparison}
\end{figure*}

In addition to the comparisons with \cite{berger2020}, we check our stellar radii against those inferred from high-resolution spectroscopy \cite[][see Figure \ref{fig:m19comparison}]{martinez2019}. \cite{martinez2019} derived atmospheric parameters from the archival spectra in the CKS sample by measuring equivalent widths for a carefully curated sample of Fe {\sc{i}} and Fe {\sc ii} lines \citep{ghezzi2010,ghezzi2018}. 
This sample is a more fair comparison to our sample in terms of precision, due to the combination of spectroscopic $\teff$, $\log g$, [Fe/H], and \gaia\ parallaxes used. We find relatively good agreement, with $(R_{\star} - R_\mathrm{M19})/R_\star = -1.1 \pm 1.4$\%, where $R_{M19}$~is the radii from \cite{martinez2019}. Thus, although there is an offset, the radii derived in \cite{martinez2019} largely agree with those derived here, and the difference is within systematic uncertainties of $\approx$2\% for radii derived from \gaia\ DR2 parallaxes \citep{zinn2019b}. 
The difference between our radii and those derived in \citeauthor{martinez2019} can likely be traced to differences in the effective temperature between the two samples. On average, the difference in $\teff$ is 108~K with a scatter of 171~K, where the effective temperatures from APOGEE are lower, explaining our smaller inferred radii (see Figure \ref{fig:m19comparison}).

\subsubsection{Planet Radii}

 We derive each of the planet radii using the reported transit depth in the DR24 KOI catalog \citep{mullally2015}. We apply the simple relationship, 
\begin{align}
    R_p = R_\star \sqrt{\delta_{tr}}
\end{align}
to calculate the planet radii in our sample, where $\delta_{tr}$~is the measured transit depth. The uncertainty in planet radius for our catalog is found by propagating the errors on $R_\star$~with the uncertainties from the \kepler\ DR24 transit depth measurement. The resulting planet radii in our sample have a median uncertainty of $\sigma_{R_p}/R_p = 3.4\%$ (mean: 3.7\%).

\section{Sample Selection and Planet Classes}\label{sec:samples}

For this study we define three individual samples that we introduce here before describing them in detail below. The first sample is the stellar planet-search sample, \ssamp.
\ssamp\ is the parent sample of stars that may have been observed by the APOGEE-KOI program. This translates to the \kepler\ field stars within the APOGEE footprint that are then down-selected based on our scientific goals.
The second sample is \csamp, or the control sample, which is a subset of \ssamp. Because we don't have detailed chemical abundances for each star in \ssamp, \csamp\ acts as a proxy from which we can infer the bulk properties (i.e., abundance-ratio distributions) of \ssamp. 
The final sample is the vetted planet sample, \psamp. \psamp\ is the sample of planets whose host stars were observed by the APOGEE-KOI Goal Program that is then further vetted to remove False Positives and ensure a well-characterized sample of planet candidates.

 \subsection{\ssamp: Stellar Planet Search Sample}

To select the appropriate planet search sample, \ssamp, we start from the catalog of stars in \cite{berger2020}. We downsample this table to replicate the selection function of the APOGEE-KOI survey. These cuts are listed below.  

\begin{enumerate}
    \item \textit{Brightness Cut}, $H<14$: This is the brightness limit in the APOGEE-KOI planet sample, chosen because it is the limit for which a one-hour integration with APOGEE yields a S/N~$\gtrsim$~10, i.e., sufficient to derive reliable radial velocities. We apply this cut to each star in the field sample. 
    \item \textit{APOGEE Field Cut}: $100\,\arcsec < d < 1.5\,^\circ$, where $d$~is the angular distance from the center of the nearest APOGEE-KOI field. The upper limit of $1.5\,^\circ$ represents the limit placed by the Sloan 2.5-meter telescope's field of view, and $100\,\arcsec$~is an instrumental limit derived from a central post that obscures targets in the center of the plate design \citep{owen1994,zasowski2017}.
\end{enumerate}

At this point, it is important to note that the individual fields for the APOGEE-KOI program were chosen to maximize the number of observable KOIs per field. If each \kepler\ tile is expected to have the same number of KOIs, the choice to maximize the number of targets in the APOGEE-KOI program may introduce a bias leading us to overestimate the planet occurrence rate. However, it is more likely that the planet yield per field is driven by a combination of the number of stars per field where transiting planets are detectable, which would favor the fields closer to the Galactic mid plane, and the quality of the light curves in the particular field, which would be diminished by crowding and favor fields farther from the Galactic mid plane. Both of these effects are accounted for in our occurrence rate methodology either directly (e.g., the number of planet-search stars) or indirectly (e.g., the expected $S/N$ for a transiting planet with a given period and radius). Therefore, we believe that the choice of observed fields does not impart a significant bias that is not already accounted for in our methodology.

We applied a further series of criteria to ensure that our sample is well suited to the ASPCAP analysis and completeness model we employ in \S\ref{subsec:completeness}, and to remove stars that are evolved or likely to be a member of a binary system. To select this sample, we make use of the stellar properties derived by \cite{berger2018,berger2020} to apply the following cuts: 

\begin{enumerate}
    \item \textit{Effective Temperature Cut,}~$4700\,\mathrm{K} < T_\mathrm{eff,\,B20} < 6360 \, \mathrm{K}$:\ We remove stars outside the temperature range well-suited to the ASPCAP analysis (4700-6200~K; see \S\ref{sec:xfetrends}). 
    However, to account for systematic offsets in the \cite{berger2020} temperature scale and the ASPCAP temperature scale, we incorporate into our selection the median $\teff$ offset for stars with ASPCAP-derived $\teff$ between 4600-4800~K, and 6100-6300~K. In the former sample there is a negligible offset (B20-ASPCAP) of $-$1~K, and in the latter there is a more significant offset of $+$160~K.
    \item \textit{Maximum Transit Duration Cut,}~$t_\mathrm{dur,\,max} < 15\,\mathrm{hr}$: Because the \kepler\ Transiting Planet Search module \citep[TPS;][]{twicken2016} doesn't include transit durations, $t_\mathrm{dur} > 15$~hr, we remove stars that can reasonably include such long duration transits from our planet-search sample. This criterion is logically analogous to removing evolved stars from the planet search sample. This is typical in \kepler\ occurrence rate studies, usually as a recommendation to removing stars with large radius, such as $R_\star\gtrsim1.25 R_\odot$, when applying empirical measurements of the \kepler\ pipeline detection efficiency \citep{christiansen2015,christiansen2016,christiansen2017, burke&catanzarite2017c}. To determine such stars, we employ the following approximation for the transit duration of a planet assuming a circular orbit and impact parameter of $b=0$, with a given period, $P$,
    \begin{align}\label{eq:tdur}
        t_\mathrm{dur} \approx 1.426 \, \mathrm{hr} \,  \left(\frac{\rho_\star}{\rho_\odot} \right)^{-1/3} \left(\frac{P}{\mathrm{days}} \right)^{1/3}   ,
    \end{align}
    where $\rho_\star$~is the mean density of the star. Finally, $t_\mathrm{dur,\,max}$ is obtained by setting $P=300$~days.  The motivation behind setting a limit of 300 days is to avoid regions of parameter space where planets would have fewer transits and as a result may introduce a higher rate of false alarms in our sample, which for this work we assume is negligible.
    \item \textit{Astrometric Noise Cut,}~$RUWE < 1.2$: We utilize the Renormalized Unit Weight Error ($RUWE$) from \gaia~DR2 provided in \cite{berger2020} to remove stars that are likely to show signs of multiplicity. The $RUWE$ parameter is a combination of goodness of fit metrics that quantifies deviations of a given star's sky motion from a 5-parameter astrometric solution. Single stars are expected to show a Gaussian distribution centered at $RUWE=1$, which suggests that sources with $RUWE$ significantly greater that that expected from a Gaussian distribution are likely to have companions that induce detectable centroid offsets in the \gaia~DR2 astrometric pipeline. Following the motivation from \cite{bryson2020a}, we choose $RUWE<1.2$ as our cutoff to be the limit above which we would reliably expect stars to be binaries. 
    \item \textit{Likely Binary Cut}, \texttt{BinFlag} $\ne$~1 \texttt{or}~3: We remove stars that are likely to be binaries, as determined by \cite{berger2018}. \cite{berger2018} use \texttt{BinFlag}=1 or \texttt{BinFlag}=3 to denote a star likely to be a binary due to its inferred radius. We do not remove stars with \texttt{BinFlag}=2, which are stars likely to be binaries as determined from high-resolution AO or speckle imaging, because those data are only available for a small subset of the planet search sample, and removing such stars is likely to create a bias. 
\end{enumerate}

After applying these cuts we are left with 22,146 stars in \ssamp. This defines our planet-search sample, with stars that have typical masses ranging from 0.7-1.3~$M_\odot$, and distances ranging from 100-2000 pc.

\subsection{\csamp: APOGEE-\kepler\ ``Control" Sample}

In addition to the KOIs that were observed in the APOGEE-KOI program, a number of stars were chosen to fill the APOGEE plates as a control sample for the purpose of comparing the chemistry of stars with and without detected transiting planets. The control sample was chosen to reflect the bulk properties of the KOI sample by matching the joint distributions of effective temperatures, $H$-band magnitudes, and $\log g$ from the \kepler\ Input Catalog \citep[KIC;][]{brown2011}. It is from this sample of stars that we construct \csamp.

 At this point, we want to emphasize the purpose of \csamp. \csamp\ is used solely to infer the abundance distributions of \ssamp. Therefore, there are two requirements needed to ensure that \csamp\ is representative of the abundances of \ssamp. First, it must broadly reflect the Galactic coordinates, distances, masses, and ages of the stars in \ssamp, properties that are known to correlate with chemical abundance distributions \cite[see e.g.,][]{hayden2015}.
 The second criterion is that there must not be systematic differences that would bias the ASPCAP analysis. 
 For example, differences in $S/N$, $T_\mathrm{eff}$, and $\log g$ may all lead to systematic offsets in the derived abundances that could lead one to conclude there are differences in the underlying distributions when that is not truly the case.

Because \csamp~ already reflects \ssamp~ in terms of Galactic coordinates, distances, and $H$-mag (and therefore $S/N$) by its very construction, we only need to apply the cuts that ensure the stars in \csamp\ are amenable to the ASPCAP analysis, and that they reflect the ages and masses of the stars of interest.
Therefore, we apply the \textit{Maximum Transit Duration Cut} and the \textit{Effective Temperature Cut}, because differences in the distribution of stellar densities (and therefore $\log g$) can be indicators of age differences, and differences in effective temperature are most likely to lead to systematic offsets in the derived abundances. After these two cuts, we are left with 72 stars in \csamp. 
Chemical abundances and other stellar parameters for the stars in \csamp\ are listed in Table \ref{tab:csamp}.

\begin{figure*}
\centering
\includegraphics{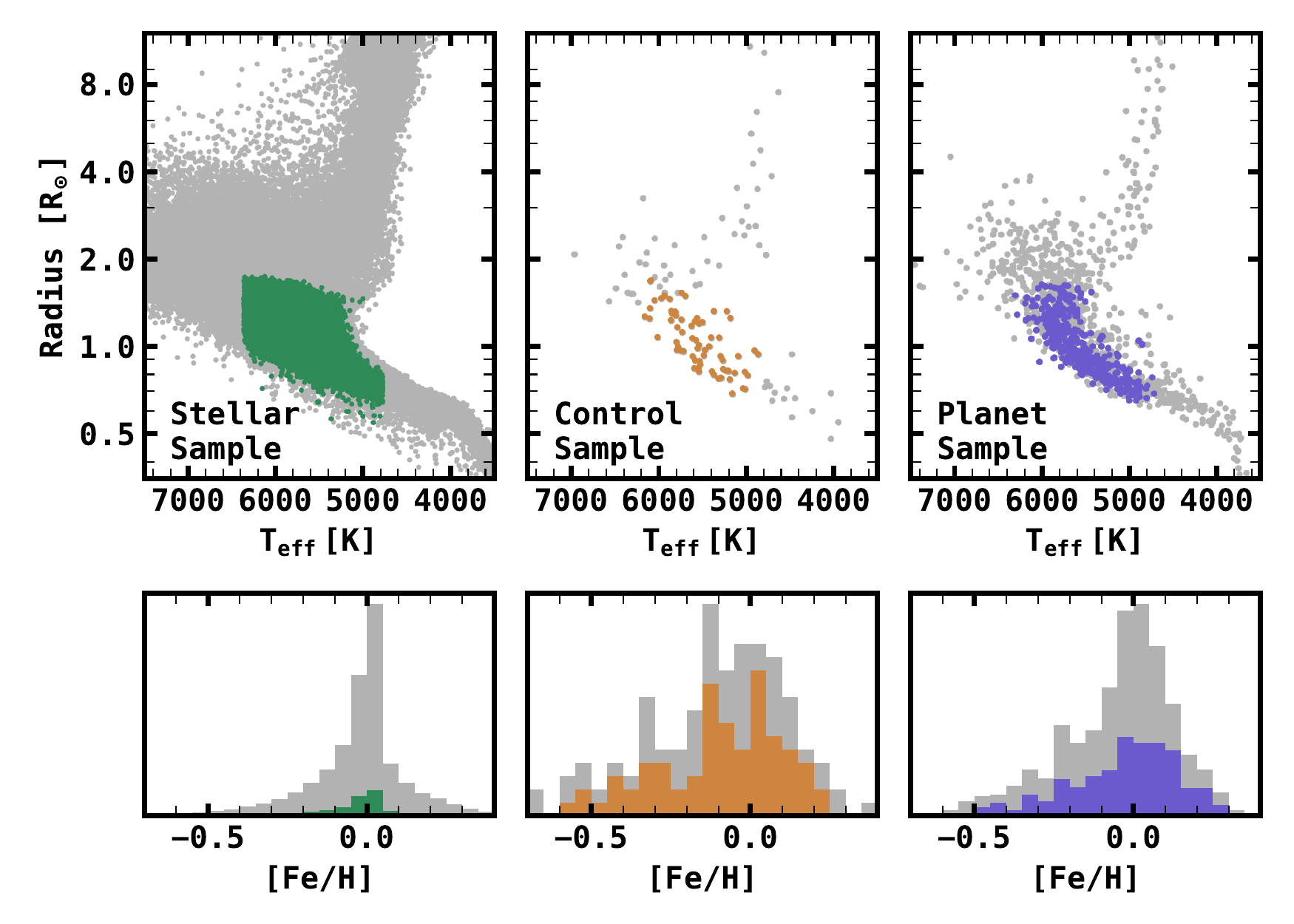}
\caption{The three samples considered in this study. The effective temperature and radii of the stars in each sample are shown along the top row, and the metallicity distribution function for each sample is shown along the bottom row. The metallicity distributions are scaled to arbitrary units. \textit{Left}: \kepler\ field stars with parameters derived in \cite{berger2020}.  The stars cut from \ssamp\ are shown in gray, and those included in \ssamp\ are shown in green. The metallicities for the stars in \ssamp\ are heterogeneous, or assumed to be solar, and thus are not as reliable for this study. \textit{Center}: The stars in the Control sample (gray), and the subset of these stars included in \csamp\ (tan). \textit{Right}: All the stars in the APOGEE-KOI program (gray) and the stars included in \psamp\  (purple). }
\label{fig:all_samples}
\end{figure*}

\begin{table*}[h]
\centering
    \caption{Derived properties and ASPCAP-derived chemical abundances for each star in \csamp. (This table is available in its entirety in machine-readable form)}
    \begin{tabular}{llll}\hline \hline
    Label & Column Description
    & Label & Column Description \\ \hline
APOGEE\_ID & Unique APOGEE Identifier
 &
Teff &    Effective Temperature in K
 \\
Teff\_e &  16th percentile of Teff posterior  
 &
Teff\_E &  84th percentile of Teff posterior  
 \\
logg &  logarithm of the surface gravity in cm/s$^2$
 &
logg\_e &  16th percentile of logg posterior  
 \\
logg\_E &  84th percentile of logg posterior  
 &
mass &   Stellar Mass in $M_\odot$
 \\
mass\_e &  16th percentile of mass posterior  
 &
mass\_E &  84th percentile of mass posterior  
 \\
radius & Stellar radius in $R_\odot$
 &
radius\_e &  16th percentile of radius posterior  
 \\
radius\_E &  84th percentile of radius posterior  
 &
Fe\_H &   [Fe/H] in dex
 \\
Fe\_H\_ERR &  Gaussian uncertainty of Fe\_H  
 &
Ni\_Fe &   [Ni/Fe] in dex
 \\
Ni\_Fe\_ERR &  Gaussian uncertainty of Ni\_Fe  
 &
Si\_Fe &    [Si/Fe] in dex
 \\
Si\_Fe\_ERR &  Gaussian uncertainty of Si\_Fe  
 &
Mg\_Fe &    [Mg/Fe] in dex
 \\
Mg\_Fe\_ERR &  Gaussian uncertainty of Mg\_Fe  
 &
C\_Fe &    [C/Fe] in dex
 \\
C\_Fe\_ERR &  Gaussian uncertainty of CI\_Fe  
 &
Al\_Fe & [Al/Fe] in dex
 \\
Al\_Fe\_ERR &  Gaussian uncertainty of Al\_Fe  
 &
Ca\_Fe &    [Ca/Fe] in dex
 \\
Ca\_Fe\_ERR &  Gaussian uncertainty of Ca\_Fe  
 &
Mn\_Fe &    [Mn/Fe] in dex
 \\
Mn\_Fe\_ERR &  Gaussian uncertainty of Mn\_Fe  
 &
S\_Fe &    [S/Fe] in dex
 \\
S\_Fe\_ERR &  Gaussian uncertainty of S\_Fe  
 &
K\_Fe &    [K/Fe] in dex
 \\
K\_Fe\_ERR &  Gaussian uncertainty of K\_Fe  
 & \\ \hline
    \end{tabular}
    \label{tab:csamp}
\end{table*}

\subsection{\psamp: Vetted Planet Sample}\label{sec:psamp}

To ensure that we have a high purity planet sample, we apply an additional series of cuts to the planet candidates designed to remove False Positive detections, remove planets where the transit depth, and therefore planet radius measurement, may not be accurate, and to restrict our sample to the parameter space well-defined by our completeness correction model (\S\ref{subsec:completeness}). We define and motivate each of these cuts below. 

\begin{enumerate}

    \item \textit{ASPCAP Solution Cut:} First, we remove planet candidates whose host stars do not have a reliable ASPCAP solution. This cut was already implicitly made when adopting the stellar and planetary radii, but we repeat it here for emphasis. Because we are interested in measuring planet occurrence rates and their change with chemical abundances, we restrict our sample to stars for which the ASPCAP pipeline has derived a reliable solution to the spectroscopic fit. Spectra that do not have such a fit will not have derived abundances and are therefore not appropriate to include in our analysis. We correct for this bias in \S\ref{subsec:completeness}. 

    \item \textit{Reliability Cut}: To remove as many contaminants from \psamp, we remove all planet candidates with a False Positive disposition in the DR24 KOI catalog.
    
    \item \textit{Impact Parameter Cut,} $b<0.9$: We remove all planet candidates with impact parameter, $b>0.9$, as measured in the DR24 KOI catalog. Modeling transits with large impact parameters leads to greater uncertainties in the transit depth and therefore planet radius of the sample. Thus, we remove planet candidates with large impact parameters to ensure that we have a sample of planets with well-measured radii. 

    \item \textit{Planet Radius Cut,}~$R_p < 23 \, R_\oplus$: We place an upper limit on the radius of a planet candidate in our sample of 23\,\earthrad\ (2.1 $R_\mathrm{Jup}$), which is consistent with the radius of the largest confirmed transiting exoplanet currently known, HAT-P-67b \citep{zhou2017}. While inflated Hot Jupiters are known to have radii as large as $\sim2 R_\mathrm{Jup}$, most objects with radii larger than $2 R_\mathrm{Jup}$ are more likely to be very low-mass stars. 

    \item \textit{Excess RV Variability Cut}, $\epsilon_{RV} < \mathbf{5.3}$: To remove EBs and eclipsing brown dwarfs from \psamp, we define a metric for excess RV variability, $\epsilon_{RV}$, as
    \begin{equation}
        \epsilon_{RV} \equiv MAD(RV)/\sigma_{RV}\;,
    \end{equation}
    where $MAD(RV)$~is $1.4826\times$ the median absolute deviation of the individual RV measurements, and $\sigma_{RV}$~is the median RV uncertainty for all epochs. To estimate $\sigma_{RV}$, we add the reported RV uncertainty for each visit in quadrature with $\sigma_{RV,\,\mathrm{min}}=72\; \mathrm{m\,s^{-1}}$, which has been noted as a reliable lower limit on the relative RV error for high S/N observations in DR16, where the reported error may be underestimated \citep{pricewhelan2020}. 
    Given the varying brightness of our targets, the RV uncertainties are highly correlated with the single epoch spectrum $S/N$. As a result, a flat cut in the scatter of the RV measurements could remove bonafide planet candidates with dim host stars, while missing astrophysical False Positives around bright host stars. $\epsilon_{RV}$, therefore, gives a more accurate assessment of whether a given star is RV-variable than a flat cut in the scatter of the RV measurements. We decide on $\epsilon_{RV}=5.3$ because that is equal to the median plus thrice the MAD in our sample. 
    APOGEE RV observations in the KOI sample are capable of placing upper limits into the planetary mass regime, typically between 1-10$\,\mathrm{M_{jup}}$, depending on the orbital period of the transiting planet, spectrum $S/N$ at each epoch, and mass of the host star. Therefore, by removing all stars with significant RV variability in our sample, we in turn remove any contaminating eclipsing binaries. APOGEE's RV precision is not quite effective enough to detect planetary mass companions without detailed modeling, so our metric for RV variability is not likely to remove any real planets, such as hot Jupiters. We justify this statement briefly with out results in \S\ref{sec:occrates_p_rp}.
\end{enumerate}

\begin{figure}
\centering
\includegraphics{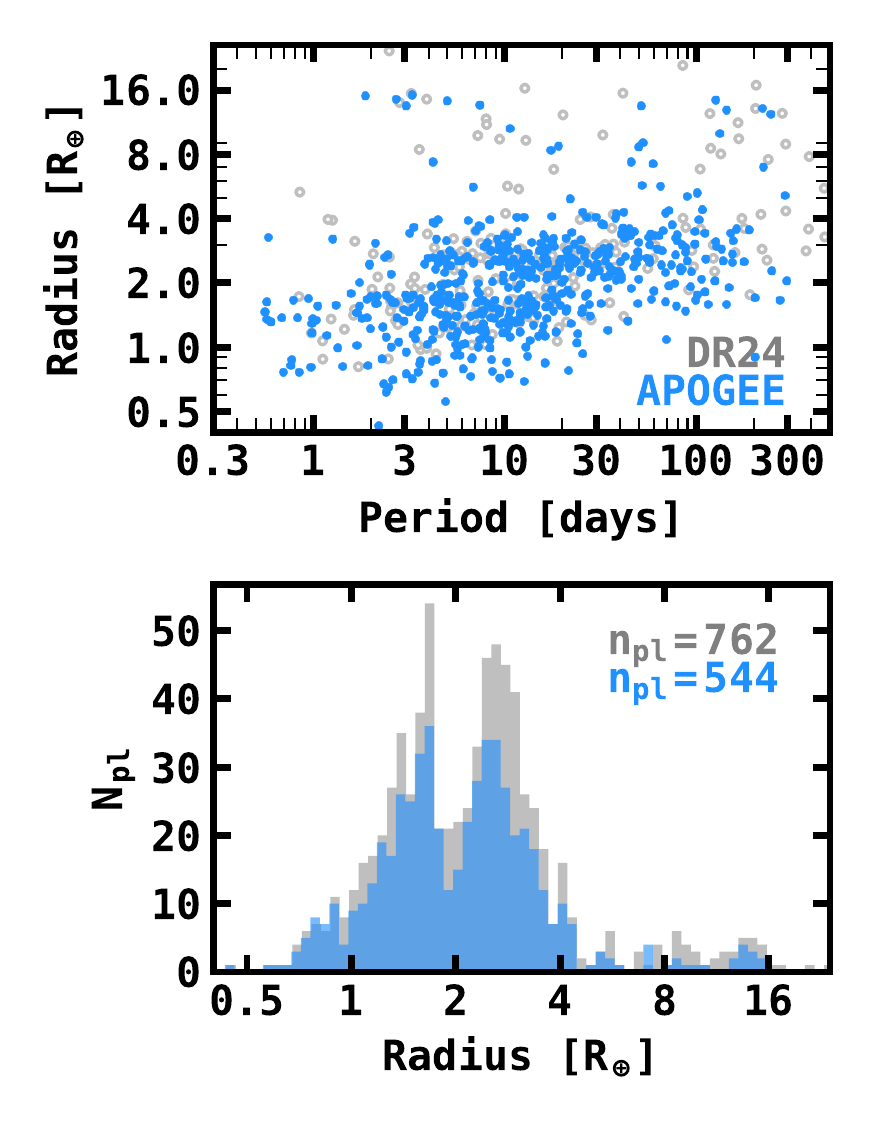}
\caption{The planets in \psamp, plotted with all the DR24 planet candidates that have a host in \ssamp. \textit{Top}: The planet radius and orbital period of all planets in \psamp. The gray points show all the planets from the DR24 KOI catalog with a host star in \ssamp\ that are not included in \psamp. \textit{Bottom}: The radius distribution of the planets in \psamp. The gray histogram shows the radii of all the planets in DR24 with a host in \ssamp, while those in \psamp\ are displayed in blue. The primary reasons for exclusion in \psamp\ are RV variability, a poor solution from ASPCAP, or pre-DR24 target selection. }
\label{fig:planetsample}
\end{figure}

After these cuts  \psamp\ consists of 544 total planet candidates. The radius and period characteristics of these candidates are shown in Figure \ref{fig:planetsample}. There are a number of features evident in this figure. For instance, the radius gap \citep{fulton2017} is clear in both the top and bottom panels of our figure, as well as a slope in orbital period in the gap measured by previous authors \citep{fulton&petigura2018, martinez2019}; these two features qualitatively validate the precision and accuracy of the radii in \psamp. Chemical abundances and planet parameters for the planet candidates in \psamp\ are listed in Table \ref{tab:psamp}.

\begin{table*}[h]
\centering
    \caption{Planet properties and ASPCAP-derived host star chemical abundances for each planet candidate in \psamp. (This table is available in its entirety in machine-readable form)}
    \begin{tabular}{llll}\hline \hline
    Label & Column Description
    & Label & Column Description \\ \hline
APOGEE\_ID &    Unique APOGEE Identifier
 &
KIC &    KIC identifier
 \\
KOI\_ID &   KOI identifier
 &
Period & Planet orbital period in days  
 \\
Rpl &   Planet radius in $R_\oplus$ 
 &
Rpl\_ERR &    Gaussian uncertainty of Rpl 
\\
 Fe\_H &    Host star [Fe/H] in dex
 &
Fe\_H\_ERR &  Gaussian uncertainty of Fe\_H  
 \\
Ni\_Fe & Host star [Ni/Fe] in dex
 &
Ni\_Fe\_ERR &  Gaussian uncertainty of Ni\_Fe  
 \\
Si\_Fe &   Host star  [Si/Fe] in dex
 &
Si\_Fe\_ERR &  Gaussian uncertainty of Si\_Fe  
 \\
Mg\_Fe &  Host star   [Mg/Fe] in dex
 &
Mg\_Fe\_ERR &  Gaussian uncertainty of Mg\_Fe  
 \\
C\_Fe &   Host star  [C/Fe]  in dex
 &
C\_Fe\_ERR &  Gaussian uncertainty of C\_Fe  
 \\
Al\_Fe & Host star    [Al/Fe]  in dex
 &
Al\_Fe\_ERR &  Gaussian uncertainty of Al\_Fe  
 \\
Ca\_Fe &  Host star   [Ca/Fe]  in dex
 &
Ca\_Fe\_ERR &  Gaussian uncertainty of Ca\_Fe  
 \\
Mn\_Fe &  Host star   [Mn/Fe] in dex
 &
Mn\_Fe\_ERR &  Gaussian uncertainty of Mn\_Fe  
 \\
S\_Fe &  Host star   [S/Fe] in dex
 &
S\_Fe\_ERR &  Gaussian uncertainty of S\_Fe  
 \\
K\_Fe &  Host star   [K/Fe] in dex
 &
K\_Fe\_ERR &  Gaussian uncertainty of K\_Fe  
 \\ \hline
    \end{tabular}
    \label{tab:psamp}
\end{table*}

\newcommand{\rowstyle}[1]{\gdef\currentrowstyle{#1}%
  #1\ignorespaces
}

\begin{table}
\centering
\begin{tabular}{cccccc}
\hline \hline
Field & $\alpha$(h:m:s) & $\delta$(d:m:s) & $n_{pl}$ & $n_\star$ & $F_\star$ \\
\hline
K04 & 19:42:47 & 49:54:07 & 72 & 3546 & 0.16 \\
K06 & 19:13:39 & 46:52:30 & 89 & 3116 & 0.141 \\
K07 & 19:00:17 & 45:12:46 & 74 & 2822 & 0.127 \\
K10 & 19:36:30 & 46:00:18 & 107 & 4297 & 0.194 \\
K16 & 19:31:05 & 42:05:24 & 93 & 4510 & 0.204 \\
K21 & 19:26:13 & 38:09:36 & 109 & 3855 & 0.174 \\
{\bf All} & N/A & N/A & {\bf 544} & {\bf 22,146} & {\bf 1.00} \\
\hline
\end{tabular}
\caption{The coordinates, number of stars in \ssamp, number of planets in \psamp, and fraction of stars in \ssamp\ for each APOGEE-KOI field.}
\label{tab:fields}
\end{table}

\subsection{Adopted Planet Classes}

We divide the planets in \psamp\ into multiple classes based on their orbital period and radius, as many previous studies have shown metallicity correlations that depend on these properties. The adopted planet size classes are motivated partially by empirical and theoretical boundaries where applicable, and partially by conventions in the literature, as explained below. 
For the planet size classes, we define the following:

\begin{enumerate}

\item \textit{Sub-Earths}, $R_p < 1 R_\oplus$: The number of planets in this class suffers particularly severely from low survey completeness, and for that reason these planets are drastically skewed toward lower orbital periods. Because of this, we don't consider these planets when measuring occurrence rates, and are hesitant to draw major conclusions when comparing the abundances of their host stars to those of stars in \csamp. There are 42 Sub-Earths in \psamp.

\item \textit{Super-Earths}, $1.0 \ R_\oplus\leq R_p < 1.9 R_\oplus$: Super-Earths are defined as planets larger than Earth, with an upper limit set by the minimum in the planet radius distribution between 1-4 $R_\oplus$ in our sample (Figure \ref{fig:planetsample}). The $1.9\,R_\oplus$~boundary we find between Super-Earths and Sub-Neptunes is slightly different than that found by \cite{fulton2017}, and closer to the boundary found by \cite{martinez2019}. 
There are 212 Super-Earths in \psamp.

\item  \textit{Sub-Neptunes}, $1.9\, R_\oplus\leq R_p < 4 \,R_\oplus$: The lower boundary is driven by the radius gap as discussed above. The upper boundary is placed as the limit where the occurrence of Sub-Neptunes tends to zero. While a more precise physically-motivated boundary is not clear, we choose 4$R_\oplus$ as an upper limit to be consistent with conventions in the literature.  There are 260 Sub-Neptunes in \psamp.

\item  \textit{Sub-Saturns}, $4\, R_\oplus\leq R_p < 8 \, R_\oplus$: The lower radius boundary for Sub-Saturns is given by the decrease in Sub-Neptune occurrence rates described above, and the upper limit is driven by the approximate radius at which planets are typically $\gtrsim100 M_\oplus$ \citep{petigura2017b}. There are 13 Sub-Saturns in \psamp.

\item \textit{Jupiters}, $8\ R_\oplus\leq R_p < 23 \,R_\oplus$: The radius range for Jupiter-sized planets is given by the upper boundary for Sub-Saturns, and by the upper limit placed by the largest known confirmed planet, as mentioned in \S\ref{sec:psamp}. There are 17 Jupiters in \psamp. 

\end{enumerate}

In addition to these size classes, we also define three different period boundaries for planets of differing orbital separations (i.e., orbital period). 

\begin{enumerate}
\item  \textit{Hot}, $P\leq10$~days\footnote{Note: For the occurrence rate analyses, our definition of hot planets doesn't include planets with $P<1$~day, due to the lack of injections used to test the \kepler\ pipeline completeness at these short periods (see \S\ref{subsec:completeness} and Figure \ref{fig:pdet_mes}).}: There is a well-documented break in the occurrence rate of planets with respect to orbital period, showing two different regimes above and below $P\sim10$~days \citep{youdin2011,howard2012,mulders2015a}. There are 248 hot planets in \psamp.

\item  \textit{Warm}, $10< P \leq100$~days: The boundary for warm planets is given by the lower bound on hot planets, and on the upper end where completeness becomes an issue for Super-Earths. This range of orbital periods is also consistently used in the literature, so we adopt it as well for ease of comparison. There are 262 warm planets in \psamp.

\item  \textit{Cool}, $100<P\leq300$~days: We define this period range  as our cool sample. The number of planets in this range suffers severely from decreased \kepler\ survey efficiency, and only contains 34 planets in \psamp. In addition, studying the population of \kepler\ planets with  $P\gtrsim300$ days requires a careful approach to modeling the \kepler\ False Alarm rate, which we assume to be negligible \citep{bryson2020}. 
 
\end{enumerate}

We refer to these classes often throughout the rest of this work.

\section{Results}\label{sec:allresults}

\subsection{Assessment of Differences Between Host Star Abundances and the Field}\label{sec:hoststars}

\begin{figure*}
\centering
\includegraphics{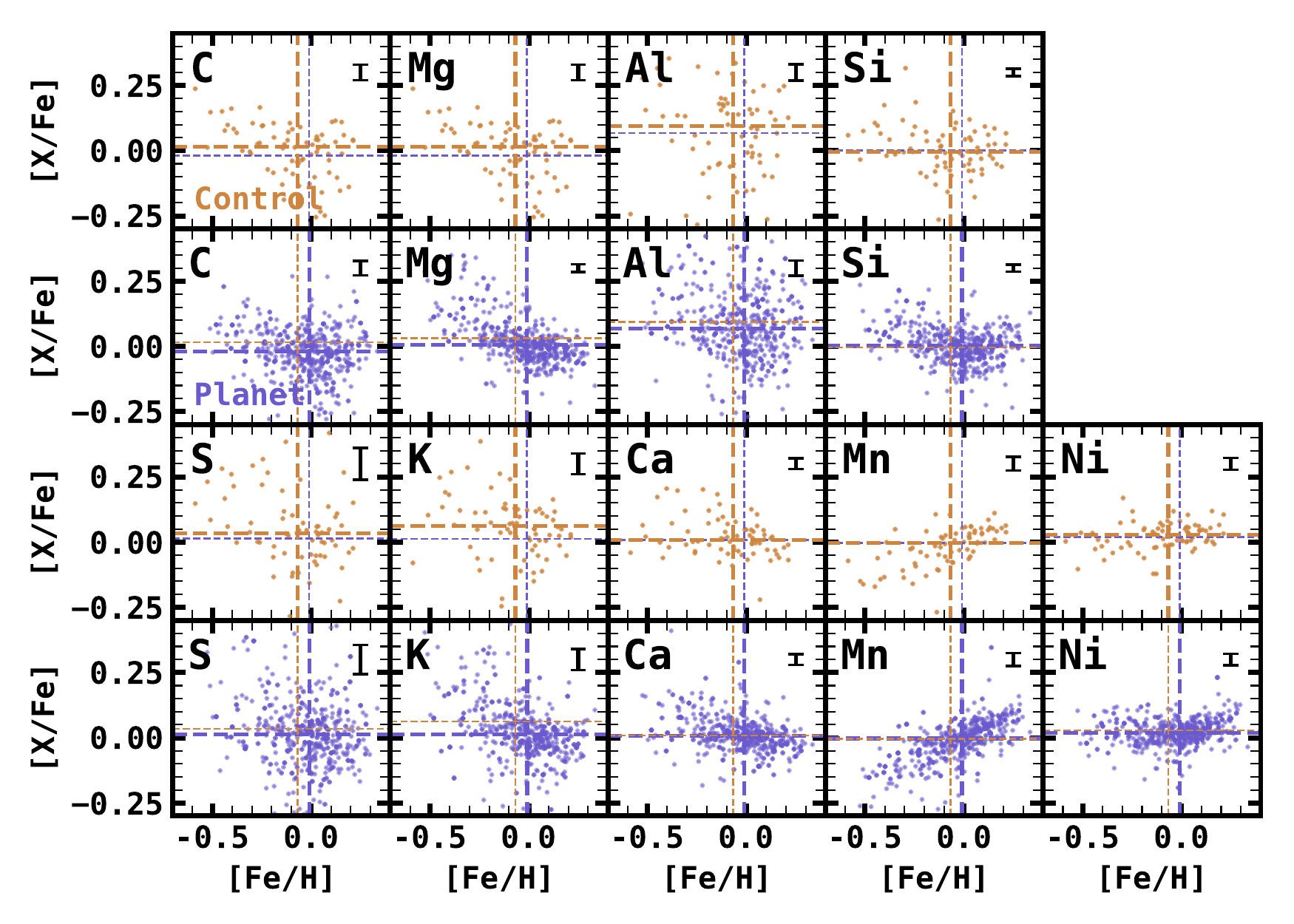}
\caption{Chemical abundances for the planet host (purple) and control (tan) samples. The chemical abundance displayed is shown in the upper left corner of each panel. The median error ($\pm 1\sigma$) for each abundance is shown by the black error bar in the top right corner of each panel, and the dashed lines indicate the median abundances for the planet host sample (purple) and the control sample (tan). }
\label{fig:abundances2d}
\end{figure*}

In this section we examine whether there are any clear correlations with planet type and host chemical abundance. We also make more detailed comparisons between the abundances of \csamp\  and \psamp. The chemical abundances of both \psamp\ and \csamp\ are shown in Figure \ref{fig:abundances2d}. For this section, we rely on the abundance ratios to Fe, [X/Fe], because there is a clear offset in [Fe/H] between \csamp\ and \psamp\ visible in Figure \ref{fig:abundances2d}, where stars in \csamp\ are more metal-poor on average. This is a well-known property of the stars with known transiting planets when compared to the stars in the \kepler\ field. Because of this difference, using [X/H] as a metric is almost certainly guaranteed to reproduce the [Fe/H] differences already known, and our goal is to search for new differences.

\begin{table}
    \centering
    \begin{tabular}{ccc} \hline  \hline
[X$_i$/Fe] & \csamp & \psamp \\ \hline 
Fe$^a$  & -0.068$\pm$0.183  & -0.010 $\pm$ 0.163 \\
C  & 0.015$\pm$0.097  & -0.019 $\pm$ 0.079 \\
Mg  & 0.031$\pm$0.082  & 0.006 $\pm$ 0.060 \\
Al  & 0.094$\pm$0.201  & 0.067 $\pm$ 0.122 \\
Si  & -0.004$\pm$0.090  & 0.002 $\pm$ 0.058 \\
S  & 0.034$\pm$0.125  & 0.013 $\pm$ 0.099 \\
K  & 0.062$\pm$0.096  & 0.012 $\pm$ 0.076 \\
Ca  & 0.008$\pm$0.059  & 0.008 $\pm$ 0.046 \\
Mn  & -0.004$\pm$0.077  & -0.003 $\pm$ 0.074 \\
Ni  & 0.028$\pm$0.044  & 0.019 $\pm$ 0.041 \\
\hline  
    \end{tabular}
    \caption{The median and mean absolute deviation of each abundance distribution in \csamp\ and \psamp. $^a$For iron, the abundance is reported with respect to Hydrogen, [Fe/H]}
    \label{tab:medians}
\end{table}

After defining the planet size and orbital period classes above, the first natural question is whether hosts of differing planet classes tend toward specific abundance patterns. Therefore, to detect any differences in the distribution of the host star abundances and the abundances of general stars in the field, we apply four unique statistical tests, considering a result significant if the $p$-value for the statistic is $<$0.001. Given the large number of tests between planet subclass and each of the ten elemental abundances considered (160 tests), $p<0.001$ should give a $\lesssim$10\% probability that a false positive is among these results. 
The results of these tests are shown in Table \ref{tab:tests}, and for the sake of brevity they are discussed further in the Appendix (\ref{sec:appstats}). In short, we find no new credible differences, according to these tests, between the chemistry of stars in \csamp\ and those in \psamp\ that are not easily explained by already known trends between planet properties and the metallicities of their host stars \citep{santos2004,valenti&fischer2005,ghezzi2010,ghezzi2018,buchhave2014,schlaufman2015, mulders2016,wilson2018,petigura2018,narang2018,owen&murrayclay2018,ghezzi2021}.

\subsection{Abundance Trends with Planet Period and Radius}

In this section, we test whether there are any correlations between the host star abundances and planet properties. While these correlations can reveal important trends, it is important to note that the trends discussed in this section do not take completeness or detection biases into account. When appropriate, we mention when we believe an effect may be a result of a lack of completeness. A more thorough investigation would include correcting for biases in the \kepler\ and APOGEE-KOI surveys, which is performed in \S\ref{sec:results}.

\begin{figure*}
\centering
\includegraphics{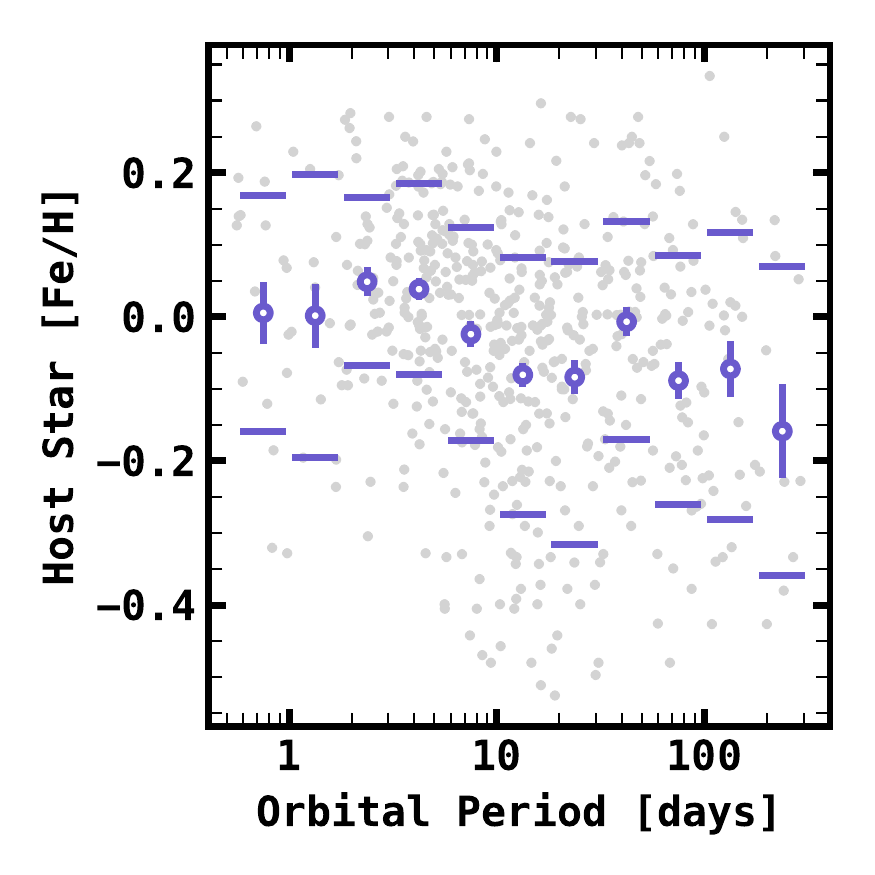}
\includegraphics{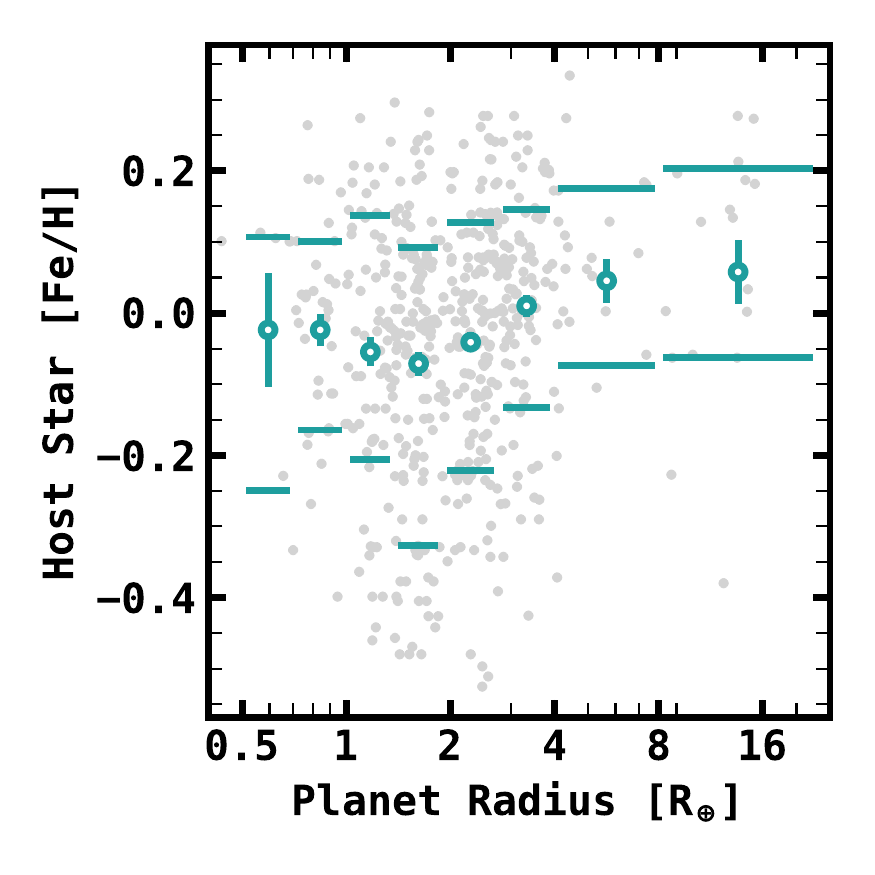}
\caption{ \textit{Left:} The average metallicity for host stars of planets in given orbital period bins. The circular points show the average metallicity, while the horizontal lines show the 68\% confidence interval on the metallicity distribution. We recover the same planet period--stellar metallicity anti-correlation reported in previous literature \citep[e.g.,][]{mulders2016, wilson2018} \textit{Right:} The average host star metallicity as a funciton of planet radius, binned for planets of given size classes, Sub-Earths, Super-Earths, Sub-Neptunes, Sub-Saturns, and Jupiters. The Sub-Earth, Super-Earth, and Sub-Neptune classes are split into two radius bins each. We find similar relations as in the literature, that there is a notable increase in the average host metallicity for planets with larger radii. In particular, there are very few planets with $R_p>4R_\oplus$~and $\langle [\mathrm{Fe/H}] \rangle < -0.2$. }
\label{fig:fehprad}
\end{figure*}

\begin{figure*}
\centering
\includegraphics{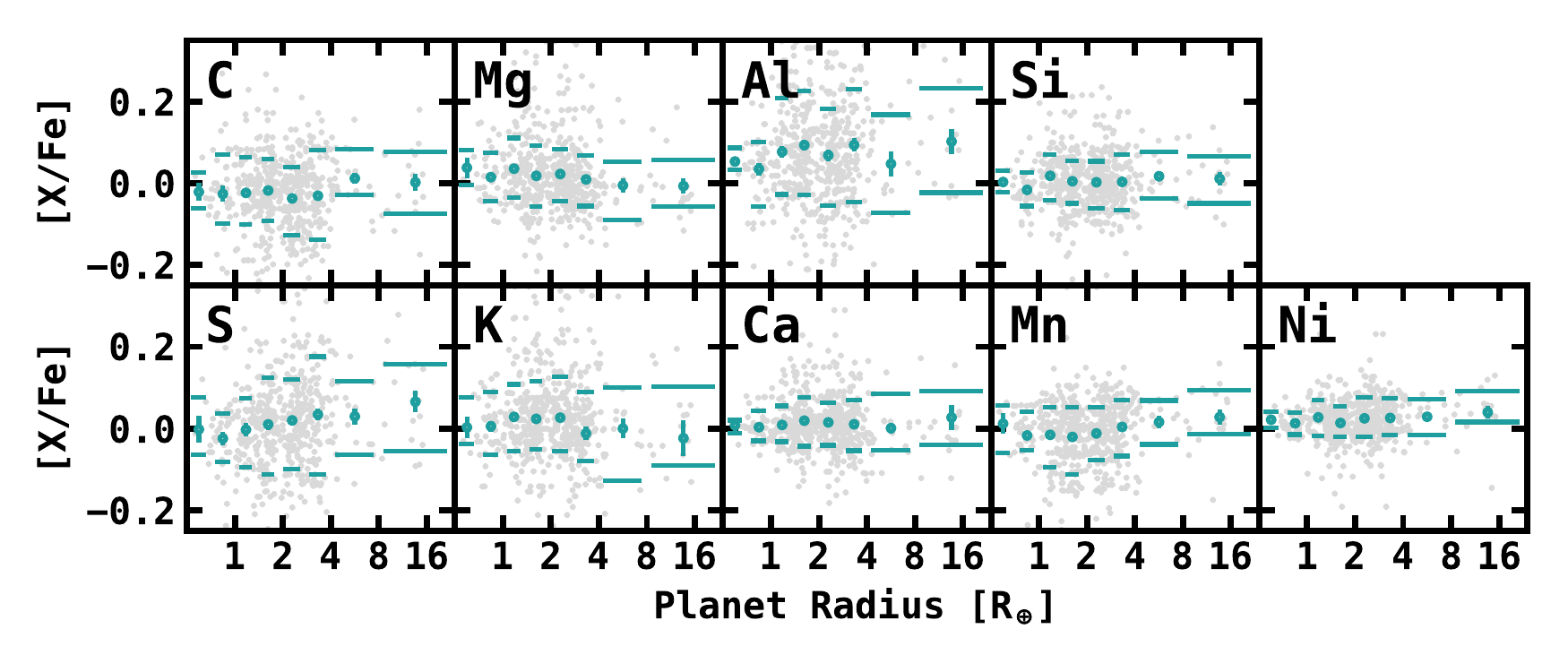}
\includegraphics{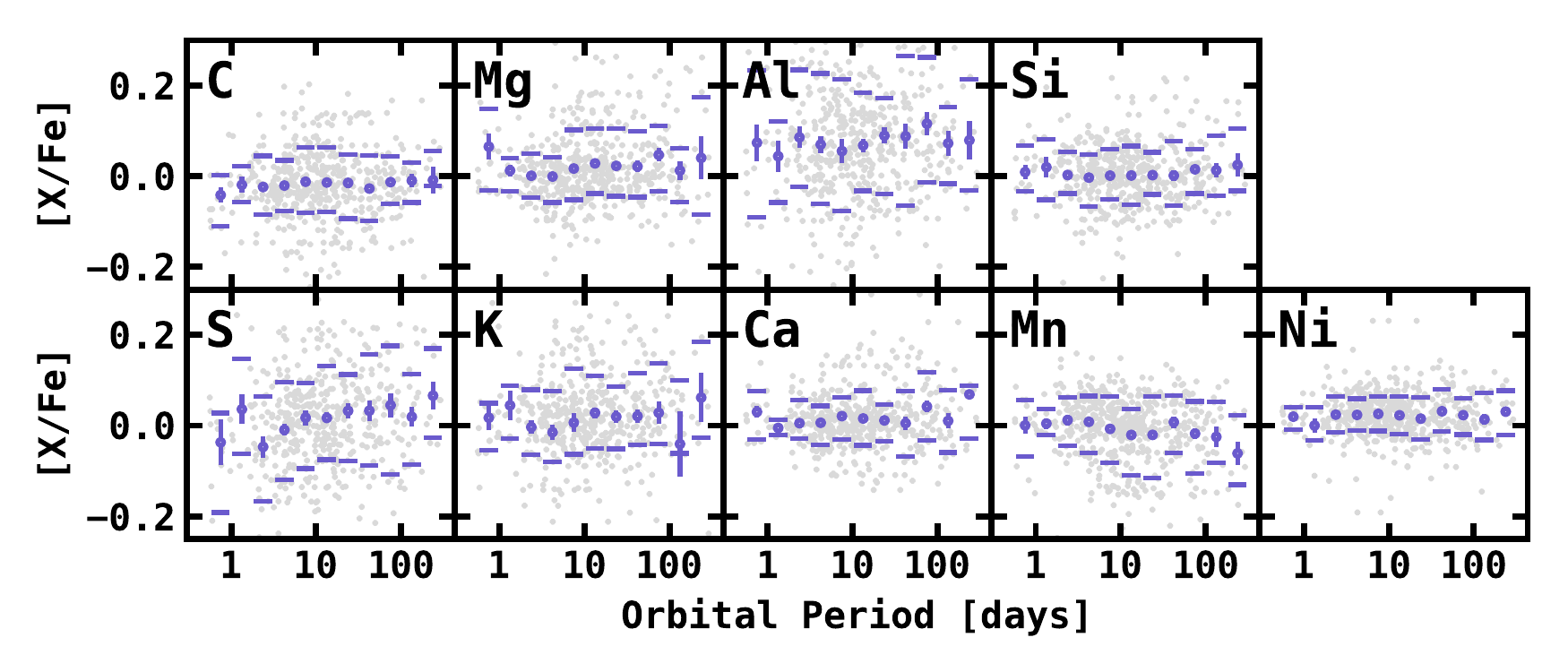}
\caption{\textit{Top:} Trends with planet radius and abundance ratios to iron. Just like with Figure \ref{fig:fehprad}, the points represent the means of each bin, with error bars representing the error on the mean [X/Fe] from bootstrapping. The horizontal lines show the 16th and 84th percentiles of the distribution in each bin to display the variance. We detect significant positive correlations between [Mn/Fe], and [S/Fe] vs. $R_p$. \textit{Bottom}: The distribution of host star abundance ratios to iron as a function of planet period. We detect a statistically significant positive correlations between [S/Fe] and $P$. Such a correlation cannot be explained with well known trends in [Fe/H]. }
\label{fig:xfe_period}
\end{figure*}

In Figures \ref{fig:fehprad} and \ref{fig:xfe_period} we plot the mean and variance of the abundance distributions for different planet radius and planet period bins. As in the literature, we recover an anti-correlation [Fe/H] of the host star and the planet orbital period. We also recover a positive correlation between the planet radius and the host star [Fe/H]. Within these broader correlations, there are a few interesting results. For instance, while there is a general anti-correlation between planet orbital period and host star [Fe/H], there is an increase in the average metallicity distribution at $P\sim30$~days. This slight increase is apparent in Figure 3 of \cite{petigura2018} as well, though to a lesser extent. This feature is also pointed out in \cite{wilson2018} as a possible transition period at $P\sim 23$~days. While the exact cause of this bump is not well-constrained by this work, we hypothesize that it is due to an increase in the relative number of Sub-Saturns at these orbital periods. Because the presence of Sub-Saturn planets are positively correlated with enhanced metallicity, and they also have increasing occurrence rate at warm orbital periods.

We also see a number of interesting trends between planet radius and host star [Fe/H]. For one, we confirm the claim made by several authors \citep{buchhave2014,schlaufman2015,wang&fischer2015,ghezzi2018,petigura2018} that larger radius planets are positively correlated with host star [Fe/H]. Digging deeper we also find a few other interesting results. For instance, there is an apparent increase in the metallicity of Sub-Earths. However, as cautioned, these planets suffer from low completeness, and are heavily skewed toward shorter periods. Thus, this bump can be explained by the stellar metallicity planet orbital period trend discussed above.

Another interesting trend we find is that Sub-Neptunes with larger radii ($R_p \sim 3$-$4\, R_\oplus$) have host stars with enhanced [Fe/H] compared to smaller Sub-Neptunes ($R_p \sim 1.9$-$3\,R_\oplus$). This is predicted by the theory of atmospheric loss via core-heating, where the radii of Sub-Neptunes are expected to increase with metallicity, $Z$, via the relation $d\log R_p / d\log Z \sim 0.1$ \citep{gupta&schlichting2019,gupta&schlichting2020}. This dependence arises from the assumption that the planet's atmospheric opacity is proportional to the metallicity of the stellar host. 
Planets with lower opacity envelopes contract on shorter timescales because these envelopes lose their residual core heat more efficiently through radiation.
As a result, one would expect that for a given age, Sub-Neptunes orbiting stars with higher metallicity will have contracted less and have larger radii on average.

To test for significant trends in our sample, we calculate the Spearman $\rho$ rank correlation coefficient between the iron normalized abundances for the planet hosts in our sample and the logarithm of the radii and periods of the planets in our sample. The results of these statistical tests are shown in Table \ref{tab:spearman}. As with the tests in the previous section, we consider a result significant if the $p$-value is $<$0.001. In this vein we uncover a few statistically significant correlations.
The most clear correlations we recover are correlations with planet radius and [Mn/Fe] and [S/Fe]. Perhaps unsurprisingly, the correlation with [Mn/Fe] is positive meaning that it is most likely influenced by statistically strong correlations with [Fe/H]. We can see from Figure \ref{fig:abundances2d}, in fact, [Mn/Fe] displays strong correlations with [Fe/H], so this is likely due to known correlations with [Fe/H].

However, the origin of the positive trend with [S/Fe] is less clear. [S/Fe] does not display the same correlation as [Mn/Fe]. Interestingly, [S/Fe] is the only abundance (though [Mn/Fe] is nearly significant for the reasons described above) that is significantly correlated with planet period as well ($p = 1.2 \times 10^{-5}$). 
Even more interesting, these correlations cannot be explained by already known trends with [Fe/H]. If that were the case, [S/Fe] would be expected to show a correlation with either planet period or radius and then must show an anti-correlation in the other, as with [Fe/H]. However, [S/Fe] shows a strong positive correlation with both planet radius and planet period.
Even more interestingly, the significant [S/Fe] trends do not appear to be the result of confounding correlations between [S/Fe] and stellar parameters that may affect the detectability of planets. [S/Fe] is not significantly correlated with $\teff$ in \psamp\ (based on a Spearman correlation test; $\rho=+0.08, p=0.065$), nor is [S/Fe] significantly correlated with $R_\star$ ($\rho=+0.10, p=0.019$). For the time being, we report this as a tentative trend, though we are still unclear of the source of this trend with S abundance-ratios.

\begin{table}[]
    \centering
    \begin{tabular}{ccccccc} 
\hline \hline 
[X/Fe] & $n_\mathrm{pl}$ & $\rho_{P}$ & $p_{P}$& $\rho_{R_p}$ & $p_{R_p}$ \\ \hline 
C & 544 &0.073&0.088& 0.021 & 0.62   \\
Mg & 544 &0.096&0.025& -0.130 & 0.0024   \\
Al & 540 &0.036&0.4& 0.046 & 0.28   \\
Si & 544 &0.037&0.39& 0.009 & 0.83   \\
S & 542 &0.187&1.2$\times$10$^{-5}$& 0.145 & 0.00069   \\
K & 540 &0.072&0.096& -0.060 & 0.16   \\
Ca & 544 &0.057&0.19& -0.032 & 0.46   \\
Mn & 544 &-0.127&0.0031& 0.161 & 0.00016   \\
Ni & 544 &-0.002&0.96& 0.117 & 0.0064   \\
\hline 
    \end{tabular}
     \caption{The results of the Spearman $\rho$ rank coefficient to test for correlations between abundance ratios to iron and $\log P$ and $\log R_p$.}
    \label{tab:spearman}
\end{table}

\subsection{Planet Occurrence as a Function of Chemical Abundance}\label{sec:results}

In this section we calculate the occurrence rates of planets as a function of $P$, $R_p$, and $[X/H]$. We fit a parametric model to describe the general trends of the planetary distribution function (PLDF) and their dependence on these properties. This analysis represents an improvement from the analysis in \S\ref{sec:hoststars}, as we are now accounting for the selection functions of \kepler\ and APOGEE; thus the conclusions we draw about the PLDF from this analysis should be independent of observational biases.

We employ a common strategy to measure the PLDF that has been used in previous studies: the number of planets per star (NPPS) is calculated over a grid of $P$ and $R_p$, utilizing the inverse detection efficiency method and a maximum likelihood approach \citep[e.g.,][]{youdin2011,fressin2013,burke2015,mulders2015a,mulders2018,petigura2018}. We give a brief description of our completeness model below, but refer the reader to the Appendix (\S\ref{sec:methods}) for details on our methodology.

\subsubsection{Completeness Model}

In this subsection we give a brief description of our completeness model, $\eta(\vect{x}, \vect{z})$, where $\vect{x}$ are planet properties and $\vect{z}$ are stellar properties, but refer the reader to the appendix for details (\ref{subsec:completeness}). 
Our approach varies slightly from most \kepler\ occurrence rate studies, because we also need to correct for biases inherent in the follow-up program. In other words, inclusion in \psamp\ is dependent on more than membership in \ssamp\ and a detected planet candidate in \kepler. There are additional biases imposed by the APOGEE selection function, instrumental setup, and spectroscopic analysis pipeline that must be considered. 
In total we account for four unique biases for a planet candidate to be included in \psamp: 
\begin{enumerate}
    \item The geometric probability that a planet with a randomly oriented orbital plane transits its host star $(p_\mathrm{tra})$
    \item The probability that a transiting planet is detected by \kepler\ $(p_{det})$,
    \item The probability that a planet candidate was observed in the APOGEE-KOI program $(p_\mathrm{apo})$
    \item The probability that ASPCAP doesn't fail to produce reliable atmospheric parameters for the host star $(1-p_\mathrm{fail})$. 
\end{enumerate}

\begin{figure*}
\centering
\includegraphics{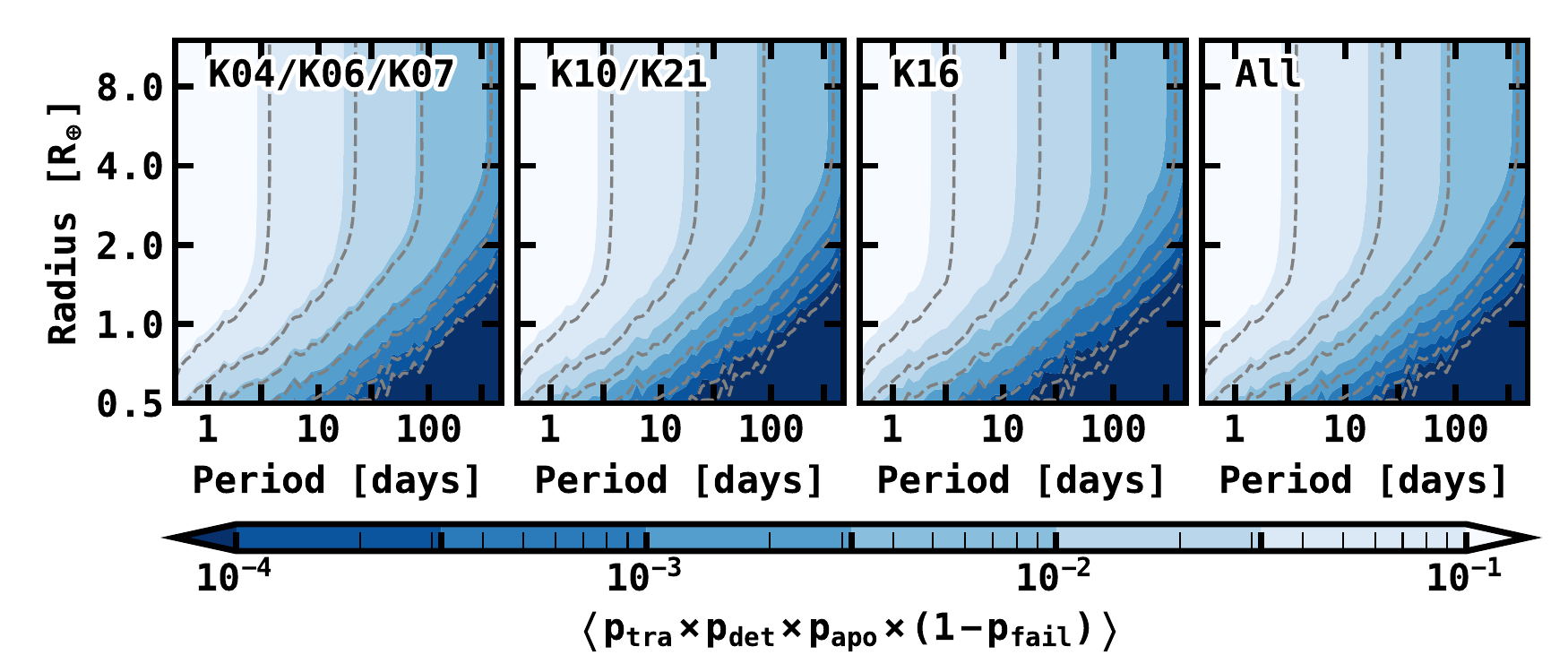}
\caption{The mean completeness model, $ \eta = p_\mathrm{tra} \times p_\mathrm{det} \times p_\mathrm{apo} \times (1-p_\mathrm{fail})$, for each APOGEE field in \ssamp, and the combined model from all fields. The blue filled contours give the survey efficiency in the $P$-$R_p$ plane, representing the probability that a given planet orbiting a star in \ssamp\ is in \psamp. The lightest shade shows where $\eta>0.1$, while the darkest shade represents survey efficiencies of $\eta<10^{-4}$. The gray dashed lines are the corresponding contours for the \kepler\ DR24 pipeline efficiency and are shown for comparison to highlight the effects of the APOGEE-KOI program selection function. The panels representing the APOGEE fields are organized from least to most divergent from the DR24 pipeline efficiency with the combined survey efficiency on the far right.}
\label{fig:pdet_all}
\end{figure*}

Assuming that each of the four terms above are independent, we calculate the total average survey efficiency for each field as the product of each term, given by
\begin{align}
    \left\langle \eta \right \rangle = \frac{1}{n_\star} \sum_i^{n_\star} p_\mathrm{tra,i} \times p_\mathrm{det,i} \times p_\mathrm{apo,i} \times (1-p_\mathrm{fail,i})\;\;,
\end{align}
where $\left \langle \eta \right \rangle$ is the average survey efficiency across \ssamp. 
The mean survey efficiency for each field is shown in Figure \ref{fig:pdet_all}. 
By marginalizing over all the stars in \ssamp\ in this way, we've removed stellar properties from our expression for survey efficiency, so that $\eta = \eta(\vect{x}) = \eta(P,R_p)$. This relies on an implicit assumption that chemical abundances are not correlated with survey efficiency. As shown in \S\ref{sec:xfetrends}, some elements show correlations between $\teff$ and abundance ratio. However, in \S\ref{sec:xfebiases} we find that this bias does not significantly affect our conclusions.

\subsubsection{Occurrence Rates in the $P$-$R_p$ Plane} \label{sec:occrates_p_rp}

We first calculate the occurrence rate of planets in the $P$-$R_p$~plane, making use of the completeness model in \S \ref{subsec:completeness}. Because we are not applying any stellar properties (i.e., abundances) for these calculations,  we calculate the occurrence rates as described in \S\ref{sec:formalism} and \S\ref{sec:fitting} for equally spaced bins in $\log P$~and $\log R_p$.

\begin{figure*}
\centering
\includegraphics[width=0.95\textwidth]{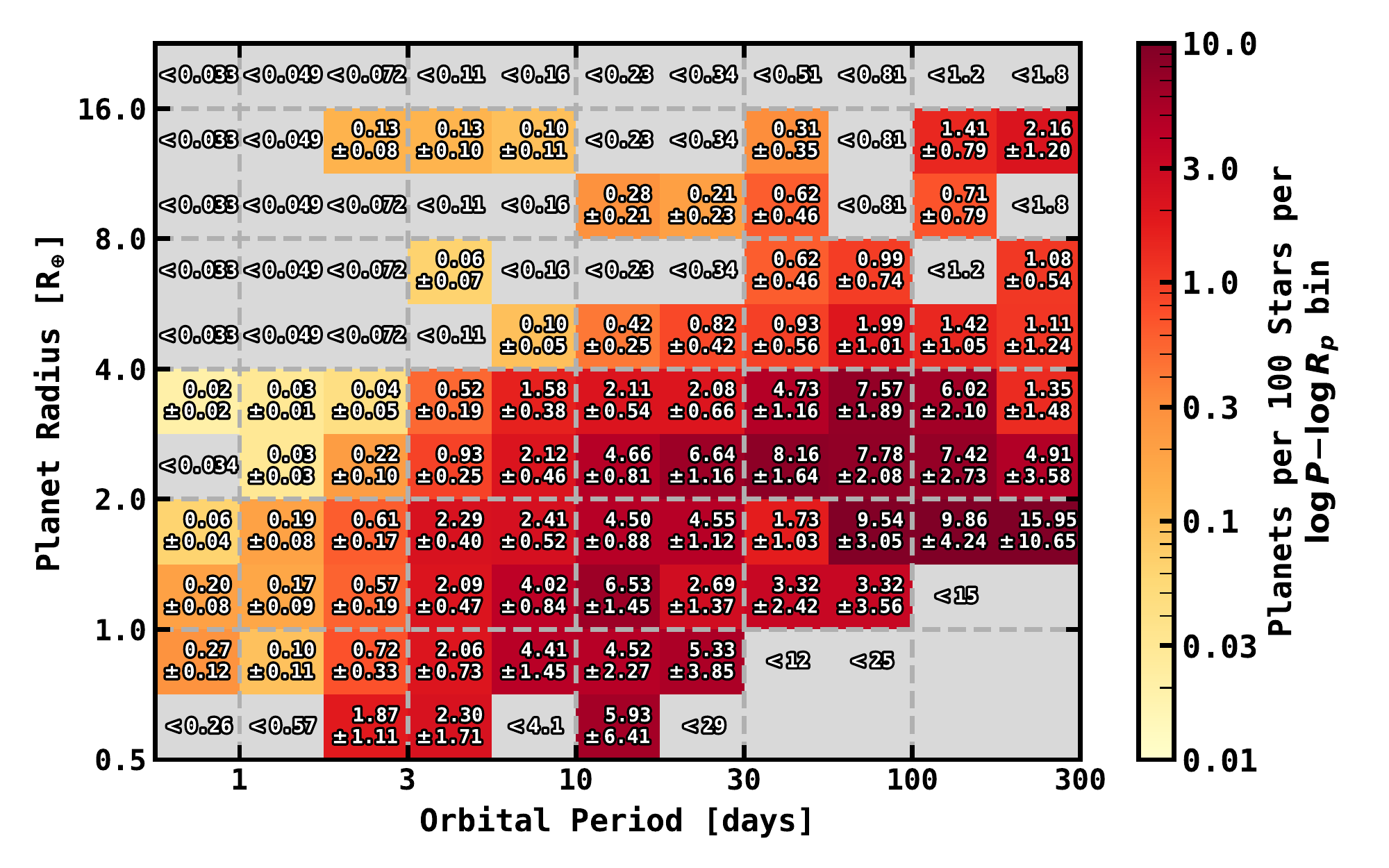}
\caption{The planet occurrence rate in the $P$-$R_p$~plane. We divide the planet into bins of size $\Delta \log P \times \Delta \log R_p = 0.25 \times 0.15$~dex. The color shows the measured occurrence rate in the bin of interest on a logarithmic scale. The gray bins do not have any detected planets. The numbers in each cell shows the occurrence rate in units of number of planets per 100 stars. The uncertainty shown is taken to half of the 68\% confidence interval range on the occurrence rate. Bins without detected planets have the upper limit displayed. Bins with no detected planets and no upper limit listed are areas of low completeness where an upper limit is not constraining. }
\label{fig:occratesgrid}
\end{figure*}

 We first divide the $P$-$R_p$ plane into logarithmic bins of $\Delta \log P \times \Delta \log R_p =$  0.25 dex $\times$ 0.15 dex, and we plot these occurrence rates in Figure \ref{fig:occratesgrid}. Each bin is shaded in accordance with its occurrence rate, and annotated with our measured occurrence rate and error, or with an upper limit on the occurrence rate in the case that a planet was not detected in that bin. For compactness, the error on the occurrence rate is taken to be half of the 68\% confidence interval around the measured value, which is why some of the errors imply a range of uncertainty with a negative occurrence rate, which is unphysical. Bins that do not have any annotations represent regions with low completeness where our derived upper limit is not restricting.

 We use the same bins as in \cite{petigura2018} for the sake of comparison, and we find that our results are qualitatively similar. For instance, we both find that the most abundant planet types are warm Sub-Neptunes, warm Super-Earths, and then cool Jupiters, in that order. For Jupiters, we find a sharp rise in occurrence rate for $P>100$~days. This trend is present in the CKS sample as well, and has been noted in RV surveys \citep{cumming2008}. This rise in occurrence rate is thought to be correlated with the water ice line at $\sim$1 au, leading to the facilitation of more massive planetary cores. There is also an island of relatively high occurrence for hot Jupiters centered on  $P\approx3$~days. From our data alone, it's not clear if this is a statistically significant increase centered at $P\approx3$~days, or if it is simply a result of declining occurrence rates below $P\approx100$~days. However, this increase in occurrence rates is also found in the California Planet Search program \citep{cumming2008}, the CKS survey \citep{petigura2018}, and other studies \citep[e.g.,][]{cumming1999,udry2003,hsu2019}, lending credibility to its existence. Overall, we find an occurrence rate for hot Jupiters of $f=0.37^{+0.13}_{-0.19}$~planets per 100 stars, compared to the CKS team's measurement of $f\approx0.57$~planets per 100 stars. This occurrence rate is more consistent with \cite{santerne2016} and \cite{masuda&winn2017} who measured $f={0.47^{+0.08}_{-0.08}}$ and $f=0.43^{+0.07}_{-0.06}$~planets per 100 stars, respectively. This agreement in the occurrence rate of hot Jupiters bolsters our claim from  \S\ref{sec:psamp} that removing RV variable sources does not remove a significant fraction of planets.

\begin{figure*}[t]
\centering
\includegraphics[width=0.95\textwidth]{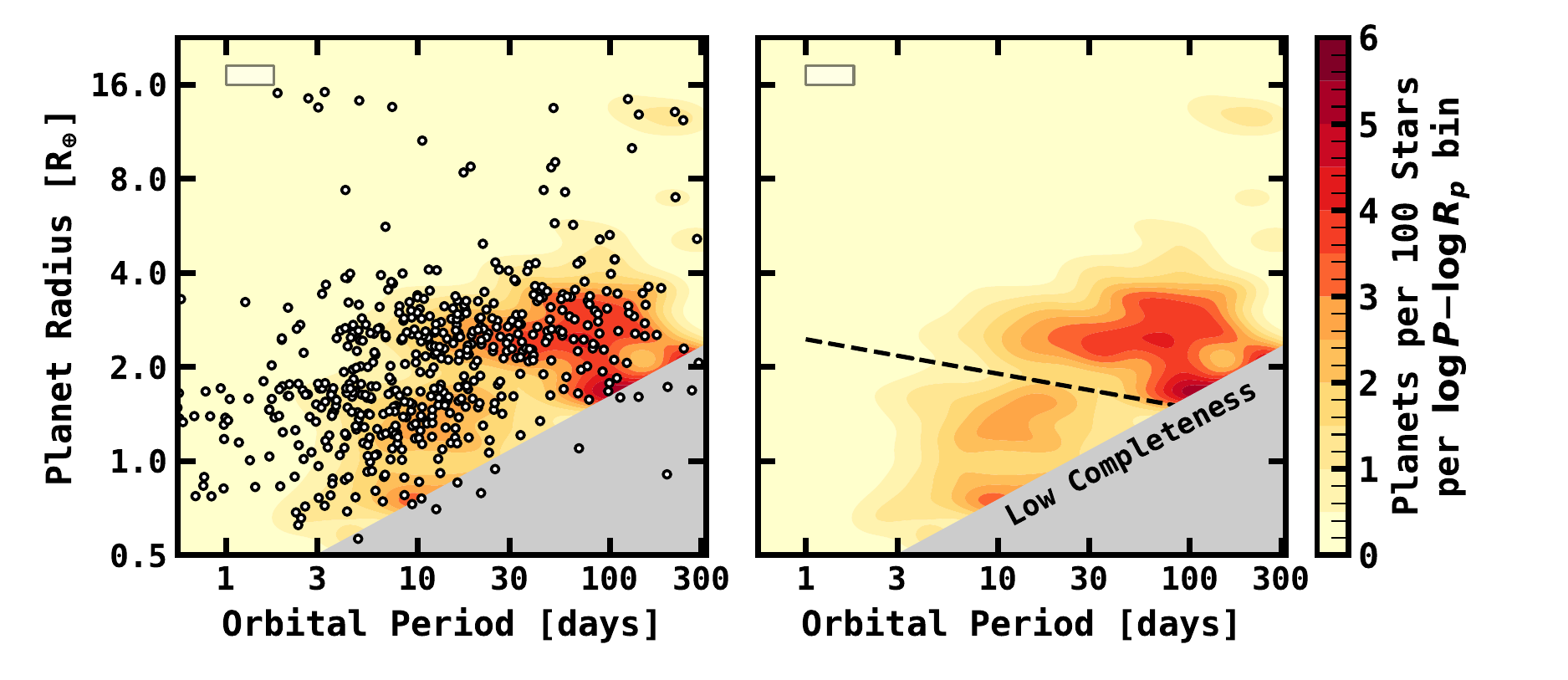}
\includegraphics[width=0.95\textwidth]{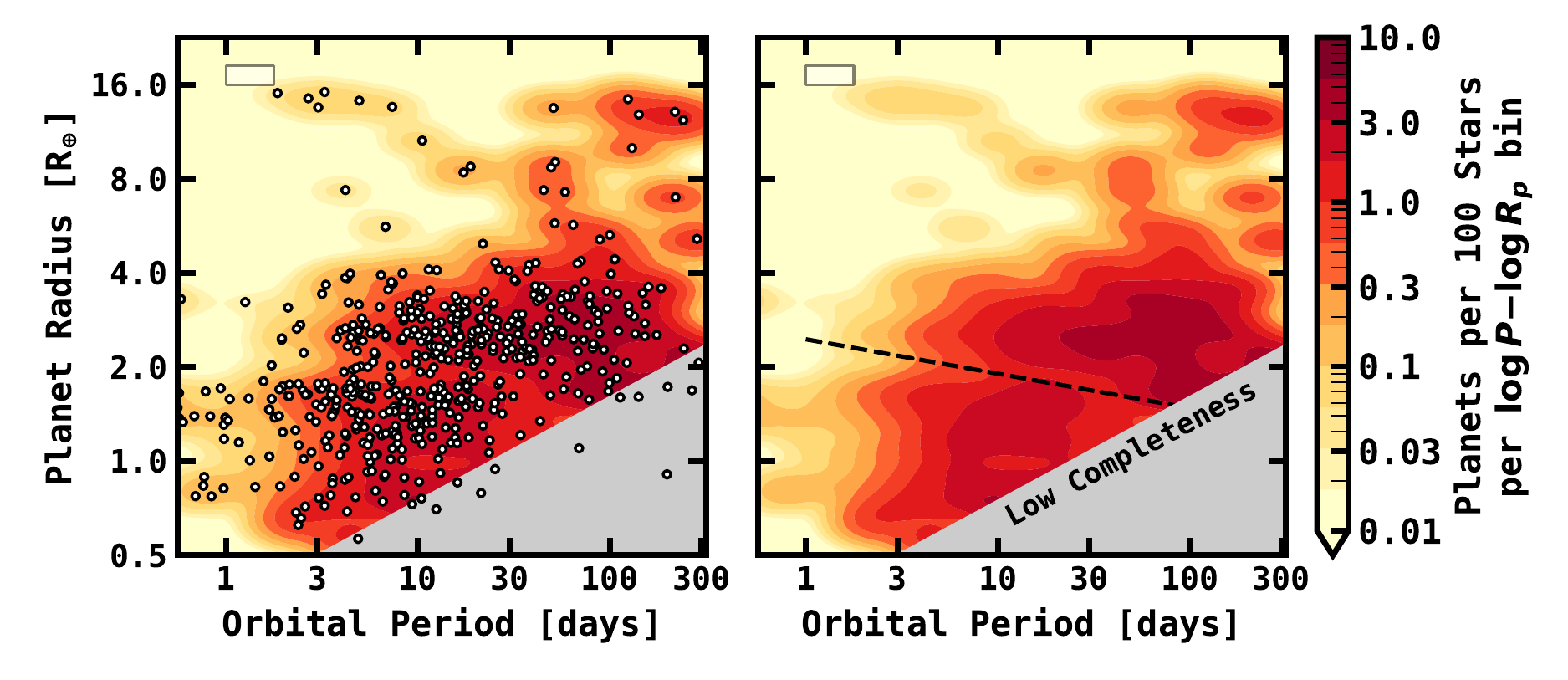}
\caption{The planet occurrence rate in the $P$-$R_p$~plane. For each row of figures, the left panel shows the planets in \psamp\ as white points, and the filled contours show the derived occurrence rate, with darker shades representing higher occurrence rates. The top row displays the occurrence rates on a linear scale, and the bottom row displays occurrence rates on a log scale, where darker shades of red indicate higher occurrence and lighter shades of yellow indicate lower occurrence. A box in the upper left corner of each panel shows the FWHM of the Gaussian kernel used to calculate the contours for this figure. The gray region denotes areas of low survey completeness.
The dashed black line shows the location and slope of the radius gap measured by \cite{martinez2019}. }
\label{fig:occratescontour}
\end{figure*}

 However, we find a few key differences with previous studies as well. For instance, the occurrence of Sub-Neptunes and Super-Earths is nearly twice as high in some of the bins as compared to that found by the CKS survey. One explanation for this apparent difference in the occurrence rates of small planets is simply a systematic difference in the planet radii. For instance, this work typically has more precisely-measured planet radii due to the inclusion of \gaia\ parallaxes in our analysis, which could cause certain bins in the $P$-$R_p$ plane to appear to have higher occurrence simply due to sharper features in the occurrence rate distribution. 
 The bins themselves were also chosen arbitrarily, so increased occurrence for a given bin can appear inflated due to the choice of bin edges. To more accurately judge this potential difference, we calculate occurrence rates in arbitrarily small bins in the $P$-$R_p$ plane, then convolve these occurrence rates with a two-dimensional Gaussian kernel of size  $\Delta \log P \times \Delta \log R_p =$  0.25 dex $\times$ 0.1 dex. The occurrence rates as a result of this smoothing are shown in Figure \ref{fig:occratescontour}. This figure gives a more intuitive understanding of the occurrence rate of planets in the $P$-$R_p$ plane, and avoids the effects of binning that may misrepresent the PLDF. We find that our occurrence rates indeed are slightly larger than in \cite{petigura2018} at the peak of the warm Sub-Neptune and Super-Earth distributions. 
 This difference may be due to the APOGEE-KOI selection function, which selects a higher fraction of lower mass stars due to its magnitude cut in the near-infrared where such stars are brighter, rather than on the optical $Kp$~magnitude. Lower mass stars are known to have increased occurrence rates for small ($R_p\lesssim 4 R_\oplus$)~planets \citep{mulders2015a}.

One feature present in our occurrence rate distribution is the radius gap \citep{fulton2017}, with a notable dependence on the location of the gap with orbital period. This trend was uncovered by an independent analysis of the CKS spectra performed by \cite{martinez2019}. We find excellent agreement between the slope they found and the planet occurrence rate distribution in our sample. This slope in the radius gap is shown as a dashed black line in Figure \ref{fig:occratescontour}.

\begin{figure*}
\centering
\includegraphics{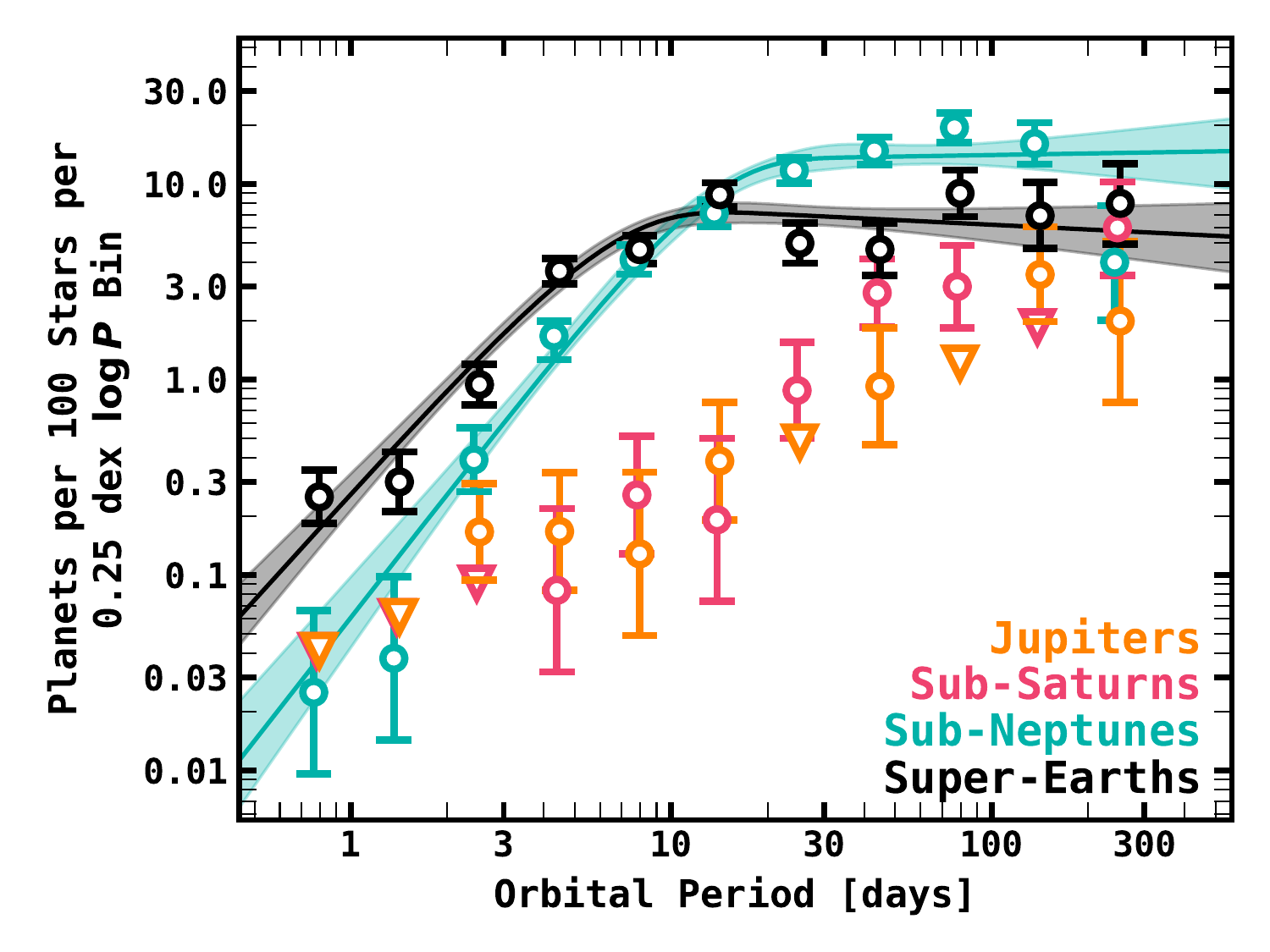}
\caption{The number of planets per star (multiplied by 100) for a given orbital period bin and planet size class. The colors denote the planet size class. The circular points denote the occurrence rates while the triangular points denote upper limits. 
We fit the Sub-Neptune and Super-Earth occurrence rates with a function of the form, $f_{P} \propto  P^\alpha (1 - e^{-(P/P0)^\gamma} )$. The lines show the adopted best fit solution, and the shaded regions denote the $1\sigma$ confidence interval of credible models. The occurrence rates shown in this figure are displayed over substantially larger bin sizes than those used to fit the model but are displayed here to guide the eye.}  
\label{fig:periodoccrates}
\end{figure*}

We also find that the occurrence rate of Sub-Neptunes and Super-Earths as a function of orbital period can be well described by a distribution of the form, 
\begin{align}
    f_{P} = C P^\alpha (1 - e^{-(P/P0)^\gamma} ) \;\; ,
\end{align}
which effectively acts as a power law distribution, with a break at $P=P_0$. At $P \ll P_0$, the distribution acts as a power law with $f_P \propto P^{\alpha + \gamma}$, and at $P\gg P_0$, the distribution acts as a power law of the form, $f_P \propto P^{\alpha}$. We fit the differential occurrence rate of small planets with respect to period using this functional form for both Sub-Neptunes and Super-Earths. We use bin sizes of $\Delta \log P = 0.005$~dex, and initialize the MCMC routine with 50 walkers, 10,000 total steps, and 1000 burn-in steps. 
Sub-Saturns and Jupiters are not well described by this functional form. The fits are displayed in Figure \ref{fig:periodoccrates}.

For Super-Earths, we find a transition period of $=P_0 = 6.5^{+1.6}_{-1.3}$~days, and for Sub-Neptunes we find a transition period of $P_0=13.0^{+5.0}_{-3.2}$~days. This is consistent with the theory of photoevaporation \citep{owen&wu2013,owen&wu2017}, as planets at shorter orbital periods are subject to higher incident FUV and XUV flux, and are thus more subject to atmospheric stripping. As a result, one would expect the occurrence rate of Sub-Neptunes to drop before the occurrence rate of Super-Earths. Super-Earths and Sub-Neptunes have a consistently steep rise in occurrence at short orbital period, with $\gamma=2.1^{+0.2}_{-0.2}$ for Sub-Neptunes and $\gamma=1.9^{+0.2}_{-0.2}$ for Super-Earths. At longer orbital periods, Sub-Neptunes level off in occurrence rate with $\alpha=0.03^{+0.16}_{-0.20}$ consistent with no change, and Super-Earths may have a slight decrease in occurrence rate at longer orbital periods with  $\alpha=-0.08^{+0.13}_{-0.14}$, though these are also consistent with no change. These parameters are all consistent with those measured by \cite{petigura2018}.

In addition, the transition period we measure for Super-Earths, is in agreement with the transition period found in \cite{wilson2018}, $P_0 = 8.3^{+0.1}_{-4.3}$~days, who analyzed planets of all size classes. In \cite{wilson2018} the transition period was measured by finding the period in which the metallicity distributions of host stars with their innermost detected planet above and below the transition period are the most statistically different.

\subsubsection{Occurrence Rates with $P$, $R_p$, and [X/H]}\label{sec:occrates_xh}

To test the significance with which each element correlates with planet occurrence, we fit a parametric function of the form
\begin{align} \label{eq:xpfit}
    f_{X,P} = C P^\alpha 10^{\beta X} \;\; ,
\end{align}
where $X=[X/H]$, using the bootstrapping monte-carlo method described in \S\ref{subsec:fstar}. 
This is an extension of the model used by \cite{petigura2018}, who modeled the correlation between planet occurrence rates and metallicity. The abundance term in the above equation is equivalent to a power law relationship with the number density of atoms in the star's photosphere, 
\begin{align}
    f_{X,P} \propto  \left ( \frac{n_X}{n_H}\right )^\beta \;\; ,
\end{align}
where $n_X$ is the number density of atoms of element $X$, and $n_H$ is the number density of hydrogen atoms in a star's photosphere. 
With this relationship in mind, a value of $\beta>0$ would indicate a correlation between the number of planets and the presence of that particular element, while a value of $\beta < 0$ would indicate an anti-correlation between the planet occurrence rate and the number density of atoms of that particular element.

If we naively assume that the abundance ratios in the stellar photosphere are the same as the abundance ratios of the protoplanetary disk in the first $\sim$1-10~Myrs during planet formation before the gas disk disperses, then a non-zero differential occurrence rate density between two independent elements may indicate that the presence of one element more efficiently facilitates or suppresses planet formation compared to the presence of the other element.
Such a result may indicate the composition of dust grains that grow to planetesimals more efficiently, or gaseous molecules that are preferentially accreted.

\begin{figure*}
\centering
\includegraphics[width=\textwidth]{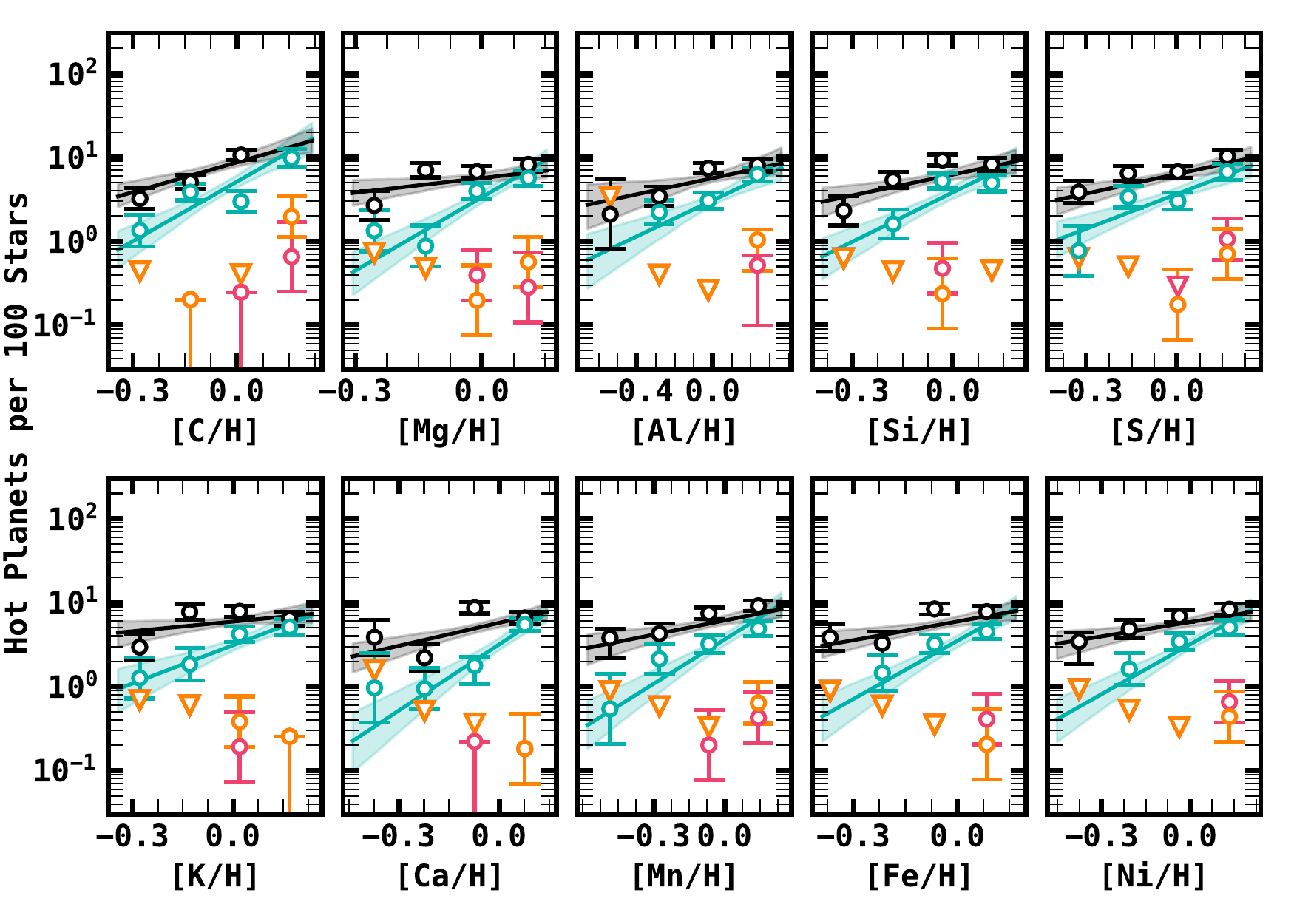}
\caption{The occurrence rate of hot planets ($P=1$-10~days) as a function of chemical abundances for ten different chemical elements. The colors represent planets of different size classes (Jupiters: Orange, Sub-Saturns: Pink, Sub-Neptunes: Teal, Super-Earths: Black). The points show the number of planets per 100 stars per bins equally spaced across the inner 90\% of the abundance distribution for each element. 
The triangles show upper limits (90th percentile) on the planet occurrence rate, and the lines and shaded regions show our best fit and 1$\sigma$ uncertainty to a power law distribution of the form, $f_{X,P} \propto P^\alpha\,10^{\beta X}$, where we've integrated over the period dependence to display the fit in one dimension. 
Models are not shown for Sub-Saturns and Jupiters when the fit is poorly constrained. We emphasize once again that the occurrence rates and upper limits displayed in this figure are for larger bin sizes than those used to fit the power law distribution, and are included to guide the eye. }
\label{fig:xfe_occrates_hot}
\end{figure*}

\begin{figure*}
\centering
\includegraphics[width=\textwidth]{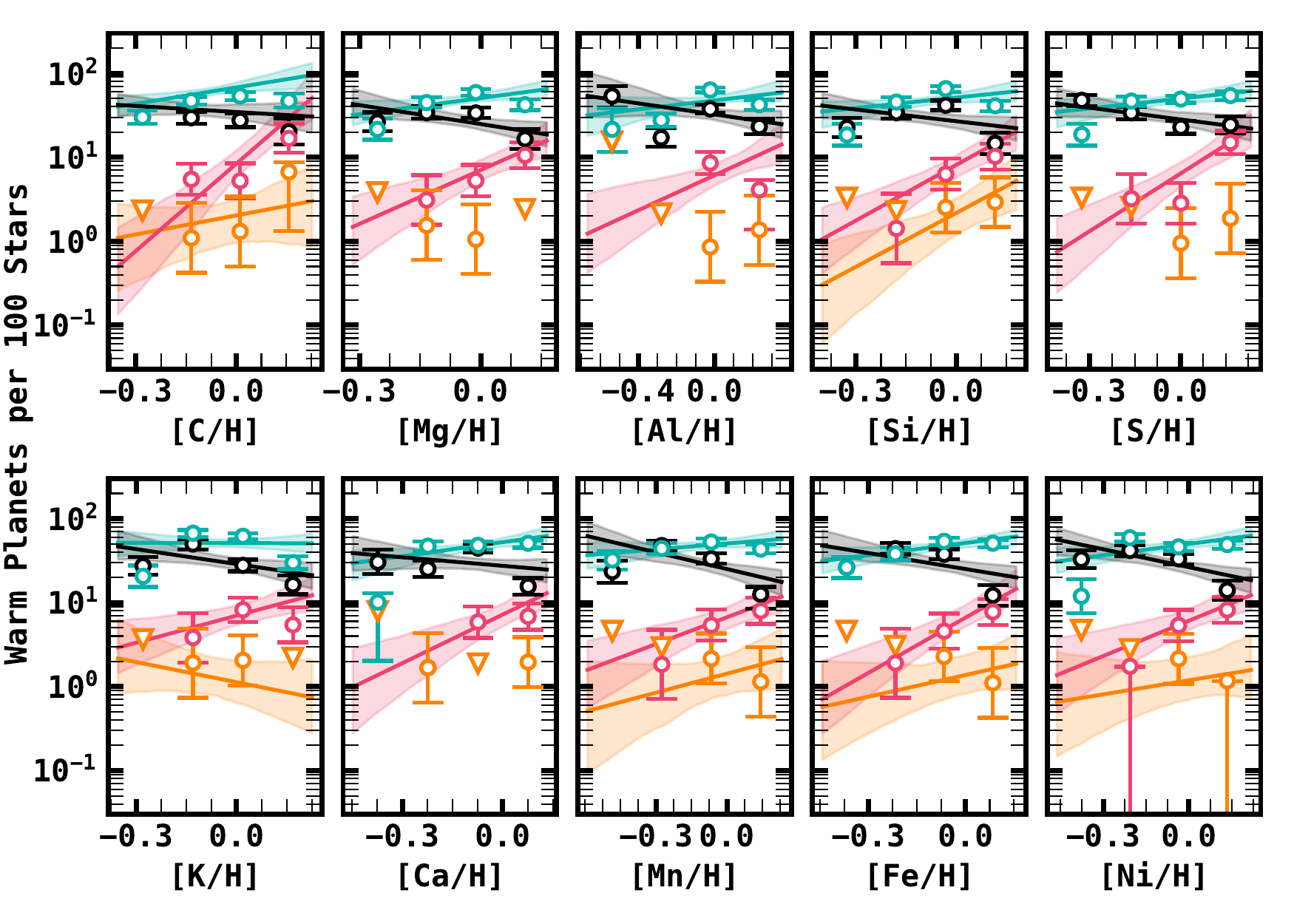}
\caption{The same as Figure \ref{fig:xfe_occrates_hot}, but for warm planets ($P=10$-100~days). We see overall weaker trends for each element and planet size class, with the possible exception of Sub-Saturns given the low number of detections at short periods.  }
\label{fig:xfe_occrates_warm}
\end{figure*}

As discussed in the appendix (\S\ref{subsec:fstar}), the conclusions stemming from this analysis are limited by uncertainties in the stellar abundance distribution function, $\fstar (X)$, rather than the Poisson error. 
In other words, the low number of stars in \csamp\ are the dominant source of uncertainty in deriving $\beta$. For each planet size and period class we observe, we find consistent values for the period dependence, $\alpha$, across all elements in a given planet size and period class. We also find no correlation between $\alpha$ and $\beta$, for any element $X$ and any planet period period or size class in the posterior distributions. The results of these parametric fits are listed in Table \ref{tab:occrate_fits}, and plotted in Figures \ref{fig:xfe_occrates_hot} and \ref{fig:xfe_occrates_warm}.

 For the hot period class of planets, we find a positive correlation with all abundances and planet size classes, except Sub-Saturns that we are not able to constrain due to the low number of detections. The fits and range of credible models for all the hot planets are plotted in Figures \ref{fig:xfe_occrates_hot} and \ref{fig:xfe_occrates_warm}. Because the model is two-dimensional, we integrate over the period dependence and only display the dependence on the elemental abundances.

 The hot Jupiters in our sample are poorly constrained, but still consistent with $\beta > 0$, with $\beta$ ranging from $\beta = +10.3^{+5.7}_{-4.3}$ for Si, to $\beta=+3.8^{+5.2}_{-3.8}$ for Fe. All of these values are consistent at the $1\sigma$ level, but not well constrained. For hot Super-Earths, the element number density correlation ranges from $\beta = +0.41^{+0.44}_{-0.42}$ at the lowest for K, and $\beta = +1.16^{+0.47}_{-0.43}$ for C at the highest. These values are consistent at the $1\sigma$ level, and there are no clear differences between each different element.

For hot Sub-Neptunes, the correlation coefficients, $\beta$, are all $>0$ and mostly consistent across all elements. However, we do find hints at variation among different chemical species. The correlation strengths range from $\beta=2.86^{+0.76}_{-0.69}$ for Mg to $\beta = 1.06^{+0.46}_{-0.42}$~for Al have a $\sim2\sigma$~discrepancy. However, we are hesitant to trust these differences due to potential non-LTE effects that may bias the Al abundance ratio in ASPCAP which are computed in LTE.  
 The dependence on other elements range between these extremes. Given the conservative uncertainties placed by our analysis, future studies are needed to determine if credible, more subtle, variations exist.
Such a difference may give rise to important mechanisms in the formation or evolution of hot Sub-Neptune systems.

 For warm planets, the correlation strength is reduced compared to the corresponding strength for hot planets for all planet size classes except possibly Sub-Saturns considering we were unable to constrain the correlation strength for hot Sub-Saturns. For warm Super-Earths, we find a tentative anti-correlation of $\beta \approx -0.6$ for most elements, though they are also consistent with no correlation. Therefore, we do not make any claims about the dependence of the warm Super-Earth occurrence rate and the abundance of any chemical species. The abundance of Sub-Neptunes gives the opposite result, and we find that the there is a slight correlation, with $\beta \approx +0.4$ across all elements, but with similarly-sized errors such that we are unable to make a claim that the occurrence of warm Sub-Neptunes is positively correlated with the abundance of any particular chemical species. Warm Jupiters also have this same result, with $\beta$ ranging from $-0.8$ to  $+2.2$ and errors ranging from 1.0 to 1.8~dex. Although our uncertainties are larger for each of these different chemical trends, these values are all consistent with the [Fe/H] dependence found by \cite{petigura2018}.

 The Sub-Saturns are the only planet size class that have a measurable correlation between planet occurrence and chemical abundances at $P=$~10-100~days. For Sub-Saturns we measure a range of correlations from $\beta = 1.0^{+0.7}_{-0.7}$ for K, to $\beta = 3.4^{+1.3}_{-1.2}$ for C. These values are all still consistent (within 1.5-2$\sigma$) across the ten elements within our uncertainties. Our measured correlation strength for Fe  ($\beta=2.1^{+1.1}_{-1.0}$) is nearly identical to that of \cite{petigura2018} who reported $\beta = 2.1^{+0.7}_{-0.7}$ for warm Sub-Saturns.

One trend we've noticed is that the magnitude of the strength of the correlation ($|\beta|$) for Mn, is lower than for Fe in most period and planet size classes, though not significantly enough to claim a distinction. The lower value for Mn may be particularly surprising considering that [Mn/H] has the strongest correlation with [Fe/H] of all the abundances, one might expect that this effect be enhanced.  
This is likely a result of our methodology to account for the uncertainty in $\fstar(X)$. In accounting for uncertainties in $\fstar(X)$, we perform a monte-carlo, boot strapping routine that is likely to reduce the overall reported correlation strength for elements with larger errors. Because $\sigma_\mathrm{[Mn/H]} > \sigma_\mathrm{[Fe/H]}$, the correlation strength for Mn is typically lower than that of Fe, but still consistent within the uncertainties.

\section{Discussion}\label{sec:discussion}

\subsection{Variations in Correlation Strength Between Different Chemical Species}

\begin{figure}
    \centering
    \includegraphics[width=0.48\textwidth]{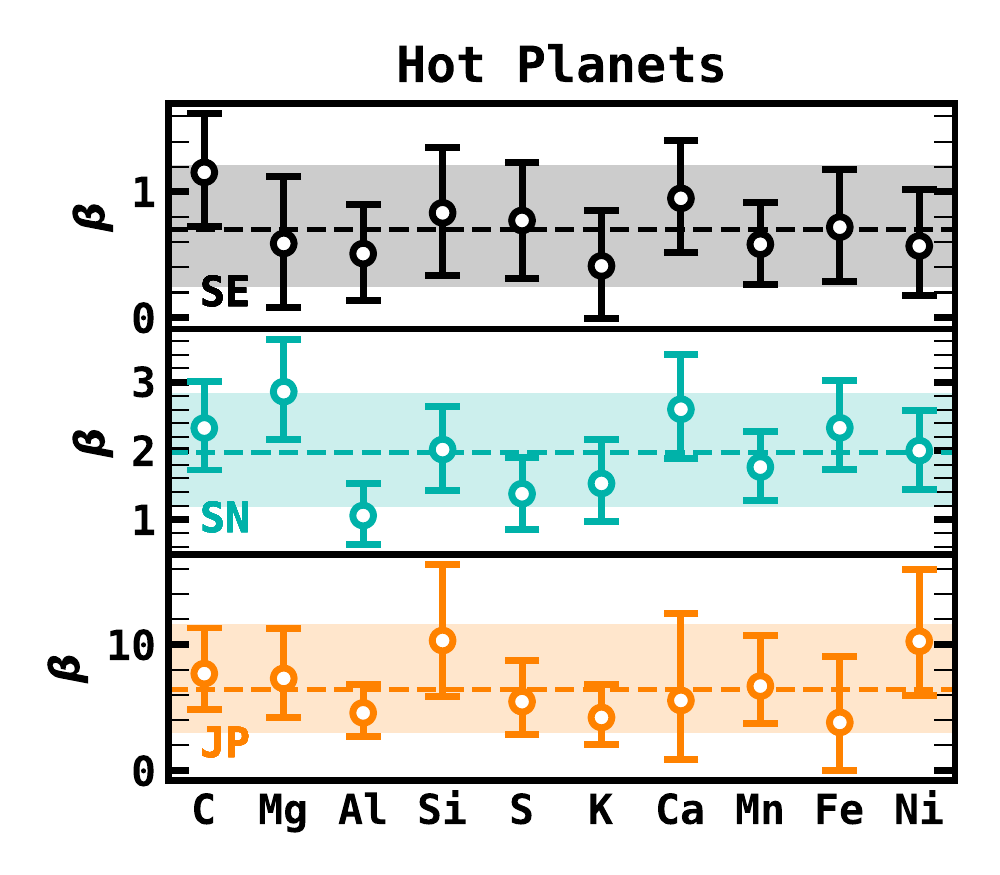}
    \caption{The derived $\beta$ and uncertainties for each chemical species, and each planet size class with periods ranging between 1-10~days. The colors show the planet size class for which $\beta$ was derived (Black: Super-Earths, Teal: Sub-Neptunes, Orange: Jupiters). The dashed line shows the mean across all elements, and the shaded region shows the inner 68\% of the posteriors to the fits performed across all elements in \S\ref{sec:occrates_xh}. In order of increasing planet size class, the averages are $\beta_\mathrm{avg} =$~+0.7, +2.0, and +6.4, though the Jupiters in our sample have a range of $\beta\sim$~3-12.}
    \label{fig:betahot}
\end{figure}

\begin{figure}
    \centering
    \includegraphics[width=0.48\textwidth]{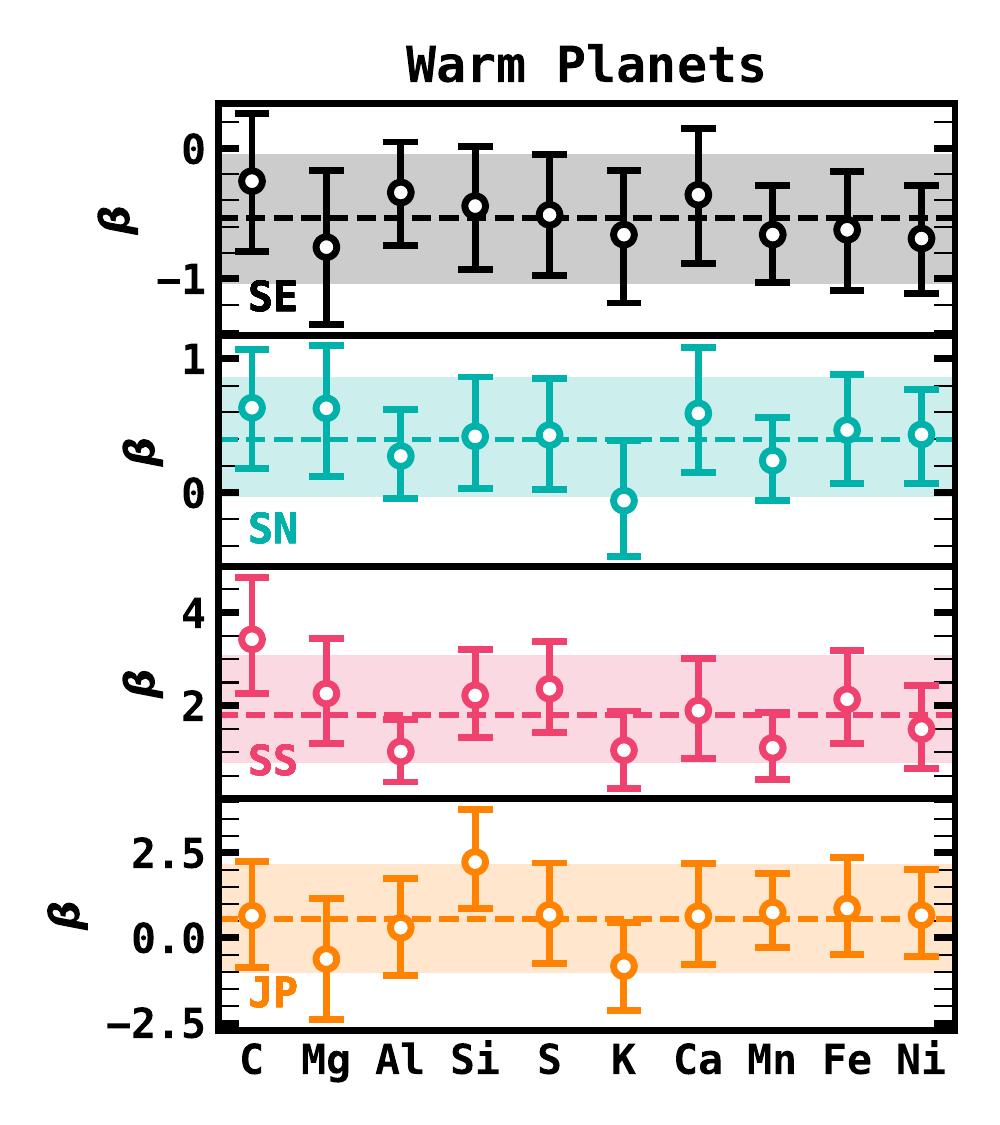}
    \caption{The derived $\beta$ and uncertainties for each chemical species, and each planet size class with $10<P<100$~days. The legend is the same as in Figure \ref{fig:betahot}, with the planet size classes represented by differing colors (Black: Super-Earths, Teal: Sub-Neptunes, Pink: Sub-Saturns, Orange: Jupiters). In order of increasing planet size class, the averages are $\beta_\mathrm{avg} =$~$-$0.5, +0.4, +1.8, and +0.6.  }
    \label{fig:betawarm}
\end{figure}

In this work we've made the first measurement of the dependence of planet occurrence as a function of detailed chemical abundances in the \kepler\ field.  The measured $\beta$ values and their uncertainties are shown in Figures \ref{fig:betahot} and \ref{fig:betawarm} for each element and planet size class. 
We are unable to confidently detect any differences in $\beta$ for different chemical species within a given planet size and period class, nor are we able to unambiguously attribute the correlation between planet occurrence and stellar chemistry to the enhancement of any one particular element. 
This lack of difference may be due to one of, or a combination of three effects. First,
the lack of difference may be intrinsic (i.e., the enhancement/depletion of all elements are equally correlated with planet occurrence); second, our null result may be due to our uncertainties, which are limited by uncertainties in $\fstar$ for Super-Earths and Sub-Neptunes, and by the lack of detections for Sub-Saturns and Jupiters (see \S\ref{subsec:fstar}), or third, 
we are unable to detect differences in this data set due to degeneracies caused by the lack of unique stellar populations probed in the \kepler\ field. 
I.e., the stars in the \kepler\ field have abundance ratios that are highly correlated for each element, making it impossible to differentiate the effects of one over another.

Due to these factors, determining the importance of unique elements in facilitating planet occurrence rates in practice can be very difficult. 
To test the dependence of each different chemical species on the planet occurrence rate separately from the known effects of enhanced bulk metallicity, we've shown that it is insufficient to simply measure the quantity $f_{X}$ for varying chemical species. 
It is equally insufficient to measure  $f_{X/Z} = df/d$[X/Fe], as the chemical abundance trends with [X/Fe] and [Fe/H] are often not linear, and vary element by element based on a complicated function of star formation history, radial migration, and nucleosynthetic yields \citep[e.g.,][]{wyse1995,mcwilliam1997,sellwood2002,hayden2014,nidever2015}. 
Disentangling such effects will rely on either more precise observations, a much larger sample where subtle differences can be detected, or targeted planet-search surveys across multiple different stellar populations with unique chemical abundance patterns, such as in the thick disk or the halo.

\subsection{Disentangling the Effects of Stellar Age, Mass, and Galactic Chemical Evolution}

Another important source of confounding variables is the relative trends with chemical abundances, stellar age, and stellar mass. Because lower metallicity stars in the thin disk were formed before the enrichment of the interstellar medium, such stars may skew toward older ages and lower masses. Disentangling these effects is particularly challenging, given that credible trends with planet occurrence and stellar mass have been unequivocally uncovered in the literature \citep[e.g.,][]{mulders2015a, dressing&charbonneau2015, fulton&petigura2018,ghezzi2018}, and estimates of stellar age are becoming more precise due to surveys such as \gaia.

For these reasons, when interpreting trends between age and planet properties, it is imperative that host star chemistry is taken into account. In short, stellar mass, age, and composition are all strong confounding variables with one another.

\subsubsection{Demonstration of an Age-Metallicity Degeneracy in Exoplanet Demographics}

There have been a number of claims relating to the demographics of planets and stellar age. For instance, \cite{berger2020a} found that the relative fraction of Super-Earths to Sub-Neptunes is lower for young ($<$1~Gyr) stars than the for old ($>$1~Gyr) stars.  
\cite{berger2020a} inferred from this that there is $\sim$Gyr evolution in the atmospheric-loss timescale for stars near the radius gap, as predicted by core-powered mass loss \citep{gupta&schlichting2019,gupta&schlichting2020}.

\begin{figure}
    \centering
    \includegraphics{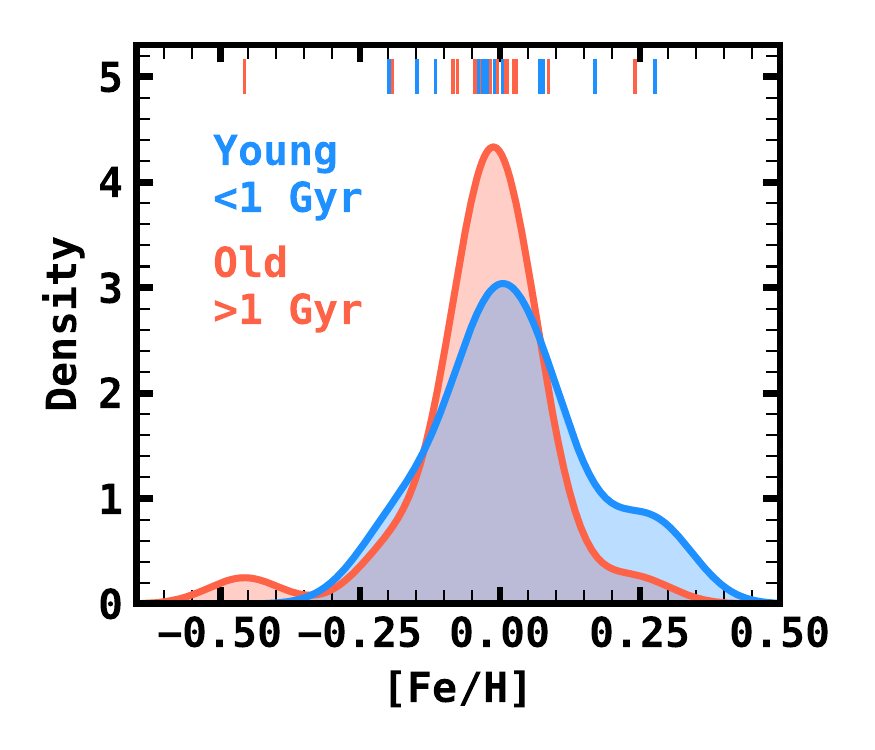}
    \caption{The metallicity distribution functions of the planet hosts in the``Young" and ``Old" samples from \cite{berger2020a} that were observed by APOGEE. The metallicities for each planet host are displayed as vertical ticks near the top of the figure. }
    \label{fig:oldyoung}
\end{figure}

While \cite{berger2020a} cite age and long-term planetary evolution as a cause for a decrease in the frequency of Sub-Neptune planets, in this study, we find that a dramatic decrease in the frequency of Sub-Neptunes can be attributed to even a small depletion of heavy elements. 
To test whether this relative decrease in the number of Sub-Neptunes can be explained by a difference in metallicity and subsequent change in occurrence rate between the ``Old" and ``Young" samples, we cross-matched our sample of all the KOIs observed in APOGEE with the ``Young" ($<$1~Gyr) and ``Old" ($>$1~Gyr) sample of planets from \cite{berger2020a}. In total, there are 25 and 23 planets with hosts in the ``Old" and ``Young" samples, respectively, with [Fe/H] measured by APOGEE. 
We then calculate the metallicity distribution function for each sample using a Gaussian kernel density estimate with a bandwidth chosen by Scott's rule \citep{scott2010}. The distribution functions are shown in Figure \ref{fig:oldyoung}. The ``Young" subsample is slightly skewed toward higher metallicities compared to the ``Old" subsample.

Using the measured metallicity distribution functions to compute the expected occurrence rates, we find that the expected occurrence in the young sample is $\approx$1.4$\times$ higher for Sub-Neptunes and $\approx$1.1$\times$ higher for Super-Earths with $P=1$-10~days. 
For comparison, \cite{berger2020a} found that $N_{SupEarth}/N_{SubNep}$ was $0.61\pm0.09$ and $1.00\pm0.10$ for the ``Young" and ``Old" samples, respectively. Thus, \cite{berger2020a} found that $N_{SupEarth}/N_{SubNep}$ was decreased by a factor of $0.61\pm 0.12$ from the ``Young" to ``Old" samples. Under the naive assumptions that $f_{SupEarth}/f_{SubNep} \propto N_{SupEarth}/N_{SubNep}$ for short periods, and that the \cite{berger2020a} KOI samples are skewed toward $P<10$~days, we would expect $N_{SupEarth}/N_{SubNep}$ to decrease by a factor of $0.80_{-0.12}^{+0.11}$. These values have only a $\approx 1.7\sigma$ discrepancy, though this is before correcting for detection biases which would lower the number of Super-Earths.
Thus, provided that the inferred metallicity distribution functions of the ``Old" and ``Young" planets observed by APOGEE is representative of the metallicity distribution functions in the sample used by \cite{berger2020a}, then the full change in the relative number of Super-Earths to Sub-Neptunes may partially be explained by slight metallicity differences between the ``Old" and ``Young" samples.

It is worth noting that \cite{berger2020a} were extremely careful to use known spectroscopic metallicities to control for differences when constructing their ``Young" and ``Old" samples. However, these metallicities came from a heterogeneous catalog, and different spectroscopic pipelines often have systematic differences up to $\sim$0.1 dex. 
Thus, with precisely and homogeneously measured metallicities from APOGEE, we are able to detect a slight difference that may bias results such as these.

We intend the above exercise to be a demonstration rather than a repudiation of the conclusions inferred in \cite{berger2020a}.
In reality, the logic behind inferring a metallicity distribution of a field sample by measuring the host star metallicity distribution of a planet sample is in conflict with the premise of this study, and should not be trusted. 
Instead, this exercise is intended to demonstrate how even small metallicity differences may bias an inferred planetary distribution function, and therefore motivate the need for  high-resolution, high-$S/N$ spectroscopic surveys to provide uniform metallicities and chemical abundances for a significant fraction of stars, so that biases from such effects can be adequately controlled.

\subsubsection{Age-Metallicity Degeneracies in the Population of Hot Jupiters}

There are also claims of demographics  for the population of hot Jupiters. For instance, by comparing the Galactic dynamics of field stars against the dynamics of hot Jupiter host stars, \cite{hamer2019} claimed that hot Jupiters are destroyed by tides while their hosts are on the main sequence. This study found that hot Jupiter host stars are kinematically cold compared to the field, implying they are younger on average. However, \cite{hamer2019} didn't have the necessary data to take into account the strong correlation with metallicity and the occurrence of hot Jupiters, and the correlation between metallicity and galactic kinematics. It has been shown by a number of investigations that populations of metal-enhanced stars have lower Galactic velocity dispersion \citep[e.g.,][]{anguiano2018}.
On the other hand, one can make a similar claim about the age of hot Jupiter host stars and their correlation with metallicity. Perhaps the Planet Metallicity Correlation can be partially explained by a correlation with stellar age.

One potentially keystone open cluster for understanding the interplay between age and chemistry on the demographics of large planets is NGC 6791. NGC 6791 is a uniquely old cluster at $7.0\pm2.5$ Gyr \citep{netopil2016}, with an enhanced metallicity, $\mathrm{[Fe/H]}=+0.35$~\citep{cunha2015, donor2020}, and a high relative alpha abundance of $[\alpha/\mathrm{Fe}] = +0.1$ \citep{linden2017}. 
An occurrence rate study within this cluster could show the relative importance of tides and enhanced heavy metals. If tides (i.e., correlations with age) are the dominant force in shaping the hot Jupiter population, then one should find an occurrence rate more similar to the field. If the dominant correlation is with enhanced abundances, the hot Jupiter population should be dramatically larger than in the field. While NGC 6791 was observed in the \kepler\ field, the main sequence turn off is very dim with a \gaia\ magnitude of $G \sim 17$. Thus, measuring a reliable planet occurrence rate may prove difficult in practice.

\subsubsection{Predictions for Planet Occurrence Rates in Nearby Open Clusters}

 An ideally perfect experiment to disentangle correlations between age and abundance ratios would be to measure the planet occurrence rate in several different open clusters with the same age but significantly varying abundance patterns, or vice versa. 
 Such an experiment is likely not possible due to the lack of existence of such clusters with a wide enough abundance spread, and with enough stars. However, nearby open clusters still provide opportunities for measuring changes in planet demographics with age.

\begin{table*}
    \centering
    \begin{tabular}{cccccc} \hline \hline 
         Cluster & Age & [Fe/H]  &  Distance  & $f_\mathrm{hot}$ & $f_\mathrm{hot}$  \\ 
         & Gyr & dex  &  pc  & $1\,$-$\,1.9 \; R_\oplus$ & 1.9$\,$-$\,4 \; R_\oplus$  \\ \hline
         Pleiades & 0.1  &  $-$0.01 & 136  & $11.8^{+1.6}_{-1.3}$  &  $6.5^{+1.1}_{-1.0}$\\
          Praesepe & 0.7 & +0.16  & 47  &  $15.6^{+4.6}_{-3.3}$ & $16.2^{+5.7}_{-3.9}$  \\
         Hyades& 0.8 &  +0.13 & 186  & $14.8^{+3.9}_{-2.9}$  & $13.8^{+4.2}_{-3.0}$ \\
         Ruprecht 147 & 2.0 & +0.12  & 310  & $12.6^{+2.0}_{-1.7}$ & $8.0^{+1.5}_{-1.3}$\\
         M67  & 3.5 & +0.03  & 880  & ${12.6}^{+2.0}_{-1.7}$  &  $8.0^{+1.5}_{-1.3}$ \\
         NGC 188 & 5.5 & +0.11  & 1990  & $14.3^{+3.4}_{-2.6}$  &  $12.4^{+3.4}_{-2.5}$ \\ 
         NGC 6791 & 7.0& +0.35 & 2300  & $21^{+12}_{-7}$ & $45^{+35}_{-18}$  \\ 
         \textbf{\kepler\ Field Stars} & -- & -- & --  & $6.6^{+0.6}_{-0.6}$ & $3.9^{+0.5}_{-0.5}$ \\
\hline 
    \end{tabular}
    \caption{Predicted occurrence rates, $f_\mathrm{hot}$, of hot planets ($P=$~1-10~days) for a few nearby open clusters in the absence of long-term planetary evolution. $f_\mathrm{hot}$ is given in units of Number of Planets per 100 Stars for each size class. Our model is extrapolated for NGC 6791, so the uncertainties for this cluster are quite large. The occurrence rates for the \kepler\ field are derived from the field sample in this study. }
    \label{tab:clusters}
\end{table*}

To date, there have been a few such studies targeting young transiting planets with \emph{K2} \citep{howell2014} which observed open clusters such as the Hyades \citep[][]{mann2016,mann2018,vanderburg2018}, the Pleiades \citep{gaidos2017}, and Praesepe \citep{rizzuto2018}, among others. However, while there have been uniform searches for transiting planets in such clusters \citep{rizzuto2017}, occurrence rates and completeness corrections have proven difficult due to effects such as crowding and the presence of strong correlated noise in the light curves of young stars.

In an attempt to facilitate planet demographics studies for such clusters, we make predictions for planet occurrence rates for a number of nearby open clusters given their metallicity. 
Deviations in the actual planet occurrence rate from our predictions may then be attributed to age or some other non-chemistry related effect. 
Though some of the clusters mentioned below are scarcely populated, measurements of the planet occurrence rate may be approachable if measured in the aggregate (i.e., the expected planet occurrence rate from stars across multiple clusters). 
An alternate approach may be to measure occurrence rates with RV surveys, which do not have as harsh of a geometric bias as transit surveys but come with other complications such as enhanced stellar activity in young stars.

We make predictions for the occurrence rate of hot planets ($f_\mathrm{hot}$) for the Pleiades, Praesepe, the Hyades, Ruprecht 147, M67, NGC 188, and NGC 6791. We make predictions for hot planets because they are more likely to be discovered with TESS \citep{ricker2015} and \emph{K2}, 
and because hot planets have the strongest correlations with enhanced heavy elements. 
These predictions are listed in Table \ref{tab:clusters}. For each of these clusters, we adopt the ages and distances of the clusters derived from \gaia\ DR2 photometry \citep{gaia2018b}, and the metallicities from APOGEE DR16 \citep{donor2020} where available, and from a homogenized catalog \citep{netopil2016} when not observed by APOGEE. 
We assume that the stars in the cluster all have equivalent metallicity and derive the expected occurrence rate using the posterior distributions of the fits to Equation \ref{eq:xpfit} with [Fe/H] performed in \S\ref{sec:results}.

There are two important caveats to these predictions. First, these predictions only hold if comparing a collection of planet-search stars with the same mass distribution as \ssamp. Second, these predictions only hold true if a power law is an accurate parameterization of the true shape function over the metallicities of interest. In fact, there is some evidence that small planet occurrence may plateau at host star metallicities greater than $\sim$0.2 dex \citep{zhu2019}.
In addition to these limitations, 
predictions for clusters with [Fe/H] less than $\sim$-0.5 dex and greater than $\sim$0.2 dex rely on extrapolation and may be suspect as a result.

These predicted occurrence rates are meant to serve as a benchmark with which to compare age trends in the planet population. While it is tempting to explain differences between the population of planets in a cluster against the field population by invoking the age of the cluster or even the cluster environment, we show here that the metallicity of a cluster is a strong confounding variable, and alone could be responsible for a 100\% increase in the planet occurrence rate for as small a difference as $\sim$0.1-0.2 dex. In this way, accounting for metallicity effects in planet occurrence rates is a crucial step in understanding the difference between the populations of young planets in clusters, and older planets in the field (or in older clusters).

\subsubsection{Diffusion as Yet Another Confounding Variable in Age-Metallicity Correlations}

One more important consideration when interpreting the correlations between planet occurrence and stellar abundances is the role of atomic diffusion. 
Atomic diffusion acts to deplete the surface abundances of certain elements by as much as $\sim$0.15 dex, depending on stellar mass and age \citep{souto2018a, souto2019}. 
These processes complicate the interpretation of results such as those from this work, and general metallicity-planet occurrence rate trends because surface metal abundances are lower than the abundances of the nebula from which the star and planetary system formed. The relative depletion of surface abundance is a complicated function of stellar age, mass, and chemical species,
complicating matters further. 
Ideally, one would correct for diffusion to estimate the initial abundances of the nebula in interpreting the role of specific elements in shaping the planetary distribution function, or estimating, e.g., initial abundances of planetary atmospheres assuming they are similar to the nebular composition. However, such a correction relies on accurate estimates of stellar masses and ages, and an accurate and precise model of the effects of diffusion.

\section{Conclusion} \label{sec:conclusion}

In this paper we investigate the trends in the distribution of \kepler\ planets with the chemical abundances of their host stars as measured for stars in the APOGEE-KOI program \citep{fleming2015}. Leveraging precise atmospheric parameters as measured by high S/N, high resolution near-infrared spectra, we derive precise planetary radii ($\sigma_{R_p}\approx 3.4\%$) for 544 \kepler\ planet candidates. Using this sample of planet hosts, along with a control sample representative of the planet-search sample, we measure the abundance distribution functions for the \kepler\ field stars and derive planet occurrence rates as a function of abundance ratios for C, Mg, Al, Si, S, K, Ca, Mn, Fe, and Ni.
In general, we find that the enhancement of any of these ten elements correlates with increased occurrence rates, and the strength of the correlation between planet occurrence rate and abundance ratio is consistent across these ten elements.

At $P<10$~days, we find that an enhancement of 0.1 dex in any of the ten elements in this study results in a $\approx$20\% increase in the occurrence of Super-Earths and a $\approx$60\% increase in the occurrence of Sub-Neptunes. The strength of these correlations are weaker for planets with $P=10$-$100$~days,
 and we can only confidently confirm a positive correlation with the occurrence rate of Sub-Saturns and the enhancement of metals in this period regime. 
 While we are unable to contribute the increase in occurrence rates to any one particular element, we argue that this is due primarily to astrophysical correlations caused by Galactic chemical evolution, and a more rigorous approach or a modified sample is needed to fully disentangle such degeneracies.

Finally, we conclude this work with a caution to the interpretation of trends in planet demographics in the context of Galactic chemical evolution and the ages and masses of planet hosting stars.

\acknowledgements

We thank the anonymous reviewer whose comments improved the quality of our work. 

CIC acknowledges support by NASA Headquarters under the NASA Earth and Space Science Fellowship Program through grant 80NSSC18K1114.
SH is supported by an NSF Astronomy and Astrophysics Postdoctoral Fellowship under award AST-1801940.
DAGH acknowledges support from the State Research Agency (AEI) of the Spanish
Ministry of Science, Innovation and Universities (MCIU) and the European
Regional Development Fund (FEDER) under grant AYA2017-88254-P.
JT acknowledges that support for this work was provided by NASA through the NASA Hubble Fellowship grant \#51424 awarded by the Space Telescope Science Institute, which is operated by the Association of Universities for Research in Astronomy, Inc., for NASA, under contract NAS5-26555. 

Funding for the Sloan Digital Sky 
Survey IV has been provided by the 
Alfred P. Sloan Foundation, the U.S. 
Department of Energy Office of 
Science, and the Participating 
Institutions. 

SDSS-IV acknowledges support and 
resources from the Center for High 
Performance Computing  at the 
University of Utah. The SDSS 
website is www.sdss.org.

SDSS-IV is managed by the 
Astrophysical Research Consortium 
for the Participating Institutions 
of the SDSS Collaboration including 
the Brazilian Participation Group, 
the Carnegie Institution for Science, 
Carnegie Mellon University, Center for 
Astrophysics | Harvard \& 
Smithsonian, the Chilean Participation 
Group, the French Participation Group, 
Instituto de Astrof\'isica de 
Canarias, The Johns Hopkins 
University, Kavli Institute for the 
Physics and Mathematics of the 
Universe (IPMU) / University of 
Tokyo, the Korean Participation Group, 
Lawrence Berkeley National Laboratory, 
Leibniz Institut f\"ur Astrophysik 
Potsdam (AIP),  Max-Planck-Institut 
f\"ur Astronomie (MPIA Heidelberg), 
Max-Planck-Institut f\"ur 
Astrophysik (MPA Garching), 
Max-Planck-Institut f\"ur 
Extraterrestrische Physik (MPE), 
National Astronomical Observatories of 
China, New Mexico State University, 
New York University, University of 
Notre Dame, Observat\'ario 
Nacional / MCTI, The Ohio State 
University, Pennsylvania State 
University, Shanghai 
Astronomical Observatory, United 
Kingdom Participation Group, 
Universidad Nacional Aut\'onoma 
de M\'exico, University of Arizona, 
University of Colorado Boulder, 
University of Oxford, University of 
Portsmouth, University of Utah, 
University of Virginia, University 
of Washington, University of 
Wisconsin, Vanderbilt University, 
and Yale University.

This research has made use of the NASA Exoplanet Archive, which is operated by the California Institute of Technology, under contract with the National Aeronautics and Space Administration under the Exoplanet Exploration Program.

This work has made use of data from the European Space Agency (ESA) mission \emph{Gaia}  (\url{https://www.cosmos.esa.int/gaia}), processed by the \emph{Gaia} Data Processing and Analysis Consortium (DPAC, \url{https://www.cosmos.esa.int/web/gaia/dpac/consortium}). Funding for the DPAC has been provided by national institutions, in particular the institutions participating in the \gaia\ Multilateral Agreement.

This research has made use of NASA’s Astrophysics Data System.

This publication makes use of data products from the Two Micron All Sky Survey, which is a joint project of the University of Massachusetts and the Infrared Processing and Analysis Center/California Institute of Technology, funded by the National Aeronautics and Space Administration and the National Science Foundation.

\software{{\texttt{astropy} \citep{astropy}}, 
         {\texttt{dustmaps} \citep{dustmaps}},
         {\texttt{emcee} \citep{emcee}},
         {\texttt{matplotlib} \citep{matplotlib}}, 
         {\texttt{numpy} \citep{numpy}}, 
         {\texttt{pandas} \citep{pandas}}, 
         {\texttt{scikit-learn} \citep{scikit-learn}}, 
         {\texttt{scipy} \citep{scipy}},
         \texttt{topcat} \citep{topcat}
         }

\facilities{Sloan (APOGEE), \kepler, 2MASS, \emph{Gaia}, NASA Exoplanet Archive} \\

\appendix
\setcounter{table}{0}
\renewcommand{\thetable}{A\arabic{table}}

\section{Description of \texttt{Isofit}}\label{sec:isofit}

\texttt{isofit} makes use of the \texttt{DFInterpolator} from the \texttt{isochrones} package \citep{morton2015} to interpolate between a grid of MESA Isochrones and Stellar Tracks (MIST) models \citep{dotter2016,choi2016}. The parent grid is defined from the MIST grid of models with solar-scaled alpha abundances and rotation and interpolated in initial [Fe/H] ([Fe/H]$_\mathrm{init}$), initial mass ($M_\mathrm{init}$), and Equivalent Evolutionary Phase ($EEP$)\footnote{For a detailed description of the $EEP$~ parameter, see \cite{dotter2016}.}. The points in the parent grid for each of these parameters are $0.1 \leq M_\mathrm{init}/M_\odot \leq 8$ in steps of 0.02, $-2 < \mathrm{[Fe/H]_{init}} < 0.5$ in steps of 0.05 dex, and $202 \leq EEP \leq $ 1710, in steps of 1. The range in $EEP$ roughly represents each step in a stellar evolutionary track from the Zero-Age Main Sequence to the beginning of the White Dwarf cooling track. 
In total, the parent grid contains $\sim$15 million valid models.

To infer model parameters for a given set of observations, in this case ${\theta}_i$ = $\{ \teff$,  $\log g$, $\mathrm{[Fe/H]}$, $\pi$, $Ks$, $E(B-V) \}$, \texttt{isofit} computes the likelihood for the model input parameters, ${x}_i$ = \{ $M_\mathrm{init}$, $\mathrm{[Fe/H]_{init}}$, $EEP$, $d$, $E(B-V)$ \} and derives an integrated posterior distribution over all likelihoods and priors \citep[e.g.,][]{serenelli2013, huber2017}.  More specifically, the posterior probability is given by
\begin{subequations}
\begin{align}
    p(\vect{x} | \vect{\theta}) &\propto  p(\vect{x}) p(\vect{\theta} | \vect{x}) \\
    &\propto p(\vect{x}) \prod_i \mathrm{exp}\left[- \frac{(\theta_i- \theta_i(\vect{x}))^2}{2 \sigma_{\theta,i}^2}  \right]
\end{align}
\end{subequations}
where $\sigma_{\theta,i}$ are the Gaussian errors on the measurement $\theta_i$, and $\theta_i(\vect{x})$ are the inferred model parameters for input vector $\vect{x}$. The likelihood for the inferred model parallaxes is given by
\begin{align}
    p(\pi | d) \propto \mathrm{exp}\left[- \frac{1}{2 \sigma_{\pi}^2}\left(\pi- \frac{1}{d} \right)^2  \right]
\end{align}
where $d$~is the model distance used to derive apparent magnitudes. For apparent magnitudes, in this case $Ks$, \texttt{isofit} calculates the inferred model apparent magnitude using the MIST grid of bolometric corrections, $BC_m$, the inferred model distance modulus, $\mu = 5 \log d -5$, and the inferred model bolometric magnitude, $M_\mathrm{bol}$,
\begin{align}
    m = M_\mathrm{bol} - BC_m + \mu + A_m
\end{align}
where $A_m$~is the extinction in band $m$, calculated from $E(B-V)$ and the extinction law from \cite{wang2019}.

For model output parameters that do not have associated observations, we assume a flat prior. The exception for this is distance, which has a decreasing density prior with a length scale, $l = 1350\,\mathrm{pc}$ \citep[as in, e.g.,][]{bailer-jones2015, huber2017}, given by
\begin{align}
    p(d) \propto  \frac{d^2}{2 l^3} e^{d/l} \; .
\end{align}
Finally, we take the natural log of each term, and sum them together to get the log-likelihood estimate for a given set of input parameters, $\vect{x}_i$.

To find the initial best fit model, \texttt{isofit} calculates the log-likelihood for all the models of an initial course grid, interpolated from the parent grid with steps of 0.05 $M_\odot$, $5\;EEP$, and $0.1$~dex in [Fe/H], that agree with the spectroscopic parameters, $\teff, \log g,$ and [Fe/H] within $\pm5\sigma$. Then, \texttt{isofit} calculates a fine grid around the course grid point that returns the maximum log-likelihood, and repeats the process but with finer step sizes of 0.01 $M_\odot$, $0.5\;EEP$, and $0.01$~dex in [Fe/H]. 
We then instantiate an MCMC routine \citep[\texttt{emcee};][]{emcee} with a Gaussian ball centered around the model parameters that return the largest log-likelihood with the observed $\pi^{-1}$ and $E(B-V)$.

\section{Results of Statistical Tests Comparing \csamp\ and \psamp\ Abundances}\label{sec:appstats}

We first test for normality in each distribution using the Shapiro-Wilkes test for normality. 
In this case, we find that the abundance distributions in \csamp\ are only consistent with a normal distribution in the case of [Fe/H], [Si/Fe], and [C/Fe], i.e., for all other abundances the $p$-value was sufficiently low $(p<0.001)$ that we reject the null hypothesis that the data were pulled from a normal distribution. This lack of normality motivates us to adopt non-parametric tests: the Kolmogorov-Smirnov (KS) test,
the Mann-Whitney U-test (MW),
and the Brown-Forsythe test (BF).  
The KS test is used to measure any difference between two cumulative distribution functions. Because of this the KS test is applicable in a variety of situations, but in general is not very sensitive. 
Therefore, we also apply the MW test, which tests for differences in the means of two samples, and the BF test which tests for differences in the variances of the two samples. For each subsample of planet type, we apply these three tests against the abundance distributions in \csamp. We do not conduct tests on sub-samples where $n_{pl}<10$ to avoid erroneous conclusions caused by small number statistics.

We find a few statistically significant differences from this methodology. First, we find that the Fe abundances for each planet size class is not significantly enhanced compared to the field,
except in the case of hot planets, and in particular, hot Sub-Neptunes. The only other elemental abundance that shows statistically significant differences between \psamp\ and \csamp\ distributions is [K/Fe]. However, [K/Fe] is highly correlated with [Fe/H], so it's most likely that this result is only tracing differences in the Fe abundances already known. There are also significant differences detected between the Si and Mg abundance distributions of \csamp\ and \psamp.


With regard to Fe, we find that all hot planet hosts are enhanced compared to \csamp, a result that agrees with the literature, and has been pointed out by a number of authors \citep{mulders2016,wilson2018,petigura2018,narang2018,owen&murrayclay2018}. 
Hot planets also seem to be correlated with [K/Fe], in that they have a significantly different mean compared to the field. 
This is likely due to hot planet hosts having [Fe/H] distributions skewed above solar, and as a result there are no hosts with high K abundances relative to iron. In other words, the apparent differences in the \psamp\ [K/Fe] distributions are driven by underlying differences in [Fe/H]. A similar conclusion can be drawn about the apparent difference in the means of the [Mg/Fe] distributions for hot sub-Neptunes.

Finally, there are differences in the [Si/Fe] distribution of \psamp\ and \ssamp. However, these correlations are likely a combination of already known metallicity correlations and correlations in Galactic chemical evolution. \psamp\ show significantly lower variance in [Si/Fe] than \csamp, which is to be expected due to the higher average, and comparatively limited range, in [Fe/H] of \psamp. 
This is caused by the higher fraction of thick disk (high-[$\alpha$/Fe], low [Fe/H]) to thin disk (low-[$\alpha$/Fe], high [Fe/H]) stars in \csamp\ compared to \psamp. 
Because there are fewer thick disk planet hosts, those stars do not contribute significantly to the variance of \psamp\ compared to \csamp. Thus the variance in the [Si/Fe] distribution is expected to be lower in \psamp\ than in \ssamp, a result confirmed by this exercise.

\section{Occurrence Rate Methodology}\label{sec:methods}

\subsection{Formalism and Definitions}\label{sec:formalism}

Our methodology treats the detection of a transiting planet as an independent random process, i.e., as a Poisson process. We use NPPS as our definition of planet occurrence, $f$. As a note, this is not equivalent to the quantity of the Fraction of Stars With Planets (FSWP) that is often used as a definition of planet occurrence. For a transit survey, a measurement of FSWP requires detailed modeling of multiplicity, mutual inclinations, and other effects that are outside the scope of this paper. We instead default to NPPS, which is blind to these properties. In the interest of comparing to other works \citep[e.g.,][]{petigura2018}, we often report our occurrence rates in units of number of planets per 100 stars.

For a given star with properties $\vect{z}$, the probability of hosting a planet with properties $\vect{x}$ can be expressed as
\begin{align}
    df = \frac{\partial f (\vect{x}, \vect{z})}{ \partial \vect{x}} d\vect{x} \;\;,
\end{align}
where integrating over the planet properties, $\vect{x}$, gives $f(\vect{z})$, the average number of planets for a star with properties $\vect{z}$. In this paper, $\vect{x}$ is some combination of $\log R_p$ and $\log P$, and $\vect{z}$ is the abundance of some chemical species. We typically adopt $Z$ as our symbol for metallicity, and use $X$ to refer to an arbitrary chemical element. 
For compactness we adopt the following notation for a partial derivative of $f$ with respect to an arbitrary variable $x_1$ and $x_2$, 
\begin{align}
    f_{x_1} \equiv \frac{ \partial f}{\partial \log x_1} \; ; \; f_{x_1,x_2} \equiv \frac{\partial^2f}{\partial \log x_1 \; \partial \log x_2} \;\;.
\end{align}
This is similarly defined for chemical abundances as, 
\begin{align}
    f_{X} \equiv \frac{ \partial f}{\partial [X/H] } \; ; \;  f_Z \equiv \frac{ \partial f}{\partial [Fe/H] } \;\;,
\end{align}
where chemical abundance ratios are always defined with respect to hydrogen. Note, this is a change from \S\ref{sec:hoststars} where we were searching for new trends independent of [Fe/H]. For the remainder of this study, we wish to compare the strength of the correlation with the enhancement of each chemical element with planet occurrence. Thus, we adopt [X/H] to express each element on a similar scale. 
We express the differential distribution for NPPS as 
\begin{align}
    f_{\vect{x}} (\vect{x}, \vect{z}) d\vect{x} \equiv C g(\vect{x}, \vect{z} ; \theta)
\end{align}
where $g(\vect{x}, \vect{z} ; \theta)$ is a shape function (i.e., some parametric prescription used to describe the PLDF) that depends on planet and/or stellar properties with shape parameters $\theta$. A functional form for $g(\vect{x},\vect{z})$ must be assumed, with as many shape parameters, $\theta_i$, as necessary.

The total number of planets orbiting $n_\star$ stars (indexed by $i$) is then
\begin{subequations}
\begin{align}
    n_{pl} &= C \sum^{n_\star}_{i} \int g(\vect{x}, \vect{z_i} ; \theta) d \vect{x} \\
    &= n_\star C \int \fstar(\vect{z}) g(\vect{x}, \vect{z} ; \theta) d \vect{x} d \vect{z} \label{eq:subeq11}
\end{align}
\end{subequations}
where the integration takes place over some range of planet properties. In equation \ref{eq:subeq11}, the sum over all stars is replaced by an integral over the probability distribution of stellar properties, $\fstar(\vect{z})$.  $\fstar$ is normalized so that $\int \fstar(\vect{z}) d\vect{z} = 1$. In practice, the summation over the known properties of each planet search star is preferable, but in principle an accurate measurement of $\fstar(\vect{z})$ gives an equivalent result.

In this work, we calculate our occurrence rates for bins with some combination of [X/H], [Fe/H], $\log P$, and/or $\log R_p$. The width of a bin is given by $\Delta \vect{x}$, where $\Delta \vect{x} = \prod_i \Delta x_i $, where $i$ indexes over the dimensions of the bin. The occurrence within a bin, $f_\mathrm{bin}$, depends on the number of independent trials, $n_\mathrm{trial}$, that yield a detected planet, and the survey efficiency, $\eta$, which may depend on both stellar and planetary properties. We compute $n_\mathrm{trial}$ as,
\begin{align}
        n_{\mathrm{trial}, \, j} &= \sum_{i}^{n_\star} \eta(\vect{x}_j, \vect{z}_i) \\
        &= n_\star \left \langle \eta (\vect{x}_j) \right \rangle
\end{align}
where $\langle . \rangle$ denotes the arithmetic mean. For a given survey efficiency, $\eta(\vect{x},\vect{z})$,
$\eta$, also includes the number of false positives in a given sample. However, because we have removed RV variable sources from \psamp, the false positive rate from astrophysical sources is negligible. We also ignore false alarms from instrumental effects. While incorporating such false alarms is important for deriving robust occurrence rates in principle, the actual false alarm rate in \kepler\ is negligible for planets with $P\lesssim 300$~days \citep{mulally2016}.

Following the examples of \cite{bowler2015} and \cite{petigura2018}, we assume that the planet occurrence is log-uniform within a given bin of size $\Delta \vect{x}_{i,j}$, which should be reasonable at small enough bin sizes. In this case, $n_\mathrm{trial}$ for a bin can be expressed as, 
\begin{align} \label{eq:ntrial}
    n_\mathrm{trial} = \frac{n_\star}{\Delta \vect{x}_{i,j}} \int \left \langle \eta (x_{i,j}) \right \rangle d\vect{x} \;\; .
\end{align}
Thus, for a given cell with $n_\mathrm{trial}$~trials and $n_\mathrm{pl}$~detected planets, the likelihood of $f_\mathrm{bin}$ can be described by a binomial distribution of the form, 
\begin{align}
    P(f_\mathrm{bin}|n_\mathrm{pl}, n_\mathrm{trial}) &= P(n_\mathrm{pl}|f_\mathrm{bin}, n_\mathrm{trial}) \\
    &=C f_\mathrm{bin}^{n_\mathrm{pl}} (1-f_\mathrm{bin})^{n_\mathrm{nd} }
\end{align}
where $n_\mathrm{nd} = n_\mathrm{trial}-n_\mathrm{pl}$ is the number of non-detections, and $C$ is a normalization constant that takes the form
\begin{align}
    C = \frac{(n_\mathrm{trial}+1)\, \Gamma (n_\mathrm{trial}+1)}{\Gamma (n_\mathrm{pl}+1)\, \Gamma(n_\mathrm{nd}+1)}
\end{align}
When analyzing occurrence rates as a function of stellar properties, we bin the planet and stellar properties in bins bounded by [$\log P_1, \log P_2], [\log R_{p,1}, \log R_{p,2}$],~and $[X_1, X_2]$. In this way, we calculate $n_\mathrm{trial}$ via equation \ref{eq:ntrial}, multiplied by the fraction of stars, $F_\star$, with abundances between $[X_1,X_2]$, 
\begin{align}\label{eq:frac}
    F_\star = \int_{X_1}^{X_2} \fstar (X) dX \;\; .
\end{align}

In the case where there are no detected planets in a given bin, we  estimate an upper limit on the occurrence rate for that bin by numerically solving for the integral, 
\begin{align}
    \int^{f_\mathrm{bin}}_0  P(f|n_{pl}, 0) df = 90 \% \;\;.
\end{align}
We note that in practice, we only use these upper limits for display purposes, and don't directly incorporate them into our analysis.

\subsection{Parametric Fits to the Differential Occurrence Rate Distributions} \label{sec:fitting}

We wish to express the strength of the correlation between a star's chemical composition and the occurrence of various types of planets. We do this via a parametric relation, often with a power law in this work, to gauge the strength of the correlation. 
We note here that a power-law prescription for describing the differential occurrence rate does not necessarily reflect the true shape function of the occurrence rate. However, such a prescription can give a precise estimate of the average strength with which the differential occurrence rate relies on the underlying abundance. So, while a power-law fit of this form gives a precise estimate of correlation strength, this prescription may not be appropriate to robustly predict $f$ for a given stellar sample.

To this end, we can estimate the differential occurrence rate for a bin of size $\Delta \vect{x}$ via the following relation, 
\begin{align}
    C g(\vect{x}, \vect{z}; \theta) = \frac{f_\mathrm{bin}}{\Delta \vect{x}} \;\; .
\end{align}
To find the best fit shape parameters,  $\theta$, we maximize the log-likelihood of the function, described for a given bin, $i$, by
\begin{align}
    \ln L_i = n_{pl,i} \ln C g \Delta \vect{x} + n_{nd,i} \ln (1- C g \Delta \vect{x})
\end{align}
where each cell is an independent constraint on $C g \Delta \vect{x}$. Therefore, by maximizing the combined log-likelihood over all bins, indexed by $i$, 
\begin{align}
   \ln L =  \sum_i \ln L_i \;\;,
\end{align}
we find the best fit shape parameters, $\theta$. To zero in on the best fit shape parameters, we apply arbitrarily small bin sizes. As a result a number of these bins have few or no detected planets. Because the log-likelihood function we apply incorporates non-detections, this methodology is stable even to few detected planets. This approach has two advantages.  
The first is that our assumption that a bin is log-uniform has more merit in smaller bins, and the second is that for a small enough bin the errors on the occurrence rate will be dominated by Poisson statistics rather than uncertainties in $\fstar$. We expand upon this assumption in \S\ref{subsec:fstar}. After finding the best-fit parameters, $\theta$, that maximize the likelihood function, we then apply an MCMC routine\footnote{as implemented in the python package \texttt{emcee} \citep{emcee}} initialized at those parameters to explore the range of credible models.

\subsection{Completeness Corrections}\label{subsec:completeness}

In this subsection we describe our completeness model, $\eta(\vect{x}, \vect{z})$. Our approach varies slightly from most previous \kepler\ occurrence rate studies, because we also correct for biases inherent in the follow-up program as well. Most \kepler\ occurrence rate studies (this one included) select some subsample of planet-search stars from the \kepler\ Input Catalog \citep{brown2011}, and then compute a detection efficiency model from the \kepler\ pipeline that is marginalized over their planet-search sample. However, this strategy alone is not adequate because we have additional biases that are not quantified by the \kepler\ detection pipeline. In other words, inclusion in \psamp\ is dependent on more than membership in \ssamp\ and a detected planet candidate in \kepler. There are additional biases imposed by limitations in the APOGEE selection function, instrumental setup, and spectroscopic analysis pipeline that must be considered.

We consider four unique biases for a planet candidate to be included in \psamp: first, the planet has to transit its host star; second, the transiting planet must be included in the DR24 KOI catalog; third, the planet's host star must have been observed by APOGEE; and fourth, ASPCAP must have returned reliable abundance and spectroscopic parameters for the host star. We take each of these criteria as their own independent process, so that we can model the total completeness as the product of the probabilities that a planet candidate passes each step. Thus, our completeness model, $\eta$, can be described by four terms: the geometric probability that a planet with a randomly oriented orbital plane transits its host star $(p_\mathrm{tra})$, the probability that a transiting planet is detected by \kepler $(p_{det})$, the probability that a candidate was observed in the APOGEE-KOI program $(p_\mathrm{apo})$, and the probability that ASPCAP doesn't fail to produce reliable atmospheric parameters for the host star $(1-p_\mathrm{fail})$. We go into more detail for each of these terms below before presenting the combined, average survey efficiency.

\subsubsection{Transit Probability ($p_\mathrm{tra}$)}

The probability that a given planet transits depends only on the geometry of the orbit. We make the assumption that the inclination of all orbits follows an isotropic distribution. Under these assumptions, the probability for a planet to transit in our sample is simply,
\begin{align}
    p_\mathrm{tra} &=  \frac{0.9 R_\star}{a(1-e^2)}  
\end{align}
where we set $e=0$ for simplicity, include a factor of 0.9 to account for our cut on impact parameter, and $a$ is calculated from $M_\star$ and $P$ using Kepler's third law.

\subsubsection{Kepler Pipeline Detection Efficiency ($p_\mathrm{det}$)}

In this section we give our model for the \kepler\ DR24 pipeline completeness. For a planet to be detected, it must have a high enough $S/N$ to be detected, and it must pass multiple levels of vetting. 
In place of transit $S/N$, the \kepler\ pipeline utilizes the Multiple Event Statistic (MES). The MES is a measure of the null hypothesis that a \kepler\ light curve does not have a transit signal at a given epoch $(t_0)$, duration $(t_\mathrm{dur}$), and period $(P)$. Under the assumption of white noise, the MES distribution is Gaussian with a mean of zero and variance of one. Under the alternative hypothesis however, i.e., that there is a transit signal, the mean of the MES distribution is shifted by a constant proportional to the $S/N$ of the transit. \kepler\ defines a threshold of MES $> 7.1$ for a detected signal.

It is common to parameterize the \kepler\ pipeline completeness with a one-dimensional model in expected MES from a putative transiting planet (MES$_\mathrm{exp}$). However, this is not an appropriate model for the DR24 pipeline completeness, because an introduced $\chi^2$ metric used to veto false alarms severely reduced the completeness for planets with $P>40$~days \citep{christiansen2016}. Therefore, we parameterize \kepler's pipeline efficiency in two dimensions, with MES$_\mathrm{exp}$ and $P$. Rather than utilizing an analytical model, we take a purely empirical approach to assess $p_\mathrm{rec}(P, \mathrm{MES_{exp}})$ \citep[as in, e.g., ][]{petigura2013,dressing&charbonneau2015}.

We apply the results of the Monte Carlo injection and recovery tests executed by \cite{christiansen2016}. To ensure that the light curves are representative of the light curves from \ssamp, we remove the injection results from stars that are inconsistent with stars in \ssamp. We applied the following cuts to the sample of stars with injected signals, $R_\star<2 R_\odot$, $4700\,\rm{K} < T_\mathrm{eff} < 6360\,\rm{K}$, and $RUWE < 1.2$. We also attempted to limit the collection of transit recoveries to stars in APOGEE fields, but we found no differences in $p_\mathrm{rec}$ between stars in and out of APOGEE fields so elected to use stars from the entire \kepler\ field to improve our statistics. In total, this resulted in 85,257 individual injection and recovery tests.

\begin{figure}
\centering
\includegraphics{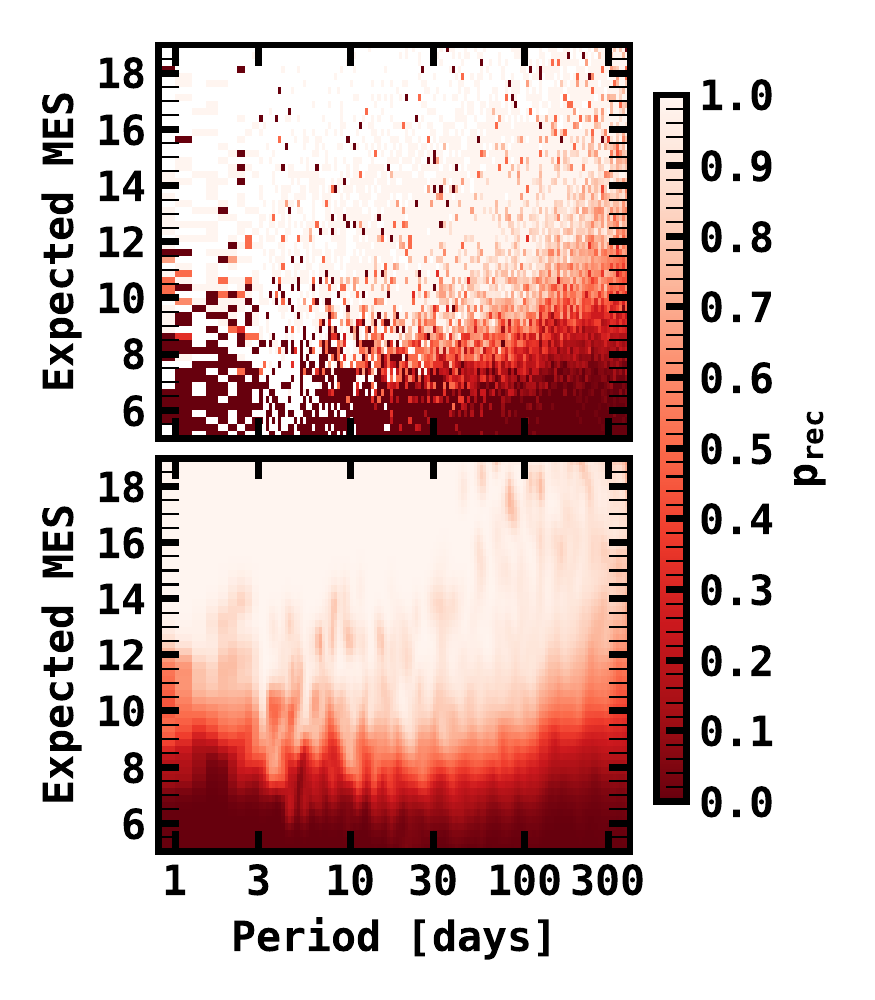}
\caption{The DR24 \kepler\ pipeline detection efficiency. The shading shows the probability of the \kepler\ TPS module to recover a transit signal ($p_{\mathrm{rec}}$) with an expected Multiple Event Statistic (MES$_\mathrm{exp}$) and Period ($P$). Darker shades of red represent a lower recovery fraction, and lighter shades represent a higher recovery fraction. \textit{Top}: The recovery probability ($p_\mathrm{rec}$) of the \kepler\ pipeline from the injection and recovery experiments in \cite{christiansen2016}. White spaces denote bins with no data. \textit{Bottom}: The interpolated, and smoothed grid of $p_\mathrm{rec}$ that we apply for our completeness corrections. 
}
\label{fig:pdet_mes}
\end{figure}

To measure $p_\mathrm{rec}(P, \mathrm{MES_{exp}})$, we defined a grid in $P$ and $\mathrm{MES_{exp}}$ and measured the fraction of recovered injections within each bin. The injections were performed using a uniform prior in $P$. However, we are interested in assessing the completeness in logarithmic bins. To account for this, we define linear bins from $P=0.25 - 10$~days, in steps of 0.25 days, and then 75 logarithmically spaced bins from 10-500 days. We binned the $\mathrm{MES_{exp}}$~of each putative signal in steps of 0.5 from 0-20, where the recovery fraction for $\mathrm{MES_{exp}}>20$ is assumed to be constant. The resulting grid is shown in the top panel of Figure \ref{fig:pdet_mes}. Due to few injections at $P\lesssim 10$~days, and at large MES, we replace any binned points that have $<3$~injections, $\mathrm{MES_{exp}}>13$, and $P<40$~days with $p_\mathrm{rec}=0.997$, the expected pipeline efficiency at arbitrarily large MES. We then convolve the $p_\mathrm{rec}$~grid with a Gaussian kernel having a width of the bin size in each dimension to smooth over any artificial features and interpolate over points with no injections, such as those at arbitrarily low $\mathrm{MES_{exp}}$ or very short $P$. The resulting grid that we apply is shown in the bottom panel of Figure \ref{fig:pdet_mes}.

For each star in \ssamp, we calculate $p_\mathrm{rec}$ over the $P$-$R_p$ grid. We calculate $\mathrm{MES_{exp}}$ for every combination of $P,R_p$ for each star in \ssamp\ and interpolate on the grid defined above to measure $p_\mathrm{rec}(P,R_p)$. We calculate $\mathrm{MES_{exp}}$ as 
\begin{align}
    \mathrm{MES_{exp}} &= \left(\frac{R_p}{R_\star}\right)^2 \frac{1}{\sigma_{cdpp}(t_\mathrm{dur})} \sqrt{\frac{T_{obs}}{P}} 
\end{align}
where $T_{obs}$ is the observation baseline and $\sigma_{cdpp}(t_\mathrm{dur})$~is the combined differential photometric precision \citep{christiansen2012} on the timescale of the transit duration, $t_\mathrm{dur}$. We calculate $t_\mathrm{dur}$ using equation \ref{eq:tdur}. The \kepler\ data products contain measurements of $\sigma_{cdpp}$ on timescales of 1.5, 2.0, 2.5, 3.0, 3.5, 4.5, 5.0, 6.0, 7.5, 9.0, 10.5, 12.0, 12.5, and 15.0 hours, which are the transit durations searched by \kepler's TPS module. We interpolate between the values provided at these timesales to estimate $\sigma_{CDPP}$ for any arbitrary $t_\mathrm{dur}$. 
The values for $T_{obs}$ and $\sigma_{cdpp}$ were taken from the DR24 stellar properties table hosted on the NExScI Exoplanet Archive\footnote{https://exoplanetarchive.ipac.caltech.edu \citep{akeson2017}}.

One last requirement in the \kepler\ pipeline is that the planet candidate must have at least three transits in the data. We quantify this probability by the window function, $p_\mathrm{win}$, as estimated in \cite{burke2015},
\begin{align}
\begin{split}
    p_\mathrm{win} &= 1- (1-f_d)^M - M f_d (1-f_d)^{M-1}\\
    &- \frac{M(M-1)}{2} f_d^2 (1-f_d)^{M-2} \; ,
\end{split}
\end{align}
where $f_d$ is the duty cycle of the observations and $M=T_{obs}/P$. For planets with $P<300$~days, this probability is typically $> 0.98$, so this is not a strong source of bias in our sample, but we include it for posterity.

\begin{figure}
\centering
\includegraphics{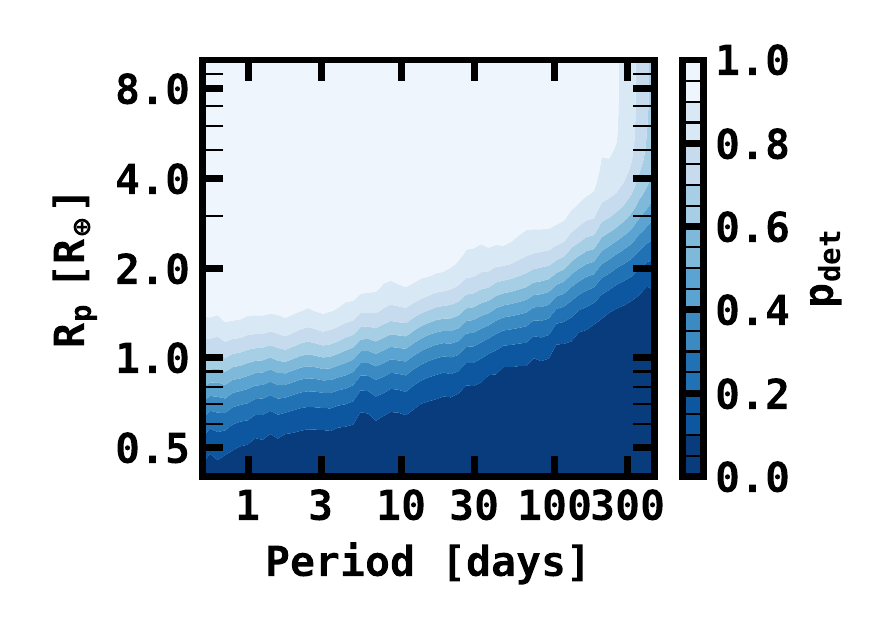}
\caption{The average DR24 \kepler\ TPS pipeline detection efficiency for the stars in \ssamp. Darker shades of blue represent a lower detection fraction, and lighter shades represent a higher detection fraction. 
}
\label{fig:pdet}
\end{figure}

The joint probability of \kepler\ detecting a transiting planet, then is given by the product $p_\mathrm{det}=p_\mathrm{rec}\times p_\mathrm{win}$. This full \kepler\ pipeline sensitivity is shown in Figure \ref{fig:pdet}.

\subsubsection{Pre-DR24 Selection Bias and Probability of Fiber Collisions ($p_\mathrm{apo}$)}

The probability that a KOI was included in the APOGEE program depends on the APOGEE field. For fields K04, K06, and K07, 
this bias is dominated by the rejection of APOGEE targets based on fiber collisions. A ``collision" occurs when two fibers, if placed on the plate, would be separated by less than the size of the protective ferrule around each fiber. For the APOGEE-N spectrograph the collision radius corresponds to 71.5\arcsec~\citep{zasowski2013, zasowski2017}. In the case of fiber collisions, one target is assigned priority, and the other is removed from the target list. We don't explicitly cut stars out of \ssamp\ for this purpose because this quantity is dynamic, and priority is assigned on a target by target basis, making it difficult to apply to a field sample. Instead we make a simplifying assumption to correct for this bias in assuming that all KOIs are equally likely not to be observed due to conflicts with other APOGEE-KOI targets. This is not strictly true, as planet candidates with low-mass host stars were given priority over competing planet candidates. However, because there are so few instances where this happens ($<$3\%) and because we remove M dwarfs from \ssamp, this is unlikely to lead to a noticeable bias in stellar host properties compared to the field. This bias is measured by selecting all the planet candidates without a ``False Positive" disposition in DR24 with a host in \ssamp\ for a given field, and comparing that to the targets in the APOGE-KOI program in that field. Note, this is not the same quantity as the number of planets in \psamp\ in that field, because we reject some candidates based on observations from APOGEE. For fields K04, K06, and K07 only five KOIs from DR24 were not observed, resulting in a flat probability of $ p_\mathrm{apo}=0.988$. This is the dominating bias for these fields.

For fields K10, K21, and K16, the most prolific bias is the selection of KOIs before a static catalog was available from \kepler. We consider this bias in conjunction with the fiber collision bias described above, as it largely dominates, and because there is not a catalog of ``expected" planets with which to compare. This is not only due to the nature of the KOI catalogs at the time of target selection, but also improvements in the \kepler\ pipeline used to reject both false positives, such as the detection of photocenter offsets and significant secondary eclipses, as well as false alarms caused by instrumental systematics such as sudden pixel sensitivity dropouts, rolling bands, and abrupt changes in the photometric noise profile between quarters \citep{coughlin2015,mulally2016}.

\begin{figure}
\centering
\includegraphics{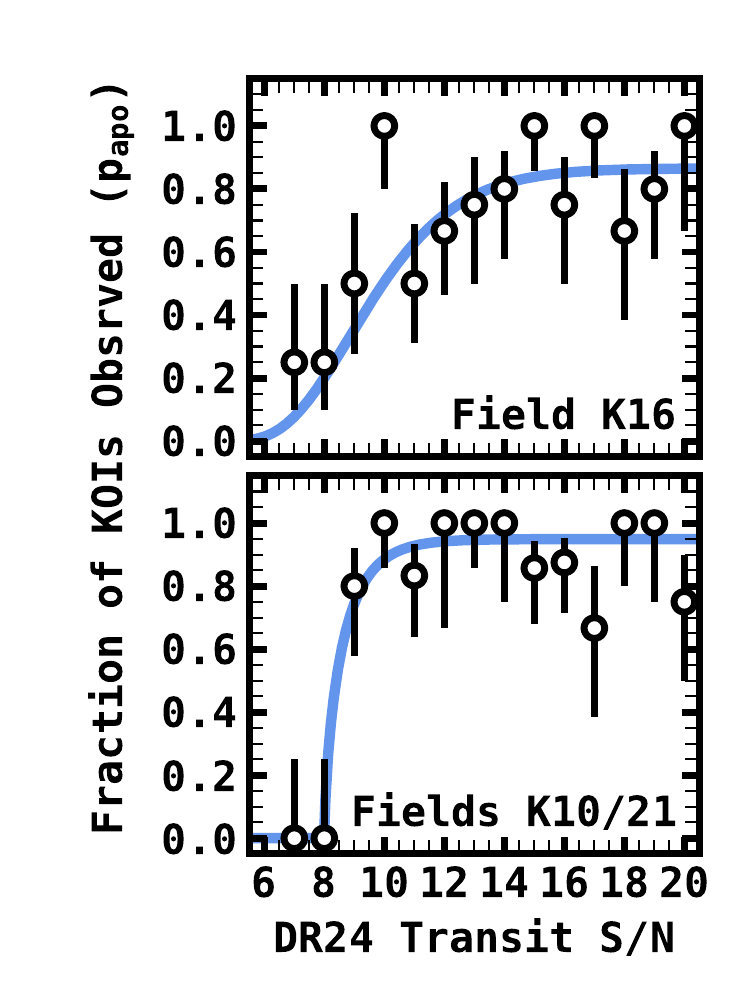}
\caption{The probability that a particular planet candidate is observed in the three fields with early target selection in the APOGEE-KOI survey. There is a bias incurred from selecting planet candidates from a pre-DR24, non-static catalog, as well as avoiding APOGEE fiber collisions. The points show the fraction of DR24 planet candidates observed in APOGEE at a given transit S/N for field K16 (top) and fields K10 and K21 (bottom). The error bars are derived assuming a binomial distribution. The blue lines show our adopted models to correct for this bias.}
\label{fig:biascorrect}
\end{figure}

     The targets for K16 were chosen as part of SDSS-III \citep{eisenstain2011}, and used as a pathfinder for the APOGEE-KOI program. This field observed 163 planet candidates with $H<14$~known at the time, in 2013 August. Of those 163 candidates, 153 had a ``Candidate" status, 4 were confirmed, and 6 were not yet dispositioned \citep{fleming2015}. The DR24 catalog, however, has 166 planet candidates in \ssamp\ with either a ``Candidate" or ``Confirmed" disposition. The APOGEE-KOI program observed 126 (75.9\%) of these planet candidates. This discrepancy is due largely to two effects: improved vetting that removed a significant fraction of the 163 original planet candidates and improvements to the \kepler\ pipeline supplemented with additional data that allowed for the discovery of planet candidates with lower S/N transits. To take this bias into account, we measure $p_\mathrm{apo}$ by taking the ratio of DR24 candidates observed by APOGEE to the total number of DR24 candidates as a function of transit S/N, and model the increasing fraction with a modified gamma cumulative distribution function of the form, 
\begin{align}
    p(S/N) =  \frac{a}{d^{b} \Gamma (b)} \int_0^{S/N}  (\xi-c)^{b-1} e^{-(\xi-c)/d} d\xi  \;\;. 
\end{align}
Measuring the fraction of observed planet candidates in S/N bins of 1.0, we find the best fit parameters to be $a=0.86$, $b=6.0$, $c=3.9$, and $d=1.0$ (See Figure \ref{fig:biascorrect}). This fit implies that at high $S/N$, only 86\% of the \kepler\ planet candidates from August 2013 would survive the more detailed vetting procedures introduced in the DR24 pipeline \citep{coughlin2015,mulally2016}. 
This bias is applied across the $P$-$R_p$~grid with the expected MES in place of the transit S/N in calculating $\left\langle \eta \right\rangle$. This fit is displayed in Figure \ref{fig:biascorrect}. Perhaps surprisingly, we didn't see any significant trend in either $P$ or $R_p$ alone.

We apply the same analysis jointly to fields K10 and K21 (as targets between these two fields were selected from the same KOI catalog) which were observed as part of the SDSS-IV bright time extension program (Beaton et al., in prep). The DR24 catalog contains 318 planet candidates in fields K10 and K21 with either a ``Candidate" or ``Confirmed" disposition. APOGEE observed 130 out of 148 (87.8\%) of these candidates in field K21 and 155 out of 170 (91.2\%) of these candidates in field K10. Combining these two fields, and fitting a modified gamma cumulative distribution function as for Field K21, we find the best fit parameters of $a=0.98$, $b=0.22$, $c=7.5$ , $d=1.7$. This fit implies that at high $S/N$, 98\% of the \kepler\ planet candidates detected at the time of the APOGEE-KOI survey target selection would survive the more detailed vetting procedures introduced in the DR24 pipeline. Therefore, the dominant factor in the discrepancy between the number of planet candidates in these fields and those observed in APOGEE is dominated by improvements to the \kepler\ pipeline that resulted in the detection of lower S/N transits. We multiply this correction over all stars in fields K10 and K21 when calculating $\left\langle \eta \right \rangle$. See Figure \ref{fig:biascorrect}.

\subsubsection{Probability of ASPCAP Failure ($p_\mathrm{fail}$)}

Although the typical star in our sample has a spectrum with $S/N>100$, our sample still includes a non-negligible fraction of dim stars near the $S/N$ limit of ASPCAP's capabilities. Because of these low $S/N$ sources, there is a non-negligible fraction of planet candidates ($\sim$10\%) that would otherwise be in \psamp\ that are not included in our analysis. To account for this bias, we assume the dominant reason for an ASPCAP failure is due to this low $S/N$~effect, although some spectra may fail due to other reasons (e.g., stray light from a bright companion). To account for this bias we model the failure probability as a function of $H$ because the number of cadences are designed to derive reliable orbital solutions, and therefore are not set by any stellar parameters or observable quantities that should bias the results of the occurrence rates.

\begin{figure}
\centering
    \includegraphics{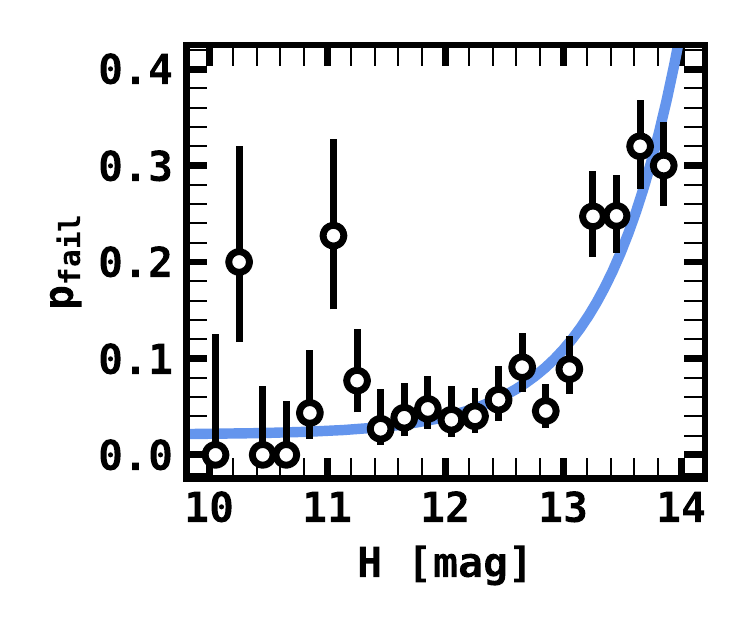}
    \caption{The ASPCAP failure rate as a function of $H$~magnitude. The data points are the fraction of stars observed in the APOGEE-KOI program where ASPCAP did not derive a solution in a given magnitude bin, with error bars assumed from a binomial distribution. The blue line shows our fitted model for the failure rate.}
    \label{fig:aspcapfail}
\end{figure}

To measure the failure probability, $p_\mathrm{fail}$, we take a similar approach to measuring $p_\mathrm{apo}$. We calculate the fraction of stars for a given $H$ bin of 0.2 magnitudes that have an ASPCAP-derived best fit solution for the input parameters (i.e., $\teff$, $\log g$, [M/H], $\xi_t$, [C/M], [N/M], and [$\alpha$/M]). We then fit the fraction of stars without a reliable ASPCAP solution using a simple modified power law of the form, 
\begin{align}
    p_\mathrm{fail}(H) = c + e^{a(H-b)}
\end{align}
where $H$ is the magnitude of the bin. We find the best fit parameters to be $a=1.70$, $b=14.45$, and $c=0.02$. The best fit model and fractions are displayed in Figure \ref{fig:aspcapfail}. These fits imply that the failure rate due to low~$S/N$~should be $\sim$1 at $H=14.45$, and that the failure rate for reasons other than low $S/N$ is 2\%, which is fairly insignificant compared to the other uncertainties considered in this study. 

\subsubsection{$\eta$: The Combined Survey Efficiency}

We calculate the total average survey efficiency for each field as the product of each term described above, given the form
\begin{align}
    \left\langle \eta \right \rangle = \frac{1}{n_\star} \sum_i^{n_\star} p_\mathrm{tra,i} \times p_\mathrm{det,i} \times p_\mathrm{apo,i} \times p_\mathrm{fail,i}\;\;,
\end{align}
where $\left \langle \eta \right \rangle$ is the average survey detection efficiency. The mean survey efficiency for each field is shown in Figure \ref{fig:pdet_all}. In this way, i.e., by marginalizing over all the stars in \ssamp, we've removed stellar properties from our expression for survey efficiency, so that $\eta = \eta(P,R_p)$. This relies on an implicit assumption that chemical abundances are not correlated with survey efficiency. Finally, for each $\log P$-$\log R_p$~bin, we calculate the combined survey efficiency for all the stars in the APOGEE-KOI program as a weighted sum of the efficiency for each APOGEE-KOI field indexed by $i$, 
\begin{align}
\langle \eta \rangle = \sum_i F_{\star,i} \times \langle \eta \rangle_i \;\;,
\end{align}
where $F_{\star,i}$~is the fraction of stars in \ssamp\ that are in field $i$ (see Table \ref{tab:fields}).

\subsection{Errors Due to Uncertainties in $\fstar(X)$}\label{subsec:fstar}

We do not have measured chemical abundances for all the stars in \ssamp, so we do not know precisely how many stars from \ssamp\ are in a given metallicity bin. However, with knowledge of the distribution function over each abundance, $\fstar(X)$, we can estimate this number by integrating the distribution over the bin and multiplying by the total number of stars in \ssamp\ (see Equation \ref{eq:subeq11}).

To define $\fstar$, we fit the abundance distributions in \ssamp\ using the measured abundances in \csamp\ with a Gaussian kernel density estimator (KDE). The choice of bandwidth for the KDE is non-trivial, as it may impart significant bias if overestimated and introduce variance if underestimated. To select the optimal bandwidth, we fit a Gaussian KDE for a large sample of stars in the entire APOGEE database. The intent is to define a realistic distribution function from this sample, draw a number of stars equal to those in \csamp, and compare how well a given bandwidth recreates the defined distribution. We select this sample of stars such that it should broadly reflect our assumptions about the true abundance distributions in \ssamp.

We remove stars from the APOGEE DR16 catalog with $\log g <3.5$, $T_\mathrm{eff}<4000$~K,  $T_\mathrm{eff}>6500$~K, $\pi / \sigma_{\pi}<10$~in \gaia~DR2, and a distance $>1$~kpc as reported by \cite{bailer-jones2018}. In addition to these sample selection cuts, we also apply a number of cuts designed to remove stars with poor quality measurements. We remove stars with $S/N<50$ and any of the following ASPCAP or Star Flags set\footnote{for a description of these flags, see \url{https://www.sdss.org/dr16/algorithms/bitmasks/}}: \texttt{TEFF\_BAD}, \texttt{LOGG\_BAD}, \texttt{METALS\_BAD}, \texttt{ALPHAFE\_BAD}, \texttt{STAR\_BAD}, \texttt{SN\_BAD}, and \texttt{VERY\_CLOSE\_NEIGHBOR}. This leaves $\sim$111,000 stars to define the parent sample.

\begin{figure*}
    \centering
    \includegraphics{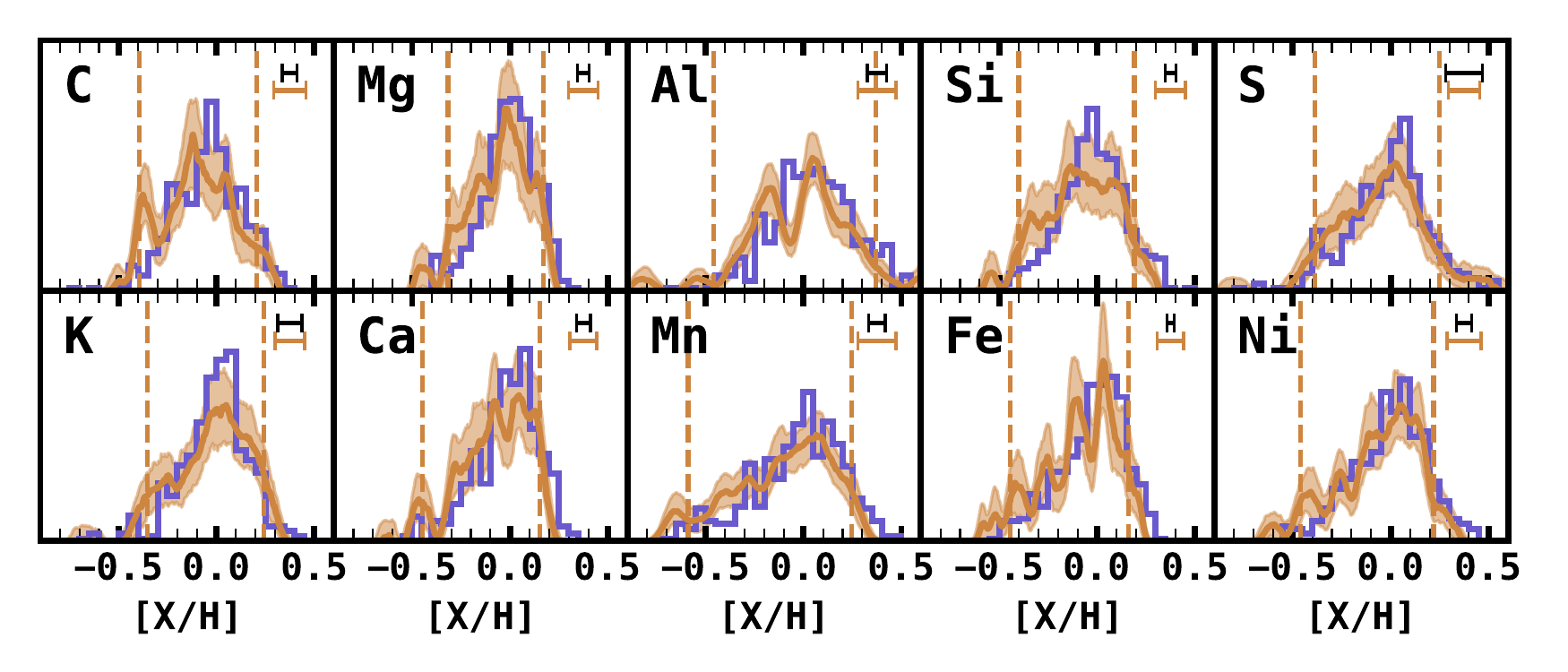}
    \caption{Our measurements for $\fstar(X_i)$, applying a Gaussian KDE. The tan line shows our derived KDE, with the shaded region showing the $1\sigma$ region of credible models obtained from Monte Carlo sampling and bootstrapping. The abundance ratio distributions in \psamp\ are shown in purple for comparison. The bandwidth used to model the abundance ratio distributions for \csamp\ are represented by the tan error bar in the upper corner, and the vertical dashed lines mark off the inner 90th percentiles used in the occurrence rate analyses. The black error bar shows the median abundance uncertainty. The element is noted in the top left of each panel. }
    \label{fig:abundance_1ddist}
\end{figure*}

From this APOGEE dwarf star sample we fit a KDE to the [Fe/H] measurements, and use this as our ground truth metallicity distribution. We then randomly sample 72 measurements from the KDE (chosen to match the number of stars in \csamp), and test each bandwidth from 0.01 -- 0.30 dex, in intervals of 0.005 dex using a Kfolds cross-validation procedure with ten folds. We repeat this experiment 1000 times, and take the mean of the 1000 iterations to be the optimal model. This same experiment is run for each of the ten elements, resulting in our choices of bandwidth for each element given in Table \ref{tab:bandwidths}. We display the resultant KDEs for each abundance in Figure \ref{fig:abundance_1ddist}. 
It's worth noting at this point that we are not assuming that \ssamp\ has the same abundance distribution as the APOGEE dwarf sample. Rather, we are assuming that optimizing our model selection for fitting the APOGEE dwarf sample will also optimize the model selection for fitting the abundance distributions of \csamp.

\begin{table}
    \centering
    \begin{tabular}{cc} \hline \hline 
        $X$& $\sigma_{X/H}$ \\ \hline
        C & 0.12 \\
        Mg & 0.10 \\
        Al & 0.13 \\
        Si & 0.11 \\
        S & 0.11 \\
        K & 0.11 \\
        Ca & 0.09 \\
        Mn & 0.14 \\
        Fe & 0.09    \\  
        Ni & 0.12 \\
        \hline 
    \end{tabular}
    \caption{The bandwidth adopted for the Gaussian kernel used to estimate the distribution for each elemental abundance in \csamp.}
    \label{tab:bandwidths}
\end{table}

To calculate the planet occurrence rate, we rely on knowing the quantity $n_\mathrm{trial}$. In most occurrence rate studies, the uncertainty on this quantity is ignored, as uncertainties in the derived occurrence rates are typically dominated by Poisson error \citep[e.g.,][]{youdin2011,fressin2013,petigura2013,burke2015,mulders2015a,petigura2018,narang2018}.
However, because the number of stars in \csamp\ is small compared to \ssamp, our strategy of extrapolating abundance distributions from \csamp\ leads to a significant uncertainty in $\fstar$, and therefore $n_\mathrm{trial}$~for a given abundance bin as compared with other studies. 
To estimate these errors and whether they are significant in relation to the Poisson error in our data, we again use the APOGEE dwarf sample described above. We perform a similar experiment to the one used to determine the optimal bandwidth, i.e., randomly drawing 72 measurements from the defined KDE for each APOGEE dwarf star abundance. 
We then fit a Gaussian KDE to the randomly drawn measurements using the optimal bandwidth determined above, and measure the difference in $F_\star$ as inferred from equation \ref{eq:frac} between the defined KDE and the experiment KDE. We repeat this procedure 1,000 times to derive the typical uncertainty, $\sigma_{F_\star}$, and the bias (i.e., mean offset from the true value), $\delta_{F_\star}$, in $F_\star$~for each abundance bin. Performing this experiment, we draw two interesting conclusions. First, $\delta_{F_\star}$ is negligible for bins within the inner 90th  percentile of the abundance distributions. Therefore, when fitting the occurrence rate distributions, we omit abundances outside of this range. Secondly, the typical error on $\sigma_{F_\star}$ is $\sim$20-30\%, and increases to $\sim$50-60\% as you move away from the median of the distribution. This error is relatively small compared to the Poisson errors of planet size classes with few ($\lesssim$~10) detected planets. Thus, this uncertainty is important to take into account for the occurrence rates of Sub-Neptunes and Super-Earths, but is less constraining when deriving occurrence rates for Sub-Saturns and Jupiters in our sample.

To account for this uncertainty, we modify our fitting procedure when deriving occurrence rates that depend on abundances. The procedure outlined in \S \ref{sec:fitting} is repeated 100 times. In each iteration we resample the abundances in \csamp\ with replacement, adding an offset randomly drawn from a Gaussian distribution with the width set by the error reported in ASPCAP. The collection of posterior distribution from each of the 100 independent MCMC routines are then used to determine the range of credible models. This bootstrapping routine is only performed when determining occurrence rates as a function of abundances where we rely on \csamp\ to measure $\fstar$. The range of credible models for $\fstar$, as determined from the Monte Carlo bootstrapping routine, are shown in Figure \ref{fig:abundance_1ddist}. 
These estimates can also be interpreted as the abundance-ratio distribution functions for each chemical species in \ssamp, providing the first such inferences for field dwarfs observed by \kepler.

\subsection{Potential Correlations Between [X/H] and $\eta(P,R_p)$} \label{sec:xfebiases}

\begin{figure*}
\centering
    \includegraphics[width=\textwidth]{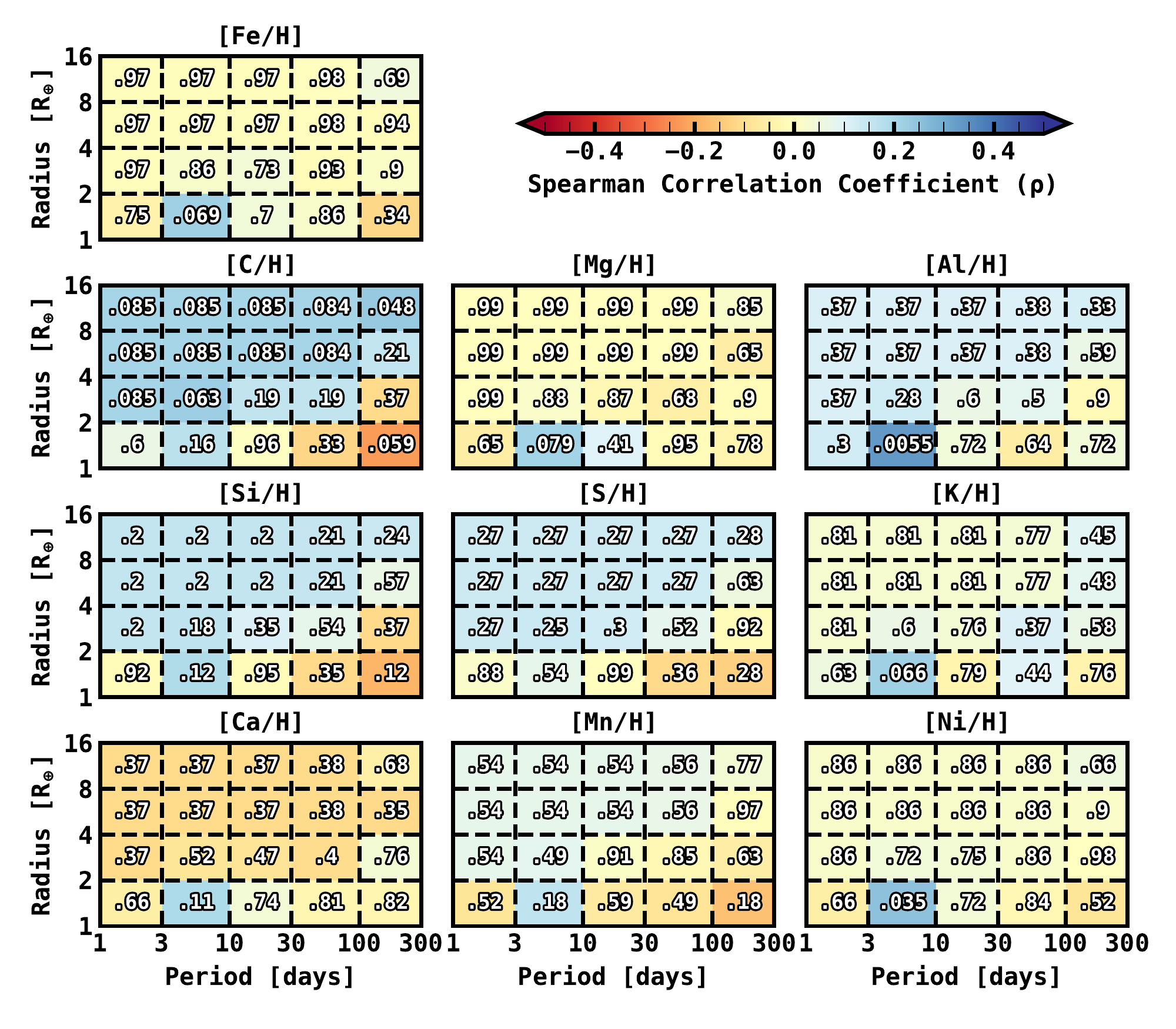}
    \caption{ The likelihood of a correlation between the survey efficiency, $\eta(P,R_p)$, and chemical abundance for stars in \csamp. In each panel, each box in the $P$-$R_p$ plane is colored by the Spearman correlation coefficient, and the numbers in each box are the $p$~values from the Spearman correlation test. With the possible exception of a correlation between Al and the detectability of Super-Earths with $P\sim$~3-10~days, there are no statistically significant correlations between abundance ratio and survey efficiency for stars in \csamp.} 
    \label{fig:eta_xh_correlation}
\end{figure*}

Our methodology for deriving occurrence rates assumes that there are no correlations between chemical abundance and survey efficiency. 
However, as shown in \S\ref{sec:xfetrends}, there are systematic trends between effective temperature and abundance ratio for C, Al, and Si which may challenge these assumptions. To determine whether these systematic trends are a significant confounding variable in our measured occurrence rates, we estimate the average survey efficiency, $\eta(P,R_p)$, as a function of [X/H] for stars in \csamp. 
We compute $\eta(P,R_p)$~for each star in \csamp\ following the prescription enumerated in \S\ref{subsec:completeness}. 
$\eta$ is calculated individually for each star in \csamp\ using the stellar parameters derived by \cite{berger2018,berger2020} over a coarse grid in $P$~and $R_p$. 
For each grid point, we determine the likelihood that there is a significant correlation between [X/H] and $\eta$ using the Spearman rank correlation coefficient. These results are shown in Figure \ref{fig:eta_xh_correlation}. Even for C, Al, and Si, the three elements whose abundance ratios showed the most significant trends, there are no statistically significant biases between abundance ratio and survey efficiency. There is a possible exception for hot Super-Earths in the case of [Al/H], which shows a correlation coefficient of $\rho=+0.33$ with a significance of $p_\rho = 0.0055$. Because this significance is only present in Al, and for a very specific range of $P$ and $R_p$, we do not make corrections in our completeness model to account for this. However, we consider this potential bias when interpreting our results. 
Aside from Al, the typical maximum and minimum Spearman correlation coefficients range from $\rho \approx \pm 0.2$ showing relatively weak correlations overall.

Although there are no statistically significant correlations between planet detection efficiency and chemical abundances,
we still wish to address the possibility that this is due to the relatively few number of stars in \csamp\ and provide context for future studies that may wish to more rigorously control for such differences. 
For some of these elements (e.g., C, Si, S) we recognize a similar pattern across the $P$-$R_p$~plane in Figure \ref{fig:eta_xh_correlation}. Abundances for planets with short periods and large radii seem more positively correlated with $\eta$, 
and abundances for planets with long periods and small radii are more negatively correlated with $\eta$. While these trends are not statistically significant, they are still worth discussion.
If there is a real positive correlation between $\teff$ and [X/H], this pattern may be explained given the biases present in our sample.

In the short-period, high transit $S/N$ regime, the survey efficiency may be positively correlated with [X/H]. This regime
is dominated by $p_\mathrm{fail}$ and $p_\mathrm{tra}$. Both of these effects reduce $\eta$ for low mass (low $\teff$) stars because stellar radii (and therefore $p_\mathrm{tra}$) is correlated with $\teff$. There is also an anti-correlation in our sample between $H$ magnitude (and therefore $p_\mathrm{fail}$) and stellar mass. As a result, a positive correlation between [X/H] and $\teff$ may explain the pattern seen at short periods and high transit~$S/N$.

In the long period, low transit~$S/N$ regime, explaining these biases is more complicated. Under the assumption of a positive correlation between [X/H] and $\teff$, a tentative anti-correlation implies that the survey efficiency is higher for lower mass stars. 
The survey efficiency is dominated by $p_\mathrm{det}$, and while it is tempting to explain this bias with the dependence of $p_\mathrm{det}$ on stellar radius, this is not the case. Because our sample selection is based on the $H$-band magnitude, there is a strong anti-correlation in our sample between $Kp$ and $\teff$. 
This results in significantly noisier light curves for lower mass stars in our sample, which negates the difference in stellar radii and results in no correlation (as measured by the Spearman correlation coefficient, $\rho=+0.01, p=0.90$) between $\teff$~and transit $S/N$, which we've modeled as $S/N \propto R_\star^{-2} \sigma_\mathrm{cdpp}^{-1}$, for the stars in our sample. 
The difference is driven by $p_\mathrm{tra}$, which depends inversely on $M_\star^{1/3}$. 
Because a planet orbiting a lower mass star has a shorter semi-major axis for a given period, the probability of a planet with a randomly oriented orbital plane transiting is higher. Thus, the survey efficiency at long periods is higher for low mass stars,
but this is primarily driven by $p_\mathrm{tra}$ and not by increased transit $S/N$~for stars with smaller radii, as one might expect.

While there are tentative explanations for why our sample may show a bias, it's important to emphasize that none of these biases were statistically significant. Therefore, we do not actually model any correlation with $\eta$ and $[X/H]$. 
It is also important to point out that we used quantities for stars in \csamp\ to determine the extent of our biases. It is appropriate to use \csamp\ for this purpose because it is an unbiased group of stars that were observed with the same strategy as the KOI sample. 
As a result, \csamp\ will show the same astrophysical and observational biases as \ssamp, and to an extent \psamp. Though it is tempting to check for trends in the planet sample directly, correcting for such a trend results in a degree of circular logic that would bias our results.

\bibliography{references}{}

\begin{thebibliography}{}
\expandafter\ifx\csname natexlab\endcsname\relax\def\natexlab#1{#1}\fi
\providecommand{\url}[1]{\href{#1}{#1}}
\providecommand{\dodoi}[1]{doi:~\href{http://doi.org/#1}{\nolinkurl{#1}}}
\providecommand{\doeprint}[1]{\href{http://ascl.net/#1}{\nolinkurl{http://ascl.net/#1}}}
\providecommand{\doarXiv}[1]{\href{https://arxiv.org/abs/#1}{\nolinkurl{https://arxiv.org/abs/#1}}}

\bibitem[{{Adibekyan} {et~al.}(2012){Adibekyan}, {Delgado Mena}, {Sousa},
  {Santos}, {Israelian}, {Gonz{\'a}lez Hern{\'a}ndez}, {Mayor}, \&
  {Hakobyan}}]{adibekyan2012a}
{Adibekyan}, V.~Z., {Delgado Mena}, E., {Sousa}, S.~G., {et~al.} 2012, \aap,
  547, A36, \dodoi{10.1051/0004-6361/201220167}

\bibitem[{{Ahumada} {et~al.}(2020){Ahumada}, {Prieto}, {Almeida}, {Anders},
  {Anderson}, {Andrews}, {Anguiano}, {Arcodia}, {Armengaud}, {Aubert}, {Avila},
  {Avila-Reese}, {Badenes}, {Balland}, {Barger}, {Barrera-Ballesteros}, {Basu},
  {Bautista}, {Beaton}, {Beers}, {Benavides}, {Bender}, {Bernardi}, {Bershady},
  {Beutler}, {Bidin}, {Bird}, {Bizyaev}, {Blanc}, {Blanton}, {Boquien},
  {Borissova}, {Bovy}, {Brandt}, {Brinkmann}, {Brownstein}, {Bundy}, {Bureau},
  {Burgasser}, {Burtin}, {Cano-D{\'\i}az}, {Capasso}, {Cappellari}, {Carrera},
  {Chabanier}, {Chaplin}, {Chapman}, {Cherinka}, {Chiappini}, {Doohyun Choi},
  {Chojnowski}, {Chung}, {Clerc}, {Coffey}, {Comerford}, {Comparat}, {da
  Costa}, {Cousinou}, {Covey}, {Crane}, {Cunha}, {Ilha}, {Dai}, {Damsted},
  {Darling}, {Davidson}, {Davies}, {Dawson}, {De}, {de la Macorra}, {De Lee},
  {Queiroz}, {Deconto Machado}, {de la Torre}, {Dell'Agli}, {du Mas des
  Bourboux}, {Diamond-Stanic}, {Dillon}, {Donor}, {Drory}, {Duckworth},
  {Dwelly}, {Ebelke}, {Eftekharzadeh}, {Davis Eigenbrot}, {Elsworth},
  {Eracleous}, {Erfanianfar}, {Escoffier}, {Fan}, {Farr},
  {Fern{\'a}ndez-Trincado}, {Feuillet}, {Finoguenov}, {Fofie},
  {Fraser-McKelvie}, {Frinchaboy}, {Fromenteau}, {Fu}, {Galbany}, {Garcia},
  {Garc{\'\i}a-Hern{\'a}ndez}, {Oehmichen}, {Ge}, {Maia}, {Geisler}, {Gelfand},
  {Goddy}, {Gonzalez-Perez}, {Grabowski}, {Green}, {Grier}, {Guo}, {Guy},
  {Harding}, {Hasselquist}, {Hawken}, {Hayes}, {Hearty}, {Hekker}, {Hogg},
  {Holtzman}, {Horta}, {Hou}, {Hsieh}, {Huber}, {Hunt}, {Chitham}, {Imig},
  {Jaber}, {Angel}, {Johnson}, {Jones}, {J{\"o}nsson}, {Jullo}, {Kim},
  {Kinemuchi}, {Kirkpatrick}, {Kite}, {Klaene}, {Kneib}, {Kollmeier}, {Kong},
  {Kounkel}, {Krishnarao}, {Lacerna}, {Lan}, {Lane}, {Law}, {Le Goff}, {Leung},
  {Lewis}, {Li}, {Lian}, {Lin}, {Long}, {Longa-Pe{\~n}a}, {Lundgren}, {Lyke},
  {Ted Mackereth}, {MacLeod}, {Majewski}, {Manchado}, {Maraston}, {Martini},
  {Masseron}, {Masters}, {Mathur}, {McDermid}, {Merloni}, {Merrifield},
  {M{\'e}sz{\'a}ros}, {Miglio}, {Minniti}, {Minsley}, {Miyaji}, {Mohammad},
  {Mosser}, {Mueller}, {Muna}, {Mu{\~n}oz-Guti{\'e}rrez}, {Myers}, {Nadathur},
  {Nair}, {Nandra}, {do Nascimento}, {Nevin}, {Newman}, {Nidever}, {Nitschelm},
  {Noterdaeme}, {O'Connell}, {Olmstead}, {Oravetz}, {Oravetz}, {Osorio},
  {Pace}, {Padilla}, {Palanque-Delabrouille}, {Palicio}, {Pan}, {Pan},
  {Parker}, {Paviot}, {Peirani}, {Ram{\'r}ez}, {Penny}, {Percival},
  {Perez-Fournon}, {P{\'e}rez-R{\`a}fols}, {Petitjean}, {Pieri},
  {Pinsonneault}, {Poovelil}, {Povick}, {Prakash}, {Price-Whelan}, {Raddick},
  {Raichoor}, {Ray}, {Rembold}, {Rezaie}, {Riffel}, {Riffel}, {Rix}, {Robin},
  {Roman-Lopes}, {Rom{\'a}n-Z{\'u}{\~n}iga}, {Rose}, {Ross}, {Rossi},
  {Rowlands}, {Rubin}, {Salvato}, {S{\'a}nchez}, {S{\'a}nchez-Menguiano},
  {S{\'a}nchez-Gallego}, {Sayres}, {Schaefer}, {Schiavon}, {Schimoia},
  {Schlafly}, {Schlegel}, {Schneider}, {Schultheis}, {Schwope}, {Seo},
  {Serenelli}, {Shafieloo}, {Shamsi}, {Shao}, {Shen}, {Shetrone}, {Shirley},
  {Aguirre}, {Simon}, {Skrutskie}, {Slosar}, {Smethurst}, {Sobeck}, {Sodi},
  {Souto}, {Stark}, {Stassun}, {Steinmetz}, {Stello}, {Stermer},
  {Storchi-Bergmann}, {Streblyanska}, {Stringfellow}, {Stutz}, {Su{\'a}rez},
  {Sun}, {Taghizadeh-Popp}, {Talbot}, {Tayar}, {Thakar}, {Theriault}, {Thomas},
  {Thomas}, {Tinker}, {Tojeiro}, {Toledo}, {Tremonti}, {Troup}, {Tuttle},
  {Unda-Sanzana}, {Valentini}, {Vargas-Gonz{\'a}lez}, {Vargas-Maga{\~n}a},
  {V{\'a}zquez-Mata}, {Vivek}, {Wake}, {Wang}, {Weaver}, {Weijmans}, {Wild},
  {Wilson}, {Wilson}, {Wolthuis}, {Wood-Vasey}, {Yan}, {Yang}, {Y{\`e}che},
  {Zamora}, {Zarrouk}, {Zasowski}, {Zhang}, {Zhao}, {Zhao}, {Zheng}, {Zheng},
  {Zhu}, \& {Zou}}]{ahumada2020}
{Ahumada}, R., {Prieto}, C.~A., {Almeida}, A., {et~al.} 2020, \apjs, 249, 3,
  \dodoi{10.3847/1538-4365/ab929e}

\bibitem[{{Akeson} {et~al.}(2017){Akeson}, {Christiansen}, {Ciardi}, {Ramirez},
  {Schlieder}, {Van Eyken}, \& {NASA Exoplanet Archive Team}}]{akeson2017}
{Akeson}, R.~L., {Christiansen}, J., {Ciardi}, D.~R., {et~al.} 2017, in
  American Astronomical Society Meeting Abstracts, Vol. 229, American
  Astronomical Society Meeting Abstracts, 146.16

\bibitem[{{Alibert} {et~al.}(2011){Alibert}, {Mordasini}, \&
  {Benz}}]{alibert2011}
{Alibert}, Y., {Mordasini}, C., \& {Benz}, W. 2011, \aap, 526, A63,
  \dodoi{10.1051/0004-6361/201014760}

\bibitem[{{Allende Prieto} {et~al.}(2006){Allende Prieto}, {Beers}, {Wilhelm},
  {Newberg}, {Rockosi}, {Yanny}, \& {Lee}}]{allendeprieto2006}
{Allende Prieto}, C., {Beers}, T.~C., {Wilhelm}, R., {et~al.} 2006, \apj, 636,
  804, \dodoi{10.1086/498131}

\bibitem[{{Anguiano} {et~al.}(2018){Anguiano}, {Majewski}, {Freeman},
  {Mitschang}, \& {Smith}}]{anguiano2018}
{Anguiano}, B., {Majewski}, S.~R., {Freeman}, K.~C., {Mitschang}, A.~W., \&
  {Smith}, M.~C. 2018, \mnras, 474, 854, \dodoi{10.1093/mnras/stx2774}

\bibitem[{{Astropy Collaboration} {et~al.}(2018){Astropy Collaboration},
  {Price-Whelan}, {Sip{\H{o}}cz}, {G{\"u}nther}, {Lim}, {Crawford}, {Conseil},
  {Shupe}, {Craig}, {Dencheva}, {Ginsburg}, {VanderPlas}, {Bradley},
  {P{\'e}rez-Su{\'a}rez}, {de Val-Borro}, {Aldcroft}, {Cruz}, {Robitaille},
  {Tollerud}, {Ardelean}, {Babej}, {Bach}, {Bachetti}, {Bakanov}, {Bamford},
  {Barentsen}, {Barmby}, {Baumbach}, {Berry}, {Biscani}, {Boquien}, {Bostroem},
  {Bouma}, {Brammer}, {Bray}, {Breytenbach}, {Buddelmeijer}, {Burke},
  {Calderone}, {Cano Rodr{\'\i}guez}, {Cara}, {Cardoso}, {Cheedella}, {Copin},
  {Corrales}, {Crichton}, {D'Avella}, {Deil}, {Depagne}, {Dietrich}, {Donath},
  {Droettboom}, {Earl}, {Erben}, {Fabbro}, {Ferreira}, {Finethy}, {Fox},
  {Garrison}, {Gibbons}, {Goldstein}, {Gommers}, {Greco}, {Greenfield},
  {Groener}, {Grollier}, {Hagen}, {Hirst}, {Homeier}, {Horton}, {Hosseinzadeh},
  {Hu}, {Hunkeler}, {Ivezi{\'c}}, {Jain}, {Jenness}, {Kanarek}, {Kendrew},
  {Kern}, {Kerzendorf}, {Khvalko}, {King}, {Kirkby}, {Kulkarni}, {Kumar},
  {Lee}, {Lenz}, {Littlefair}, {Ma}, {Macleod}, {Mastropietro}, {McCully},
  {Montagnac}, {Morris}, {Mueller}, {Mumford}, {Muna}, {Murphy}, {Nelson},
  {Nguyen}, {Ninan}, {N{\"o}the}, {Ogaz}, {Oh}, {Parejko}, {Parley}, {Pascual},
  {Patil}, {Patil}, {Plunkett}, {Prochaska}, {Rastogi}, {Reddy Janga},
  {Sabater}, {Sakurikar}, {Seifert}, {Sherbert}, {Sherwood-Taylor}, {Shih},
  {Sick}, {Silbiger}, {Singanamalla}, {Singer}, {Sladen}, {Sooley},
  {Sornarajah}, {Streicher}, {Teuben}, {Thomas}, {Tremblay}, {Turner},
  {Terr{\'o}n}, {van Kerkwijk}, {de la Vega}, {Watkins}, {Weaver}, {Whitmore},
  {Woillez}, {Zabalza}, \& {Astropy Contributors}}]{astropy}
{Astropy Collaboration}, {Price-Whelan}, A.~M., {Sip{\H{o}}cz}, B.~M., {et~al.}
  2018, \aj, 156, 123, \dodoi{10.3847/1538-3881/aabc4f}

\bibitem[{{Bailer-Jones}(2015)}]{bailer-jones2015}
{Bailer-Jones}, C. A.~L. 2015, \pasp, 127, 994, \dodoi{10.1086/683116}

\bibitem[{{Bailer-Jones} {et~al.}(2018){Bailer-Jones}, {Rybizki}, {Fouesneau},
  {Mantelet}, \& {Andrae}}]{bailer-jones2018}
{Bailer-Jones}, C.~A.~L., {Rybizki}, J., {Fouesneau}, M., {Mantelet}, G., \&
  {Andrae}, R. 2018, \aj, 156, 58, \dodoi{10.3847/1538-3881/aacb21}

\bibitem[{{Berger} {et~al.}(2018){Berger}, {Huber}, {Gaidos}, \& {van
  Saders}}]{berger2018}
{Berger}, T.~A., {Huber}, D., {Gaidos}, E., \& {van Saders}, J.~L. 2018, \apj,
  866, 99, \dodoi{10.3847/1538-4357/aada83}

\bibitem[{{Berger} {et~al.}(2020{\natexlab{a}}){Berger}, {Huber}, {Gaidos},
  {van Saders}, \& {Weiss}}]{berger2020a}
{Berger}, T.~A., {Huber}, D., {Gaidos}, E., {van Saders}, J.~L., \& {Weiss},
  L.~M. 2020{\natexlab{a}}, \aj, 160, 108, \dodoi{10.3847/1538-3881/aba18a}

\bibitem[{{Berger} {et~al.}(2020{\natexlab{b}}){Berger}, {Huber}, {van Saders},
  {Gaidos}, {Tayar}, \& {Kraus}}]{berger2020}
{Berger}, T.~A., {Huber}, D., {van Saders}, J.~L., {et~al.} 2020{\natexlab{b}},
  \aj, 159, 280, \dodoi{10.3847/1538-3881/159/6/280}

\bibitem[{Blanton {et~al.}(2017)Blanton, Bershady, Abolfathi, Albareti, Prieto,
  Almeida, Alonso-García, Anders, Anderson, Andrews, Aquino-Ortíz,
  Aragón-Salamanca, Argudo-Fernández, Armengaud, Aubourg, Avila-Reese,
  Badenes, Bailey, Barger, Barrera-Ballesteros, Bartosz, Bates, Baumgarten,
  Bautista, Beaton, Beers, Belfiore, Bender, Berlind, Bernardi, Beutler, Bird,
  Bizyaev, Blanc, Blomqvist, Bolton, Boquien, Borissova, van~den Bosch, Bovy,
  Brandt, Brinkmann, Brownstein, Bundy, Burgasser, Burtin, Busca, Cappellari,
  Carigi, Carlberg, Rosell, Carrera, Chanover, Cherinka, Cheung, Chew,
  Chiappini, Choi, Chojnowski, Chuang, Chung, Cirolini, Clerc, Cohen, Comparat,
  da~Costa, Cousinou, Covey, Crane, Croft, Cruz-Gonzalez, Cuadra, Cunha, Damke,
  Darling, Davies, Dawson, de~la Macorra, Dell’Agli, Lee, Delubac, Mille,
  Diamond-Stanic, Cano-Díaz, Donor, Downes, Drory, du~Mas~des Bourboux,
  Duckworth, Dwelly, Dyer, Ebelke, Eigenbrot, Eisenstein, Emsellem, Eracleous,
  Escoffier, Evans, Fan, Fernández-Alvar, Fernandez-Trincado, Feuillet,
  Finoguenov, Fleming, Font-Ribera, Fredrickson, Freischlad, Frinchaboy,
  Fuentes, Galbany, Garcia-Dias, García-Hernández, Gaulme, Geisler, Gelfand,
  Gil-Marín, Gillespie, Goddard, Gonzalez-Perez, Grabowski, Green, Grier,
  Gunn, Guo, Guy, Hagen, Hahn, Hall, Harding, Hasselquist, Hawley, Hearty,
  Hernández, Ho, Hogg, Holley-Bockelmann, Holtzman, Holzer, Huehnerhoff,
  Hutchinson, Hwang, Ibarra-Medel, da~Silva~Ilha, Ivans, Ivory, Jackson,
  Jensen, Johnson, Jones, Jönsson, Jullo, Kamble, Kinemuchi, Kirkby, Kitaura,
  Klaene, Knapp, Kneib, Kollmeier, Lacerna, Lane, Lang, Law, Lazarz, Lee, Goff,
  Liang, Li, Li, Lian, Lima, Lin, Lin, de~Lis, Liu, de~Icaza~Lizaola, Long,
  Lucatello, Lundgren, MacDonald, Machado, MacLeod, Mahadevan, Maia, Maiolino,
  Majewski, Malanushenko, Malanushenko, Manchado, Mao, Maraston,
  Marques-Chaves, Masseron, Masters, McBride, McDermid, McGrath, McGreer,
  Peña, Melendez, Merloni, Merrifield, Meszaros, Meza, Minchev, Minniti,
  Miyaji, More, Mulchaey, Müller-Sánchez, Muna, Munoz, Myers, Nair, Nandra,
  do~Nascimento, Negrete, Ness, Newman, Nichol, Nidever, Nitschelm, Ntelis,
  O’Connell, Oelkers, Oravetz, Oravetz, Pace, Padilla, Palanque-Delabrouille,
  Palicio, Pan, Parejko, Parikh, Pâris, Park, Patten, Peirani,
  Pellejero-Ibanez, Penny, Percival, Perez-Fournon, Petitjean, Pieri,
  Pinsonneault, Pisani, Poleski, Prada, Prakash, de~Andrade~Queiroz, Raddick,
  Raichoor, Rembold, Richstein, Riffel, Riffel, Rix, Robin, Rockosi,
  Rodríguez-Torres, Roman-Lopes, Román-Zúñiga, Rosado, Ross, Rossi, Ruan,
  Ruggeri, Rykoff, Salazar-Albornoz, Salvato, Sánchez, Aguado,
  Sánchez-Gallego, Santana, Santiago, Sayres, Schiavon, da~Silva~Schimoia,
  Schlafly, Schlegel, Schneider, Schultheis, Schuster, Schwope, Seo, Shao,
  Shen, Shetrone, Shull, Simon, Skinner, Skrutskie, Slosar, Smith, Sobeck,
  Sobreira, Somers, Souto, Stark, Stassun, Stauffer, Steinmetz,
  Storchi-Bergmann, Streblyanska, Stringfellow, Suárez, Sun, Suzuki, Szigeti,
  Taghizadeh-Popp, Tang, Tao, Tayar, Tembe, Teske, Thakar, Thomas, Thompson,
  Tinker, Tissera, Tojeiro, Toledo, de~la Torre, Tremonti, Troup, Valenzuela,
  Valpuesta, Vargas-González, Vargas-Magaña, Vazquez, Villanova, Vivek, Vogt,
  Wake, Walterbos, Wang, Weaver, Weijmans, Weinberg, Westfall, Whelan, Wild,
  Wilson, Wood-Vasey, Wylezalek, Xiao, Yan, Yang, Ybarra, Yèche, Zakamska,
  Zamora, Zarrouk, Zasowski, Zhang, Zhao, Zheng, Zheng, Zhou, Zhou, Zhu,
  Zoccali, \& Zou}]{blanton2017}
Blanton, M.~R., Bershady, M.~A., Abolfathi, B., {et~al.} 2017, The Astronomical
  Journal, 154, 28.
\newblock \url{http://stacks.iop.org/1538-3881/154/i=1/a=28}

\bibitem[{{Bowler} {et~al.}(2015){Bowler}, {Shkolnik}, {Liu}, {Schlieder},
  {Mann}, {Dupuy}, {Hinkley}, {Crepp}, {Johnson}, {Howard}, {Flagg},
  {Weinberger}, {Aller}, {Allers}, {Best}, {Kotson}, {Montet}, {Herczeg},
  {Baranec}, {Riddle}, {Law}, {Nielsen}, {Wahhaj}, {Biller}, \&
  {Hayward}}]{bowler2015}
{Bowler}, B.~P., {Shkolnik}, E.~L., {Liu}, M.~C., {et~al.} 2015, \apj, 806, 62,
  \dodoi{10.1088/0004-637X/806/1/62}

\bibitem[{{Brewer} {et~al.}(2016){Brewer}, {Fischer}, {Valenti}, \&
  {Piskunov}}]{brewer2016}
{Brewer}, J.~M., {Fischer}, D.~A., {Valenti}, J.~A., \& {Piskunov}, N. 2016,
  \apjs, 225, 32, \dodoi{10.3847/0067-0049/225/2/32}

\bibitem[{{Brown} {et~al.}(2011){Brown}, {Latham}, {Everett}, \&
  {Esquerdo}}]{brown2011}
{Brown}, T.~M., {Latham}, D.~W., {Everett}, M.~E., \& {Esquerdo}, G.~A. 2011,
  \aj, 142, 112, \dodoi{10.1088/0004-6256/142/4/112}

\bibitem[{{Brugamyer} {et~al.}(2011){Brugamyer}, {Dodson-Robinson}, {Cochran},
  \& {Sneden}}]{brugamyer2011}
{Brugamyer}, E., {Dodson-Robinson}, S.~E., {Cochran}, W.~D., \& {Sneden}, C.
  2011, \apj, 738, 97, \dodoi{10.1088/0004-637X/738/1/97}

\bibitem[{{Bruntt} {et~al.}(2012){Bruntt}, {Basu}, {Smalley}, {Chaplin},
  {Verner}, {Bedding}, {Catala}, {Gazzano}, {Molenda-{\.Z}akowicz}, {Thygesen},
  {Uytterhoeven}, {Hekker}, {Huber}, {Karoff}, {Mathur}, {Mosser},
  {Appourchaux}, {Campante}, {Elsworth}, {Garc{\'{\i}}a}, {Handberg},
  {Metcalfe}, {Quirion}, {R{\'e}gulo}, {Roxburgh}, {Stello},
  {Christensen-Dalsgaard}, {Kawaler}, {Kjeldsen}, {Morris}, {Quintana}, \&
  {Sanderfer}}]{bruntt2012}
{Bruntt}, H., {Basu}, S., {Smalley}, B., {et~al.} 2012, \mnras, 423, 122,
  \dodoi{10.1111/j.1365-2966.2012.20686.x}

\bibitem[{{Bryson} {et~al.}(2020{\natexlab{a}}){Bryson}, {Coughlin}, {Batalha},
  {Berger}, {Huber}, {Burke}, {Dotson}, \& {Mullally}}]{bryson2020a}
{Bryson}, S., {Coughlin}, J., {Batalha}, N.~M., {et~al.} 2020{\natexlab{a}},
  \aj, 159, 279, \dodoi{10.3847/1538-3881/ab8a30}

\bibitem[{{Bryson} {et~al.}(2020{\natexlab{b}}){Bryson}, {Coughlin},
  {Kunimoto}, \& {Mullally}}]{bryson2020}
{Bryson}, S., {Coughlin}, J.~L., {Kunimoto}, M., \& {Mullally}, S.~E.
  2020{\natexlab{b}}, \aj, 160, 200, \dodoi{10.3847/1538-3881/abb316}

\bibitem[{{Buchhave} {et~al.}(2012){Buchhave}, {Latham}, {Johansen},
  {Bizzarro}, {Torres}, {Rowe}, {Batalha}, {Borucki}, {Brugamyer}, {Caldwell},
  {Bryson}, {Ciardi}, {Cochran}, {Endl}, {Esquerdo}, {Ford}, {Geary},
  {Gilliland}, {Hansen}, {Isaacson}, {Laird}, {Lucas}, {Marcy}, {Morse},
  {Robertson}, {Shporer}, {Stefanik}, {Still}, \& {Quinn}}]{buchhave2012}
{Buchhave}, L.~A., {Latham}, D.~W., {Johansen}, A., {et~al.} 2012, \nat, 486,
  375, \dodoi{10.1038/nature11121}

\bibitem[{{Buchhave} {et~al.}(2014){Buchhave}, {Bizzarro}, {Latham},
  {Sasselov}, {Cochran}, {Endl}, {Isaacson}, {Juncher}, \&
  {Marcy}}]{buchhave2014}
{Buchhave}, L.~A., {Bizzarro}, M., {Latham}, D.~W., {et~al.} 2014, \nat, 509,
  593, \dodoi{10.1038/nature13254}

\bibitem[{{Burke} \& {Catanzarite}(2017)}]{burke&catanzarite2017c}
{Burke}, C.~J., \& {Catanzarite}, J. 2017, {Planet Detection Metrics:
  Per-Target Flux-Level Transit Injection Tests of TPS for Data Release 25},
  Kepler Science Document KSCI-19109-002

\bibitem[{{Burke} {et~al.}(2015){Burke}, {Christiansen}, {Mullally}, {Seader},
  {Huber}, {Rowe}, {Coughlin}, {Thompson}, {Catanzarite}, {Clarke}, {Morton},
  {Caldwell}, {Bryson}, {Haas}, {Batalha}, {Jenkins}, {Tenenbaum}, {Twicken},
  {Li}, {Quintana}, {Barclay}, {Henze}, {Borucki}, {Howell}, \&
  {Still}}]{burke2015}
{Burke}, C.~J., {Christiansen}, J.~L., {Mullally}, F., {et~al.} 2015, \apj,
  809, 8, \dodoi{10.1088/0004-637X/809/1/8}

\bibitem[{{Choi} {et~al.}(2016){Choi}, {Dotter}, {Conroy}, {Cantiello},
  {Paxton}, \& {Johnson}}]{choi2016}
{Choi}, J., {Dotter}, A., {Conroy}, C., {et~al.} 2016, \apj, 823, 102,
  \dodoi{10.3847/0004-637X/823/2/102}

\bibitem[{{Christiansen}(2017)}]{christiansen2017}
{Christiansen}, J.~L. 2017, {Planet Detection Metrics: Pixel-Level Transit
  Injection Tests of Pipeline Detection Efficiency for Data Release 25}, Tech.
  rep.

\bibitem[{{Christiansen} {et~al.}(2012){Christiansen}, {Jenkins}, {Caldwell},
  {Burke}, {Tenenbaum}, {Seader}, {Thompson}, {Barclay}, {Clarke}, {Li},
  {Smith}, {Stumpe}, {Twicken}, \& {Van Cleve}}]{christiansen2012}
{Christiansen}, J.~L., {Jenkins}, J.~M., {Caldwell}, D.~A., {et~al.} 2012,
  \pasp, 124, 1279, \dodoi{10.1086/668847}

\bibitem[{{Christiansen} {et~al.}(2015){Christiansen}, {Clarke}, {Burke},
  {Seader}, {Jenkins}, {Twicken}, {Catanzarite}, {Smith}, {Batalha}, {Haas},
  {Thompson}, {Campbell}, {Sabale}, \& {Kamal Uddin}}]{christiansen2015}
{Christiansen}, J.~L., {Clarke}, B.~D., {Burke}, C.~J., {et~al.} 2015, \apj,
  810, 95, \dodoi{10.1088/0004-637X/810/2/95}

\bibitem[{{Christiansen} {et~al.}(2016){Christiansen}, {Clarke}, {Burke},
  {Jenkins}, {Bryson}, {Coughlin}, {Mullally}, {Thompson}, {Twicken},
  {Batalha}, {Haas}, {Catanzarite}, {Campbell}, {Kamal Uddin}, {Zamudio},
  {Smith}, \& {Henze}}]{christiansen2016}
---. 2016, \apj, 828, 99, \dodoi{10.3847/0004-637X/828/2/99}

\bibitem[{{Coughlin}(2015)}]{coughlin2015}
{Coughlin}, J. 2015, IAU General Assembly, 22, 2257510

\bibitem[{{Cumming} {et~al.}(2008){Cumming}, {Butler}, {Marcy}, {Vogt},
  {Wright}, \& {Fischer}}]{cumming2008}
{Cumming}, A., {Butler}, R.~P., {Marcy}, G.~W., {et~al.} 2008, \pasp, 120, 531,
  \dodoi{10.1086/588487}

\bibitem[{{Cumming} {et~al.}(1999){Cumming}, {Marcy}, \&
  {Butler}}]{cumming1999}
{Cumming}, A., {Marcy}, G.~W., \& {Butler}, R.~P. 1999, \apj, 526, 890,
  \dodoi{10.1086/308020}

\bibitem[{{Cunha} {et~al.}(2015){Cunha}, {Smith}, {Johnson}, {Bergemann},
  {M{\'e}sz{\'a}ros}, {Shetrone}, {Souto}, {Allende Prieto}, {Schiavon},
  {Frinchaboy}, {Zasowski}, {Bizyaev}, {Holtzman}, {Garc{\'{\i}}a P{\'e}rez},
  {Majewski}, {Nidever}, {Beers}, {Carrera}, {Geisler}, {Gunn}, {Hearty},
  {Ivans}, {Martell}, {Pinsonneault}, {Schneider}, {Sobeck}, {Stello},
  {Stassun}, {Skrutskie}, \& {Wilson}}]{cunha2015}
{Cunha}, K., {Smith}, V.~V., {Johnson}, J.~A., {et~al.} 2015, \apjl, 798, L41,
  \dodoi{10.1088/2041-8205/798/2/L41}

\bibitem[{{Dong} {et~al.}(2014){Dong}, {Zheng}, {Zhu}, {De Cat}, {Fu}, {Yang},
  {Zhang}, {Jin}, \& {Zhang}}]{dong2014}
{Dong}, S., {Zheng}, Z., {Zhu}, Z., {et~al.} 2014, \apjl, 789, L3,
  \dodoi{10.1088/2041-8205/789/1/L3}

\bibitem[{{Donor} {et~al.}(2020){Donor}, {Frinchaboy}, {Cunha}, {O'Connell},
  {Allende Prieto}, {Almeida}, {Anders}, {Beaton}, {Bizyaev}, {Brownstein},
  {Carrera}, {Chiappini}, {Cohen}, {Garc{\'\i}a-Hern{\'a}ndez}, {Geisler},
  {Hasselquist}, {J{\"o}nsson}, {Lane}, {Majewski}, {Minniti}, {Bidin}, {Pan},
  {Roman-Lopes}, {Sobeck}, \& {Zasowski}}]{donor2020}
{Donor}, J., {Frinchaboy}, P.~M., {Cunha}, K., {et~al.} 2020, \aj, 159, 199,
  \dodoi{10.3847/1538-3881/ab77bc}

\bibitem[{{Dorn} {et~al.}(2017{\natexlab{a}}){Dorn}, {Hinkel}, \&
  {Venturini}}]{dorn2017a}
{Dorn}, C., {Hinkel}, N.~R., \& {Venturini}, J. 2017{\natexlab{a}}, \aap, 597,
  A38, \dodoi{10.1051/0004-6361/201628749}

\bibitem[{{Dorn} {et~al.}(2017{\natexlab{b}}){Dorn}, {Venturini}, {Khan},
  {Heng}, {Alibert}, {Helled}, {Rivoldini}, \& {Benz}}]{dorn2017b}
{Dorn}, C., {Venturini}, J., {Khan}, A., {et~al.} 2017{\natexlab{b}}, \aap,
  597, A37, \dodoi{10.1051/0004-6361/201628708}

\bibitem[{{Dotter}(2016)}]{dotter2016}
{Dotter}, A. 2016, \apjs, 222, 8, \dodoi{10.3847/0067-0049/222/1/8}

\bibitem[{{Dressing} \& {Charbonneau}(2015)}]{dressing&charbonneau2015}
{Dressing}, C.~D., \& {Charbonneau}, D. 2015, \apj, 807, 45,
  \dodoi{10.1088/0004-637X/807/1/45}

\bibitem[{Eisenstein {et~al.}(2011)Eisenstein, Weinberg, Agol, Aihara, Prieto,
  Anderson, Arns, Éric Aubourg, Bailey, Balbinot, Barkhouser, Beers, Berlind,
  Bickerton, Bizyaev, Blanton, Bochanski, Bolton, Bosman, Bovy, Brandt,
  Breslauer, Brewington, Brinkmann, Brown, Brownstein, Burger, Busca, Campbell,
  Cargile, Carithers, Carlberg, Carr, Chang, Chen, Chiappini, Comparat,
  Connolly, Cortes, Croft, Cunha, da~Costa, Davenport, Dawson, Lee, de~Mello,
  de~Simoni, Dean, Dhital, Ealet, Ebelke, Edmondson, Eiting, Escoffier,
  Esposito, Evans, Fan, Castellá, Ferreira, Fitzgerald, Fleming, Font-Ribera,
  Ford, Frinchaboy, Pérez, Gaudi, Ge, Ghezzi, Gillespie, Gilmore, Girardi,
  Gott, Gould, Grebel, Gunn, Hamilton, Harding, Harris, Hawley, Hearty,
  Hennawi, Hernández, Ho, Hogg, Holtzman, Honscheid, Inada, Ivans, Jiang,
  Jiang, Johnson, Jordan, Jordan, Kauffmann, Kazin, Kirkby, Klaene, Knapp,
  Kneib, Kochanek, Koesterke, Kollmeier, Kron, Lampeitl, Lang, Lawler, Goff,
  Lee, Lee, Leisenring, Lin, Liu, Long, Loomis, Lucatello, Lundgren, Lupton,
  Ma, Ma, MacDonald, Mack, Mahadevan, Maia, Majewski, Makler, Malanushenko,
  Malanushenko, Mandelbaum, Maraston, Margala, Maseman, Masters, McBride,
  McDonald, McGreer, McMahon, Requejo, Ménard, Miralda-Escudé, Morrison,
  Mullally, Muna, Murayama, Myers, Naugle, Neto, Nguyen, Nichol, Nidever,
  O’Connell, Ogando, Olmstead, Oravetz, Padmanabhan, Paegert,
  Palanque-Delabrouille, Pan, Pandey, Parejko, Pâris, Pellegrini, Pepper,
  Percival, Petitjean, Pfaffenberger, Pforr, Phleps, Pichon, Pieri, Prada,
  Price-Whelan, Raddick, Ramos, Reid, Reyle, Rich, Richards, Rieke, Rieke, Rix,
  Robin, Rocha-Pinto, Rockosi, Roe, Rollinde, Ross, Ross, Rossetto, Sánchez,
  Santiago, Sayres, Schiavon, Schlegel, Schlesinger, Schmidt, Schneider,
  Sellgren, Shelden, Sheldon, Shetrone, Shu, Silverman, Simmerer, Simmons,
  Sivarani, Skrutskie, Slosar, Smee, Smith, Snedden, Stassun, Steele,
  Steinmetz, Stockett, Stollberg, Strauss, Szalay, Tanaka, Thakar, Thomas,
  Tinker, Tofflemire, Tojeiro, Tremonti, Magaña, Verde, Vogt, Wake, Wan, Wang,
  Weaver, White, White, Wilson, Wisniewski, Wood-Vasey, Yanny, Yasuda, Yèche,
  York, Young, Zasowski, Zehavi, \& Zhao}]{eisenstain2011}
Eisenstein, D.~J., Weinberg, D.~H., Agol, E., {et~al.} 2011, The Astronomical
  Journal, 142, 72.
\newblock \url{http://stacks.iop.org/1538-3881/142/i=3/a=72}

\bibitem[{{Eistrup} {et~al.}(2018){Eistrup}, {Walsh}, \& {van
  Dishoeck}}]{eistrup2018}
{Eistrup}, C., {Walsh}, C., \& {van Dishoeck}, E.~F. 2018, \aap, 613, A14,
  \dodoi{10.1051/0004-6361/201731302}

\bibitem[{{Everett} {et~al.}(2013){Everett}, {Howell}, {Silva}, \&
  {Szkody}}]{everett2013}
{Everett}, M.~E., {Howell}, S.~B., {Silva}, D.~R., \& {Szkody}, P. 2013, \apj,
  771, 107, \dodoi{10.1088/0004-637X/771/2/107}

\bibitem[{{Fischer} \& {Valenti}(2005)}]{fischer2005}
{Fischer}, D.~A., \& {Valenti}, J. 2005, \apj, 622, 1102,
  \dodoi{10.1086/428383}

\bibitem[{{Fleming} {et~al.}(2015){Fleming}, {Mahadevan}, {Deshpande},
  {Bender}, {Terrien}, {Marchwinski}, {Wang}, {Roy}, {Stassun}, {Allende
  Prieto}, {Cunha}, {Smith}, {Agol}, {Ak}, {Bastien}, {Bizyaev}, {Crepp},
  {Ford}, {Frinchaboy}, {Garc{\'{\i}}a-Hern{\'a}ndez}, {Garc{\'{\i}}a
  P{\'e}rez}, {Gaudi}, {Ge}, {Hearty}, {Ma}, {Majewski}, {M{\'e}sz{\'a}ros},
  {Nidever}, {Pan}, {Pepper}, {Pinsonneault}, {Schiavon}, {Schneider},
  {Wilson}, {Zamora}, \& {Zasowski}}]{fleming2015}
{Fleming}, S.~W., {Mahadevan}, S., {Deshpande}, R., {et~al.} 2015, \aj, 149,
  143, \dodoi{10.1088/0004-6256/149/4/143}

\bibitem[{{Foreman-Mackey} {et~al.}(2013){Foreman-Mackey}, {Hogg}, {Lang}, \&
  {Goodman}}]{emcee}
{Foreman-Mackey}, D., {Hogg}, D.~W., {Lang}, D., \& {Goodman}, J. 2013, \pasp,
  125, 306, \dodoi{10.1086/670067}

\bibitem[{{Fressin} {et~al.}(2013){Fressin}, {Torres}, {Charbonneau}, {Bryson},
  {Christiansen}, {Dressing}, {Jenkins}, {Walkowicz}, \&
  {Batalha}}]{fressin2013}
{Fressin}, F., {Torres}, G., {Charbonneau}, D., {et~al.} 2013, \apj, 766, 81,
  \dodoi{10.1088/0004-637X/766/2/81}

\bibitem[{{Fulton} \& {Petigura}(2018)}]{fulton&petigura2018}
{Fulton}, B.~J., \& {Petigura}, E.~A. 2018, \aj, 156, 264,
  \dodoi{10.3847/1538-3881/aae828}

\bibitem[{{Fulton} {et~al.}(2017){Fulton}, {Petigura}, {Howard}, {Isaacson},
  {Marcy}, {Cargile}, {Hebb}, {Weiss}, {Johnson}, {Morton}, {Sinukoff},
  {Crossfield}, \& {Hirsch}}]{fulton2017}
{Fulton}, B.~J., {Petigura}, E.~A., {Howard}, A.~W., {et~al.} 2017, ArXiv
  e-prints.
\newblock \doarXiv{1703.10375}

\bibitem[{{Gaia Collaboration} {et~al.}(2018{\natexlab{a}}){Gaia
  Collaboration}, {Brown, A. G. A.}, {Vallenari, A.}, {Prusti, T.}, {de
  Bruijne, J. H. J.}, {Babusiaux, C.}, {Bailer-Jones, C. A. L.}, {Biermann,
  M.}, {Evans, D. W.}, {Eyer, L.}, {Jansen, F.}, {Jordi, C.}, {Klioner, S. A.},
  {Lammers, U.}, {Lindegren, L.}, {Luri, X.}, {Mignard, F.}, {Panem, C.},
  {Pourbaix, D.}, {Randich, S.}, {Sartoretti, P.}, {Siddiqui, H. I.},
  {Soubiran, C.}, {van Leeuwen, F.}, {Walton, N. A.}, {Arenou, F.}, {Bastian,
  U.}, {Cropper, M.}, {Drimmel, R.}, {Katz, D.}, {Lattanzi, M. G.}, {Bakker,
  J.}, {Cacciari, C.}, {Casta\~neda, J.}, {Chaoul, L.}, {Cheek, N.}, {De
  Angeli, F.}, {Fabricius, C.}, {Guerra, R.}, {Holl, B.}, {Masana, E.},
  {Messineo, R.}, {Mowlavi, N.}, {Nienartowicz, K.}, {Panuzzo, P.}, {Portell,
  J.}, {Riello, M.}, {Seabroke, G. M.}, {Tanga, P.}, {Th\'evenin, F.},
  {Gracia-Abril, G.}, {Comoretto, G.}, {Garcia-Reinaldos, M.}, {Teyssier, D.},
  {Altmann, M.}, {Andrae, R.}, {Audard, M.}, {Bellas-Velidis, I.}, {Benson,
  K.}, {Berthier, J.}, {Blomme, R.}, {Burgess, P.}, {Busso, G.}, {Carry, B.},
  {Cellino, A.}, {Clementini, G.}, {Clotet, M.}, {Creevey, O.}, {Davidson, M.},
  {De Ridder, J.}, {Delchambre, L.}, {Dell\'{}Oro, A.}, {Ducourant, C.},
  {Fern\'andez-Hern\'andez, J.}, {Fouesneau, M.}, {Fr\'emat, Y.}, {Galluccio,
  L.}, {Garc\'{\i}a-Torres, M.}, {Gonz\'alez-N\'u\~nez, J.}, {Gonz\'alez-Vidal,
  J. J.}, {Gosset, E.}, {Guy, L. P.}, {Halbwachs, J.-L.}, {Hambly, N. C.},
  {Harrison, D. L.}, {Hern\'andez, J.}, {Hestroffer, D.}, {Hodgkin, S. T.},
  {Hutton, A.}, {Jasniewicz, G.}, {Jean-Antoine-Piccolo, A.}, {Jordan, S.},
  {Korn, A. J.}, {Krone-Martins, A.}, {Lanzafame, A. C.}, {Lebzelter, T.},
  {L\"offler, W.}, {Manteiga, M.}, {Marrese, P. M.}, {Mart\'{\i}n-Fleitas, J.
  M.}, {Moitinho, A.}, {Mora, A.}, {Muinonen, K.}, {Osinde, J.}, {Pancino, E.},
  {Pauwels, T.}, {Petit, J.-M.}, {Recio-Blanco, A.}, {Richards, P. J.},
  {Rimoldini, L.}, {Robin, A. C.}, {Sarro, L. M.}, {Siopis, C.}, {Smith, M.},
  {Sozzetti, A.}, {S\"uveges, M.}, {Torra, J.}, {van Reeven, W.}, {Abbas, U.},
  {Abreu Aramburu, A.}, {Accart, S.}, {Aerts, C.}, {Altavilla, G.}, {\'Alvarez,
  M. A.}, {Alvarez, R.}, {Alves, J.}, {Anderson, R. I.}, {Andrei, A. H.},
  {Anglada Varela, E.}, {Antiche, E.}, {Antoja, T.}, {Arcay, B.},
  {Astraatmadja, T. L.}, {Bach, N.}, {Baker, S. G.}, {Balaguer-N\'u\~nez, L.},
  {Balm, P.}, {Barache, C.}, {Barata, C.}, {Barbato, D.}, {Barblan, F.},
  {Barklem, P. S.}, {Barrado, D.}, {Barros, M.}, {Barstow, M. A.},
  {Bartholom\'e Mu\~noz, S.}, {Bassilana, J.-L.}, {Becciani, U.}, {Bellazzini,
  M.}, {Berihuete, A.}, {Bertone, S.}, {Bianchi, L.}, {Bienaym\'e, O.},
  {Blanco-Cuaresma, S.}, {Boch, T.}, {Boeche, C.}, {Bombrun, A.}, {Borrachero,
  R.}, {Bossini, D.}, {Bouquillon, S.}, {Bourda, G.}, {Bragaglia, A.},
  {Bramante, L.}, {Breddels, M. A.}, {Bressan, A.}, {Brouillet, N.},
  {Br\"usemeister, T.}, {Brugaletta, E.}, {Bucciarelli, B.}, {Burlacu, A.},
  {Busonero, D.}, {Butkevich, A. G.}, {Buzzi, R.}, {Caffau, E.}, {Cancelliere,
  R.}, {Cannizzaro, G.}, {Cantat-Gaudin, T.}, {Carballo, R.}, {Carlucci, T.},
  {Carrasco, J. M.}, {Casamiquela, L.}, {Castellani, M.}, {Castro-Ginard, A.},
  {Charlot, P.}, {Chemin, L.}, {Chiavassa, A.}, {Cocozza, G.}, {Costigan, G.},
  {Cowell, S.}, {Crifo, F.}, {Crosta, M.}, {Crowley, C.}, {Cuypers+, J.},
  {Dafonte, C.}, {Damerdji, Y.}, {Dapergolas, A.}, {David, P.}, {David, M.},
  {de Laverny, P.}, {De Luise, F.}, {De March, R.}, {de Martino, D.}, {de
  Souza, R.}, {de Torres, A.}, {Debosscher, J.}, {del Pozo, E.}, {Delbo, M.},
  {Delgado, A.}, {Delgado, H. E.}, {Di Matteo, P.}, {Diakite, S.}, {Diener,
  C.}, {Distefano, E.}, {Dolding, C.}, {Drazinos, P.}, {Dur\'an, J.},
  {Edvardsson, B.}, {Enke, H.}, {Eriksson, K.}, {Esquej, P.}, {Eynard Bontemps,
  G.}, {Fabre, C.}, {Fabrizio, M.}, {Faigler, S.}, {Falc\~ao, A. J.}, {Farr\`as
  Casas, M.}, {Federici, L.}, {Fedorets, G.}, {Fernique, P.}, {Figueras, F.},
  {Filippi, F.}, {Findeisen, K.}, {Fonti, A.}, {Fraile, E.}, {Fraser, M.},
  {Fr\'ezouls, B.}, {Gai, M.}, {Galleti, S.}, {Garabato, D.},
  {Garc\'{\i}a-Sedano, F.}, {Garofalo, A.}, {Garralda, N.}, {Gavel, A.},
  {Gavras, P.}, {Gerssen, J.}, {Geyer, R.}, {Giacobbe, P.}, {Gilmore, G.},
  {Girona, S.}, {Giuffrida, G.}, {Glass, F.}, {Gomes, M.}, {Granvik, M.},
  {Gueguen, A.}, {Guerrier, A.}, {Guiraud, J.}, {Guti\'errez-S\'anchez, R.},
  {Haigron, R.}, {Hatzidimitriou, D.}, {Hauser, M.}, {Haywood, M.}, {Heiter,
  U.}, {Helmi, A.}, {Heu, J.}, {Hilger, T.}, {Hobbs, D.}, {Hofmann, W.},
  {Holland, G.}, {Huckle, H. E.}, {Hypki, A.}, {Icardi, V.}, {Jan\ss{}en, K.},
  {Jevardat de Fombelle, G.}, {Jonker, P. G.}, {Juh\'asz, \'A. L.}, {Julbe,
  F.}, {Karampelas, A.}, {Kewley, A.}, {Klar, J.}, {Kochoska, A.}, {Kohley,
  R.}, {Kolenberg, K.}, {Kontizas, M.}, {Kontizas, E.}, {Koposov, S. E.},
  {Kordopatis, G.}, {Kostrzewa-Rutkowska, Z.}, {Koubsky, P.}, {Lambert, S.},
  {Lanza, A. F.}, {Lasne, Y.}, {Lavigne, J.-B.}, {Le Fustec, Y.}, {Le
  Poncin-Lafitte, C.}, {Lebreton, Y.}, {Leccia, S.}, {Leclerc, N.},
  {Lecoeur-Taibi, I.}, {Lenhardt, H.}, {Leroux, F.}, {Liao, S.}, {Licata, E.},
  {Lindstr\o{}m, H. E. P.}, {Lister, T. A.}, {Livanou, E.}, {Lobel, A.},
  {L\'opez, M.}, {Managau, S.}, {Mann, R. G.}, {Mantelet, G.}, {Marchal, O.},
  {Marchant, J. M.}, {Marconi, M.}, {Marinoni, S.}, {Marschalk\'o, G.},
  {Marshall, D. J.}, {Martino, M.}, {Marton, G.}, {Mary, N.}, {Massari, D.},
  {Matijevic, G.}, {Mazeh, T.}, {McMillan, P. J.}, {Messina, S.}, {Michalik,
  D.}, {Millar, N. R.}, {Molina, D.}, {Molinaro, R.}, {Moln\'ar, L.},
  {Montegriffo, P.}, {Mor, R.}, {Morbidelli, R.}, {Morel, T.}, {Morris, D.},
  {Mulone, A. F.}, {Muraveva, T.}, {Musella, I.}, {Nelemans, G.}, {Nicastro,
  L.}, {Noval, L.}, {O\'{}Mullane, W.}, {Ord\'enovic, C.}, {Ord\'o\~nez-Blanco,
  D.}, {Osborne, P.}, {Pagani, C.}, {Pagano, I.}, {Pailler, F.}, {Palacin, H.},
  {Palaversa, L.}, {Panahi, A.}, {Pawlak, M.}, {Piersimoni, A. M.}, {Pineau,
  F.-X.}, {Plachy, E.}, {Plum, G.}, {Poggio, E.}, {Poujoulet, E.}, {Prsa, A.},
  {Pulone, L.}, {Racero, E.}, {Ragaini, S.}, {Rambaux, N.}, {Ramos-Lerate, M.},
  {Regibo, S.}, {Reyl\'e, C.}, {Riclet, F.}, {Ripepi, V.}, {Riva, A.}, {Rivard,
  A.}, {Rixon, G.}, {Roegiers, T.}, {Roelens, M.}, {Romero-G\'omez, M.},
  {Rowell, N.}, {Royer, F.}, {Ruiz-Dern, L.}, {Sadowski, G.}, {Sagrist\`a
  Sell\'es, T.}, {Sahlmann, J.}, {Salgado, J.}, {Salguero, E.}, {Sanna, N.},
  {Santana-Ros, T.}, {Sarasso, M.}, {Savietto, H.}, {Schultheis, M.}, {Sciacca,
  E.}, {Segol, M.}, {Segovia, J. C.}, {S\'egransan, D.}, {Shih, I-C.},
  {Siltala, L.}, {Silva, A. F.}, {Smart, R. L.}, {Smith, K. W.}, {Solano, E.},
  {Solitro, F.}, {Sordo, R.}, {Soria Nieto, S.}, {Souchay, J.}, {Spagna, A.},
  {Spoto, F.}, {Stampa, U.}, {Steele, I. A.}, {Steidelm\"uller, H.},
  {Stephenson, C. A.}, {Stoev, H.}, {Suess, F. F.}, {Surdej, J.}, {Szabados,
  L.}, {Szegedi-Elek, E.}, {Tapiador, D.}, {Taris, F.}, {Tauran, G.}, {Taylor,
  M. B.}, {Teixeira, R.}, {Terrett, D.}, {Teyssandier, P.}, {Thuillot, W.},
  {Titarenko, A.}, {Torra Clotet, F.}, {Turon, C.}, {Ulla, A.}, {Utrilla, E.},
  {Uzzi, S.}, {Vaillant, M.}, {Valentini, G.}, {Valette, V.}, {van Elteren,
  A.}, {Van Hemelryck, E.}, {van Leeuwen, M.}, {Vaschetto, M.}, {Vecchiato,
  A.}, {Veljanoski, J.}, {Viala, Y.}, {Vicente, D.}, {Vogt, S.}, {von Essen,
  C.}, {Voss, H.}, {Votruba, V.}, {Voutsinas, S.}, {Walmsley, G.}, {Weiler,
  M.}, {Wertz, O.}, {Wevers, T.}, {Wyrzykowski, L.}, {Yoldas, A.}, {Zerjal,
  M.}, {Ziaeepour, H.}, {Zorec, J.}, {Zschocke, S.}, {Zucker, S.}, {Zurbach,
  C.}, \& {Zwitter, T.}}]{gaiadr2}
{Gaia Collaboration}, {Brown, A. G. A.}, {Vallenari, A.}, {et~al.}
  2018{\natexlab{a}}, A\&A, 616, A1, \dodoi{10.1051/0004-6361/201833051}

\bibitem[{{Gaia Collaboration} {et~al.}(2018{\natexlab{b}}){Gaia
  Collaboration}, {Babusiaux}, {van Leeuwen}, {Barstow}, {Jordi}, {Vallenari},
  {Bossini}, {Bressan}, {Cantat-Gaudin}, {van Leeuwen}, {Brown}, {Prusti}, {de
  Bruijne}, {Bailer-Jones}, {Biermann}, {Evans}, {Eyer}, {Jansen}, {Klioner},
  {Lammers}, {Lindegren}, {Luri}, {Mignard}, {Panem}, {Pourbaix}, {Randich},
  {Sartoretti}, {Siddiqui}, {Soubiran}, {Walton}, {Arenou}, {Bastian},
  {Cropper}, {Drimmel}, {Katz}, {Lattanzi}, {Bakker}, {Cacciari},
  {Casta{\~n}eda}, {Chaoul}, {Cheek}, {De Angeli}, {Fabricius}, {Guerra},
  {Holl}, {Masana}, {Messineo}, {Mowlavi}, {Nienartowicz}, {Panuzzo},
  {Portell}, {Riello}, {Seabroke}, {Tanga}, {Th{\'e}venin}, {Gracia-Abril},
  {Comoretto}, {Garcia-Reinaldos}, {Teyssier}, {Altmann}, {Andrae}, {Audard},
  {Bellas-Velidis}, {Benson}, {Berthier}, {Blomme}, {Burgess}, {Busso},
  {Carry}, {Cellino}, {Clementini}, {Clotet}, {Creevey}, {Davidson}, {De
  Ridder}, {Delchambre}, {Dell'Oro}, {Ducourant},
  {Fern{\'a}ndez-Hern{\'a}ndez}, {Fouesneau}, {Fr{\'e}mat}, {Galluccio},
  {Garc{\'\i}a-Torres}, {Gonz{\'a}lez-N{\'u}{\~n}ez}, {Gonz{\'a}lez-Vidal},
  {Gosset}, {Guy}, {Halbwachs}, {Hambly}, {Harrison}, {Hern{\'a}ndez},
  {Hestroffer}, {Hodgkin}, {Hutton}, {Jasniewicz}, {Jean-Antoine-Piccolo},
  {Jordan}, {Korn}, {Krone-Martins}, {Lanzafame}, {Lebzelter}, {L{\"o}ffler},
  {Manteiga}, {Marrese}, {Mart{\'\i}n-Fleitas}, {Moitinho}, {Mora}, {Muinonen},
  {Osinde}, {Pancino}, {Pauwels}, {Petit}, {Recio-Blanco}, {Richards},
  {Rimoldini}, {Robin}, {Sarro}, {Siopis}, {Smith}, {Sozzetti}, {S{\"u}veges},
  {Torra}, {van Reeven}, {Abbas}, {Abreu Aramburu}, {Accart}, {Aerts},
  {Altavilla}, {{\'A}lvarez}, {Alvarez}, {Alves}, {Anderson}, {Andrei},
  {Anglada Varela}, {Antiche}, {Antoja}, {Arcay}, {Astraatmadja}, {Bach},
  {Baker}, {Balaguer-N{\'u}{\~n}ez}, {Balm}, {Barache}, {Barata}, {Barbato},
  {Barblan}, {Barklem}, {Barrado}, {Barros}, {Bartholom{\'e} Mu{\~n}oz},
  {Bassilana}, {Becciani}, {Bellazzini}, {Berihuete}, {Bertone}, {Bianchi},
  {Bienaym{\'e}}, {Blanco-Cuaresma}, {Boch}, {Boeche}, {Bombrun}, {Borrachero},
  {Bouquillon}, {Bourda}, {Bragaglia}, {Bramante}, {Breddels}, {Brouillet},
  {Br{\"u}semeister}, {Brugaletta}, {Bucciarelli}, {Burlacu}, {Busonero},
  {Butkevich}, {Buzzi}, {Caffau}, {Cancelliere}, {Cannizzaro}, {Carballo},
  {Carlucci}, {Carrasco}, {Casamiquela}, {Castellani}, {Castro-Ginard},
  {Charlot}, {Chemin}, {Chiavassa}, {Cocozza}, {Costigan}, {Cowell}, {Crifo},
  {Crosta}, {Crowley}, {Cuypers}, {Dafonte}, {Damerdji}, {Dapergolas}, {David},
  {David}, {de Laverny}, {De Luise}, {De March}, {de Martino}, {de Souza}, {de
  Torres}, {Debosscher}, {del Pozo}, {Delbo}, {Delgado}, {Delgado}, {Diakite},
  {Diener}, {Distefano}, {Dolding}, {Drazinos}, {Dur{\'a}n}, {Edvardsson},
  {Enke}, {Eriksson}, {Esquej}, {Eynard Bontemps}, {Fabre}, {Fabrizio},
  {Faigler}, {Falc{\~a}o}, {Farr{\`a}s Casas}, {Federici}, {Fedorets},
  {Fernique}, {Figueras}, {Filippi}, {Findeisen}, {Fonti}, {Fraile}, {Fraser},
  {Fr{\'e}zouls}, {Gai}, {Galleti}, {Garabato}, {Garc{\'\i}a-Sedano},
  {Garofalo}, {Garralda}, {Gavel}, {Gavras}, {Gerssen}, {Geyer}, {Giacobbe},
  {Gilmore}, {Girona}, {Giuffrida}, {Glass}, {Gomes}, {Granvik}, {Gueguen},
  {Guerrier}, {Guiraud}, {Guti{\'e}}, {Haigron}, {Hatzidimitriou}, {Hauser},
  {Haywood}, {Heiter}, {Helmi}, {Heu}, {Hilger}, {Hobbs}, {Hofmann}, {Holland},
  {Huckle}, {Hypki}, {Icardi}, {Jan{\ss}en}, {Jevardat de Fombelle}, {Jonker},
  {Juh{\'a}sz}, {Julbe}, {Karampelas}, {Kewley}, {Klar}, {Kochoska}, {Kohley},
  {Kolenberg}, {Kontizas}, {Kontizas}, {Koposov}, {Kordopatis},
  {Kostrzewa-Rutkowska}, {Koubsky}, {Lambert}, {Lanza}, {Lasne}, {Lavigne}, {Le
  Fustec}, {Le Poncin-Lafitte}, {Lebreton}, {Leccia}, {Leclerc},
  {Lecoeur-Taibi}, {Lenhardt}, {Leroux}, {Liao}, {Licata}, {Lindstr{\o}m},
  {Lister}, {Livanou}, {Lobel}, {L{\'o}pez}, {Managau}, {Mann}, {Mantelet},
  {Marchal}, {Marchant}, {Marconi}, {Marinoni}, {Marschalk{\'o}}, {Marshall},
  {Martino}, {Marton}, {Mary}, {Massari}, {Matijevi{\v{c}}}, {Mazeh},
  {McMillan}, {Messina}, {Michalik}, {Millar}, {Molina}, {Molinaro},
  {Moln{\'a}r}, {Montegriffo}, {Mor}, {Morbidelli}, {Morel}, {Morris},
  {Mulone}, {Muraveva}, {Musella}, {Nelemans}, {Nicastro}, {Noval},
  {O'Mullane}, {Ord{\'e}novic}, {Ord{\'o}{\~n}ez-Blanco}, {Osborne}, {Pagani},
  {Pagano}, {Pailler}, {Palacin}, {Palaversa}, {Panahi}, {Pawlak},
  {Piersimoni}, {Pineau}, {Plachy}, {Plum}, {Poggio}, {Poujoulet},
  {Pr{\v{s}}a}, {Pulone}, {Racero}, {Ragaini}, {Rambaux}, {Ramos-Lerate},
  {Regibo}, {Reyl{\'e}}, {Riclet}, {Ripepi}, {Riva}, {Rivard}, {Rixon},
  {Roegiers}, {Roelens}, {Romero-G{\'o}mez}, {Rowell}, {Royer}, {Ruiz-Dern},
  {Sadowski}, {Sagrist{\`a} Sell{\'e}s}, {Sahlmann}, {Salgado}, {Salguero},
  {Sanna}, {Santana-Ros}, {Sarasso}, {Savietto}, {Schultheis}, {Sciacca},
  {Segol}, {Segovia}, {S{\'e}gransan}, {Shih}, {Siltala}, {Silva}, {Smart},
  {Smith}, {Solano}, {Solitro}, {Sordo}, {Soria Nieto}, {Souchay}, {Spagna},
  {Spoto}, {Stampa}, {Steele}, {Steidelm{\"u}ller}, {Stephenson}, {Stoev},
  {Suess}, {Surdej}, {Szabados}, {Szegedi-Elek}, {Tapiador}, {Taris}, {Tauran},
  {Taylor}, {Teixeira}, {Terrett}, {Teyssandier}, {Thuillot}, {Titarenko},
  {Torra Clotet}, {Turon}, {Ulla}, {Utrilla}, {Uzzi}, {Vaillant}, {Valentini},
  {Valette}, {van Elteren}, {Van Hemelryck}, {Vaschetto}, {Vecchiato},
  {Veljanoski}, {Viala}, {Vicente}, {Vogt}, {von Essen}, {Voss}, {Votruba},
  {Voutsinas}, {Walmsley}, {Weiler}, {Wertz}, {Wevers}, {Wyrzykowski},
  {Yoldas}, {{\v{Z}}erjal}, {Ziaeepour}, {Zorec}, {Zschocke}, {Zucker},
  {Zurbach}, \& {Zwitter}}]{gaia2018b}
{Gaia Collaboration}, {Babusiaux}, C., {van Leeuwen}, F., {et~al.}
  2018{\natexlab{b}}, \aap, 616, A10, \dodoi{10.1051/0004-6361/201832843}

\bibitem[{{Gaidos} {et~al.}(2017){Gaidos}, {Mann}, {Rizzuto}, {Nofi}, {Mace},
  {Vanderburg}, {Feiden}, {Narita}, {Takeda}, {Esposito}, {De Rosa}, {Ansdell},
  {Hirano}, {Graham}, {Kraus}, \& {Jaffe}}]{gaidos2017}
{Gaidos}, E., {Mann}, A.~W., {Rizzuto}, A., {et~al.} 2017, \mnras, 464, 850,
  \dodoi{10.1093/mnras/stw2345}

\bibitem[{{Garc{\'{\i}}a P{\'e}rez} {et~al.}(2016){Garc{\'{\i}}a P{\'e}rez},
  {Allende Prieto}, {Holtzman}, {Shetrone}, {M{\'e}sz{\'a}ros}, {Bizyaev},
  {Carrera}, {Cunha}, {Garc{\'{\i}}a-Hern{\'a}ndez}, {Johnson}, {Majewski},
  {Nidever}, {Schiavon}, {Shane}, {Smith}, {Sobeck}, {Troup}, {Zamora},
  {Weinberg}, {Bovy}, {Eisenstein}, {Feuillet}, {Frinchaboy}, {Hayden},
  {Hearty}, {Nguyen}, {O'Connell}, {Pinsonneault}, {Wilson}, \&
  {Zasowski}}]{aspcap}
{Garc{\'{\i}}a P{\'e}rez}, A.~E., {Allende Prieto}, C., {Holtzman}, J.~A.,
  {et~al.} 2016, \aj, 151, 144, \dodoi{10.3847/0004-6256/151/6/144}

\bibitem[{{Ghezzi} {et~al.}(2010){Ghezzi}, {Cunha}, {Smith}, {de Ara{\'u}jo},
  {Schuler}, \& {de la Reza}}]{ghezzi2010}
{Ghezzi}, L., {Cunha}, K., {Smith}, V.~V., {et~al.} 2010, \apj, 720, 1290,
  \dodoi{10.1088/0004-637X/720/2/1290}

\bibitem[{{Ghezzi} {et~al.}(2021){Ghezzi}, {Martinez}, {Wilson}, {Cunha},
  {Smith}, \& {Majewski}}]{ghezzi2021}
{Ghezzi}, L., {Martinez}, C.~F., {Wilson}, R.~F., {et~al.} 2021, arXiv
  e-prints, arXiv:2107.04153.
\newblock \doarXiv{2107.04153}

\bibitem[{{Ghezzi} {et~al.}(2018){Ghezzi}, {Montet}, \& {Johnson}}]{ghezzi2018}
{Ghezzi}, L., {Montet}, B.~T., \& {Johnson}, J.~A. 2018, \apj, 860, 109,
  \dodoi{10.3847/1538-4357/aac37c}

\bibitem[{{Gonzalez}(1997)}]{gonzalez1997}
{Gonzalez}, G. 1997, \mnras, 285, 403, \dodoi{10.1093/mnras/285.2.403}

\bibitem[{{Gonz{\'a}lez Hern{\'a}ndez} \&
  {Bonifacio}(2009)}]{gonzalezhernandez2009}
{Gonz{\'a}lez Hern{\'a}ndez}, J.~I., \& {Bonifacio}, P. 2009, \aap, 497, 497,
  \dodoi{10.1051/0004-6361/200810904}

\bibitem[{{Green}(2018)}]{dustmaps}
{Green}, G.~M. 2018, The Journal of Open Source Software, 3, 695,
  \dodoi{10.21105/joss.00695}

\bibitem[{{Green} {et~al.}(2019){Green}, {Schlafly}, {Zucker}, {Speagle}, \&
  {Finkbeiner}}]{green2019}
{Green}, G.~M., {Schlafly}, E., {Zucker}, C., {Speagle}, J.~S., \&
  {Finkbeiner}, D. 2019, \apj, 887, 93, \dodoi{10.3847/1538-4357/ab5362}

\bibitem[{{Gunn} {et~al.}(2006){Gunn}, {Siegmund}, {Mannery}, {Owen}, {Hull},
  {Leger}, {Carey}, {Knapp}, {York}, {Boroski}, {Kent}, {Lupton}, {Rockosi},
  {Evans}, {Waddell}, {Anderson}, {Annis}, {Barentine}, {Bartoszek}, {Bastian},
  {Bracker}, {Brewington}, {Briegel}, {Brinkmann}, {Brown}, {Carr},
  {Czarapata}, {Drennan}, {Dombeck}, {Federwitz}, {Gillespie}, {Gonzales},
  {Hansen}, {Harvanek}, {Hayes}, {Jordan}, {Kinney}, {Klaene}, {Kleinman},
  {Kron}, {Kresinski}, {Lee}, {Limmongkol}, {Lindenmeyer}, {Long}, {Loomis},
  {McGehee}, {Mantsch}, {Neilsen}, {Neswold}, {Newman}, {Nitta}, {Peoples},
  {Pier}, {Prieto}, {Prosapio}, {Rivetta}, {Schneider}, {Snedden}, \&
  {Wang}}]{gunn2006}
{Gunn}, J.~E., {Siegmund}, W.~A., {Mannery}, E.~J., {et~al.} 2006, \aj, 131,
  2332, \dodoi{10.1086/500975}

\bibitem[{{Gupta} \& {Schlichting}(2019)}]{gupta&schlichting2019}
{Gupta}, A., \& {Schlichting}, H.~E. 2019, \mnras, 487, 24,
  \dodoi{10.1093/mnras/stz1230}

\bibitem[{{Gupta} \& {Schlichting}(2020)}]{gupta&schlichting2020}
---. 2020, \mnras, 493, 792, \dodoi{10.1093/mnras/staa315}

\bibitem[{{Hamer} \& {Schlaufman}(2019)}]{hamer2019}
{Hamer}, J.~H., \& {Schlaufman}, K.~C. 2019, \aj, 158, 190,
  \dodoi{10.3847/1538-3881/ab3c56}

\bibitem[{Harris {et~al.}(2020)Harris, Millman, van~der Walt, Gommers,
  Virtanen, Cournapeau, Wieser, Taylor, Berg, Smith, Kern, Picus, Hoyer, van
  Kerkwijk, Brett, Haldane, del R{'{\i}}o, Wiebe, Peterson,
  G{'{e}}rard-Marchant, Sheppard, Reddy, Weckesser, Abbasi, Gohlke, \&
  Oliphant}]{numpy}
Harris, C.~R., Millman, K.~J., van~der Walt, S.~J., {et~al.} 2020, Nature, 585,
  357, \dodoi{10.1038/s41586-020-2649-2}

\bibitem[{{Hayden} {et~al.}(2014){Hayden}, {Holtzman}, {Bovy}, {Majewski},
  {Johnson}, {Allende Prieto}, {Beers}, {Cunha}, {Frinchaboy}, {Garc{\'{\i}}a
  P{\'e}rez}, {Girardi}, {Hearty}, {Lee}, {Nidever}, {Schiavon}, {Schlesinger},
  {Schneider}, {Schultheis}, {Shetrone}, {Smith}, {Zasowski}, {Bizyaev},
  {Feuillet}, {Hasselquist}, {Kinemuchi}, {Malanushenko}, {Malanushenko},
  {O'Connell}, {Pan}, \& {Stassun}}]{hayden2014}
{Hayden}, M.~R., {Holtzman}, J.~A., {Bovy}, J., {et~al.} 2014, \aj, 147, 116,
  \dodoi{10.1088/0004-6256/147/5/116}

\bibitem[{{Hayden} {et~al.}(2015){Hayden}, {Bovy}, {Holtzman}, {Nidever},
  {Bird}, {Weinberg}, {Andrews}, {Majewski}, {Allende Prieto}, {Anders},
  {Beers}, {Bizyaev}, {Chiappini}, {Cunha}, {Frinchaboy},
  {Garc{\'\i}a-Her{\'n}andez}, {Garc{\'\i}a P{\'e}rez}, {Girardi}, {Harding},
  {Hearty}, {Johnson}, {M{\'e}sz{\'a}ros}, {Minchev}, {O'Connell}, {Pan},
  {Robin}, {Schiavon}, {Schneider}, {Schultheis}, {Shetrone}, {Skrutskie},
  {Steinmetz}, {Smith}, {Wilson}, {Zamora}, \& {Zasowski}}]{hayden2015}
{Hayden}, M.~R., {Bovy}, J., {Holtzman}, J.~A., {et~al.} 2015, \apj, 808, 132,
  \dodoi{10.1088/0004-637X/808/2/132}

\bibitem[{{Heiter} \& {Luck}(2003)}]{heiter&luck2003}
{Heiter}, U., \& {Luck}, R.~E. 2003, \aj, 126, 2015, \dodoi{10.1086/378366}

\bibitem[{{Hinkel} \& {Unterborn}(2018)}]{hinkel2018}
{Hinkel}, N.~R., \& {Unterborn}, C.~T. 2018, \apj, 853, 83,
  \dodoi{10.3847/1538-4357/aaa5b4}

\bibitem[{{Holtzman} {et~al.}(2018){Holtzman}, {Hasselquist}, {Shetrone},
  {Cunha}, {Allende Prieto}, {Anguiano}, {Bizyaev}, {Bovy}, {Casey},
  {Edvardsson}, {Johnson}, {J{\"o}nsson}, {Meszaros}, {Smith}, {Sobeck},
  {Zamora}, {Chojnowski}, {Fernandez-Trincado}, {Garcia-Hernandez}, {Majewski},
  {Pinsonneault}, {Souto}, {Stringfellow}, {Tayar}, {Troup}, \&
  {Zasowski}}]{holtzman2018}
{Holtzman}, J.~A., {Hasselquist}, S., {Shetrone}, M., {et~al.} 2018, \aj, 156,
  125, \dodoi{10.3847/1538-3881/aad4f9}

\bibitem[{{Howard} {et~al.}(2012){Howard}, {Marcy}, {Bryson}, {Jenkins},
  {Rowe}, {Batalha}, {Borucki}, {Koch}, {Dunham}, {Gautier}, {Van Cleve},
  {Cochran}, {Latham}, {Lissauer}, {Torres}, {Brown}, {Gilliland}, {Buchhave},
  {Caldwell}, {Christensen-Dalsgaard}, {Ciardi}, {Fressin}, {Haas}, {Howell},
  {Kjeldsen}, {Seager}, {Rogers}, {Sasselov}, {Steffen}, {Basri},
  {Charbonneau}, {Christiansen}, {Clarke}, {Dupree}, {Fabrycky}, {Fischer},
  {Ford}, {Fortney}, {Tarter}, {Girouard}, {Holman}, {Johnson}, {Klaus},
  {Machalek}, {Moorhead}, {Morehead}, {Ragozzine}, {Tenenbaum}, {Twicken},
  {Quinn}, {Isaacson}, {Shporer}, {Lucas}, {Walkowicz}, {Welsh}, {Boss},
  {Devore}, {Gould}, {Smith}, {Morris}, {Prsa}, {Morton}, {Still}, {Thompson},
  {Mullally}, {Endl}, \& {MacQueen}}]{howard2012}
{Howard}, A.~W., {Marcy}, G.~W., {Bryson}, S.~T., {et~al.} 2012, \apjs, 201,
  15, \dodoi{10.1088/0067-0049/201/2/15}

\bibitem[{{Howell} {et~al.}(2014){Howell}, {Sobeck}, {Haas}, {Still},
  {Barclay}, {Mullally}, {Troeltzsch}, {Aigrain}, {Bryson}, {Caldwell},
  {Chaplin}, {Cochran}, {Huber}, {Marcy}, {Miglio}, {Najita}, {Smith},
  {Twicken}, \& {Fortney}}]{howell2014}
{Howell}, S.~B., {Sobeck}, C., {Haas}, M., {et~al.} 2014, \pasp, 126, 398,
  \dodoi{10.1086/676406}

\bibitem[{{Hsu} {et~al.}(2019){Hsu}, {Ford}, {Ragozzine}, \& {Ashby}}]{hsu2019}
{Hsu}, D.~C., {Ford}, E.~B., {Ragozzine}, D., \& {Ashby}, K. 2019, \aj, 158,
  109, \dodoi{10.3847/1538-3881/ab31ab}

\bibitem[{{Huber} {et~al.}(2017){Huber}, {Zinn}, {Bojsen-Hansen},
  {Pinsonneault}, {Sahlholdt}, {Serenelli}, {Silva Aguirre}, {Stassun},
  {Stello}, {Tayar}, {Bastien}, {Bedding}, {Buchhave}, {Chaplin}, {Davies},
  {Garc{\'\i}a}, {Latham}, {Mathur}, {Mosser}, \& {Sharma}}]{huber2017}
{Huber}, D., {Zinn}, J., {Bojsen-Hansen}, M., {et~al.} 2017, \apj, 844, 102,
  \dodoi{10.3847/1538-4357/aa75ca}

\bibitem[{Hunter(2007)}]{matplotlib}
Hunter, J.~D. 2007, Computing in Science \& Engineering, 9, 90,
  \dodoi{10.1109/MCSE.2007.55}

\bibitem[{{Ida} \& {Lin}(2004)}]{ida&lin2004}
{Ida}, S., \& {Lin}, D.~N.~C. 2004, \apj, 616, 567, \dodoi{10.1086/424830}

\bibitem[{{Johnson} {et~al.}(2017){Johnson}, {Petigura}, {Fulton}, {Marcy},
  {Howard}, {Isaacson}, {Hebb}, {Cargile}, {Morton}, {Weiss}, {Winn}, {Rogers},
  {Sinukoff}, \& {Hirsch}}]{johnson2017}
{Johnson}, J.~A., {Petigura}, E.~A., {Fulton}, B.~J., {et~al.} 2017, \aj, 154,
  108, \dodoi{10.3847/1538-3881/aa80e7}

\bibitem[{{J{\"o}nsson} {et~al.}(2020){J{\"o}nsson}, {Holtzman}, {Allende
  Prieto}, {Cunha}, {Garc{\'\i}a-Hern{\'a}ndez}, {Hasselquist}, {Masseron},
  {Osorio}, {Shetrone}, {Smith}, {Stringfellow}, {Bizyaev}, {Edvardsson},
  {Majewski}, {M{\'e}sz{\'a}ros}, {Souto}, {Zamora}, {Beaton}, {Bovy}, {Donor},
  {Pinsonneault}, {Poovelil}, \& {Sobeck}}]{jonsson2020}
{J{\"o}nsson}, H., {Holtzman}, J.~A., {Allende Prieto}, C., {et~al.} 2020, \aj,
  160, 120, \dodoi{10.3847/1538-3881/aba592}

\bibitem[{{Linden} {et~al.}(2017){Linden}, {Pryal}, {Hayes}, {Troup},
  {Majewski}, {Andrews}, {Beers}, {Carrera}, {Cunha}, {Fern{\'a}ndez-Trincado},
  {Frinchaboy}, {Geisler}, {Lane}, {Nitschelm}, {Pan}, {Allende Prieto},
  {Roman-Lopes}, {Smith}, {Sobeck}, {Tang}, {Villanova}, \&
  {Zasowski}}]{linden2017}
{Linden}, S.~T., {Pryal}, M., {Hayes}, C.~R., {et~al.} 2017, \apj, 842, 49,
  \dodoi{10.3847/1538-4357/aa6f17}

\bibitem[{{Lodders}(2003)}]{lodders2003}
{Lodders}, K. 2003, \apj, 591, 1220, \dodoi{10.1086/375492}

\bibitem[{Majewski {et~al.}(2017)Majewski, Schiavon, Frinchaboy, Prieto,
  Barkhouser, Bizyaev, Blank, Brunner, Burton, Carrera, Chojnowski, Cunha,
  Epstein, Fitzgerald, Perez, Hearty, Henderson, Holtzman, Johnson, Lam,
  Lawler, Maseman, Meszaros, Nelson, Nguyen, Nidever, Pinsonneault, Shetrone,
  Smee, Smith, Stolberg, Skrutskie, Walker, Wilson, Zasowski, Anders, Basu,
  Beland, Blanton, Bovy, Brownstein, Carlberg, Chaplin, Chiappini, Eisenstein,
  Elsworth, Feuillet, Fleming, Galbraith-Frew, Garcia, Garcia-Hernandez,
  Gillespie, Girardi, Gunn, Hasselquist, Hayden, Hekker, Ivans, Kinemuchi,
  Klaene, Mahadevan, Mathur, Mosser, Muna, Munn, Nichol, O'Connell, Parejko,
  Robin, Rocha-Pinto, Schultheis, Serenelli, Shane, Aguirre, Sobeck, Thompson,
  Troup, Weinberg, \& Zamora}]{majewski2017}
Majewski, S.~R., Schiavon, R.~P., Frinchaboy, P.~M., {et~al.} 2017, \aj, 154,
  94.
\newblock \url{http://stacks.iop.org/1538-3881/154/i=3/a=94}

\bibitem[{{Maldonado} {et~al.}(2019){Maldonado}, {Villaver}, {Eiroa}, \&
  {Micela}}]{maldonado2019}
{Maldonado}, J., {Villaver}, E., {Eiroa}, C., \& {Micela}, G. 2019, \aap, 624,
  A94, \dodoi{10.1051/0004-6361/201833827}

\bibitem[{{Mann} {et~al.}(2015){Mann}, {Feiden}, {Gaidos}, {Boyajian}, \& {von
  Braun}}]{mann2015}
{Mann}, A.~W., {Feiden}, G.~A., {Gaidos}, E., {Boyajian}, T., \& {von Braun},
  K. 2015, \apj, 804, 64, \dodoi{10.1088/0004-637X/804/1/64}

\bibitem[{{Mann} {et~al.}(2016){Mann}, {Gaidos}, {Mace}, {Johnson}, {Bowler},
  {LaCourse}, {Jacobs}, {Vanderburg}, {Kraus}, {Kaplan}, \& {Jaffe}}]{mann2016}
{Mann}, A.~W., {Gaidos}, E., {Mace}, G.~N., {et~al.} 2016, \apj, 818, 46,
  \dodoi{10.3847/0004-637X/818/1/46}

\bibitem[{{Mann} {et~al.}(2018){Mann}, {Vanderburg}, {Rizzuto}, {Kraus},
  {Berlind}, {Bieryla}, {Calkins}, {Esquerdo}, {Latham}, {Mace}, {Morris},
  {Quinn}, {Sokal}, \& {Stefanik}}]{mann2018}
{Mann}, A.~W., {Vanderburg}, A., {Rizzuto}, A.~C., {et~al.} 2018, \aj, 155, 4,
  \dodoi{10.3847/1538-3881/aa9791}

\bibitem[{{Mann} {et~al.}(2019){Mann}, {Dupuy}, {Kraus}, {Gaidos}, {Ansdell},
  {Ireland}, {Rizzuto}, {Hung}, {Dittmann}, {Factor}, {Feiden}, {Martinez},
  {Ru{\'\i}z-Rodr{\'\i}guez}, \& {Thao}}]{mann2019}
{Mann}, A.~W., {Dupuy}, T., {Kraus}, A.~L., {et~al.} 2019, \apj, 871, 63,
  \dodoi{10.3847/1538-4357/aaf3bc}

\bibitem[{{Marboeuf} {et~al.}(2014){Marboeuf}, {Thiabaud}, {Alibert}, {Cabral},
  \& {Benz}}]{marboeuf2014}
{Marboeuf}, U., {Thiabaud}, A., {Alibert}, Y., {Cabral}, N., \& {Benz}, W.
  2014, \aap, 570, A36, \dodoi{10.1051/0004-6361/201423431}

\bibitem[{{Martinez} {et~al.}(2019){Martinez}, {Cunha}, {Ghezzi}, \&
  {Smith}}]{martinez2019}
{Martinez}, C.~F., {Cunha}, K., {Ghezzi}, L., \& {Smith}, V.~V. 2019, \apj,
  875, 29, \dodoi{10.3847/1538-4357/ab0d93}

\bibitem[{{Masuda} \& {Winn}(2017)}]{masuda&winn2017}
{Masuda}, K., \& {Winn}, J.~N. 2017, \aj, 153, 187,
  \dodoi{10.3847/1538-3881/aa647c}

\bibitem[{{McWilliam}(1997)}]{mcwilliam1997}
{McWilliam}, A. 1997, \araa, 35, 503, \dodoi{10.1146/annurev.astro.35.1.503}

\bibitem[{{Mordasini} {et~al.}(2012){Mordasini}, {Alibert}, {Georgy},
  {Dittkrist}, {Klahr}, \& {Henning}}]{mordasini2012}
{Mordasini}, C., {Alibert}, Y., {Georgy}, C., {et~al.} 2012, \aap, 547, A112,
  \dodoi{10.1051/0004-6361/201118464}

\bibitem[{{Morton}(2015)}]{morton2015}
{Morton}, T.~D. 2015, {isochrones: Stellar model grid package}.
\newblock \doeprint{1503.010}

\bibitem[{{Mulders} {et~al.}(2015){Mulders}, {Pascucci}, \&
  {Apai}}]{mulders2015a}
{Mulders}, G.~D., {Pascucci}, I., \& {Apai}, D. 2015, \apj, 798, 112,
  \dodoi{10.1088/0004-637X/798/2/112}

\bibitem[{{Mulders} {et~al.}(2018){Mulders}, {Pascucci}, {Apai}, \&
  {Ciesla}}]{mulders2018}
{Mulders}, G.~D., {Pascucci}, I., {Apai}, D., \& {Ciesla}, F.~J. 2018, \aj,
  156, 24, \dodoi{10.3847/1538-3881/aac5ea}

\bibitem[{{Mulders} {et~al.}(2016){Mulders}, {Pascucci}, {Apai}, {Frasca}, \&
  {Molenda-{\.Z}akowicz}}]{mulders2016}
{Mulders}, G.~D., {Pascucci}, I., {Apai}, D., {Frasca}, A., \&
  {Molenda-{\.Z}akowicz}, J. 2016, \aj, 152, 187,
  \dodoi{10.3847/0004-6256/152/6/187}

\bibitem[{{Mullally} {et~al.}(2016){Mullally}, {Coughlin}, {Thompson},
  {Christiansen}, {Burke}, {Clarke}, \& {Haas}}]{mulally2016}
{Mullally}, F., {Coughlin}, J.~L., {Thompson}, S.~E., {et~al.} 2016, \pasp,
  128, 074502, \dodoi{10.1088/1538-3873/128/965/074502}

\bibitem[{{Mullally} {et~al.}(2015){Mullally}, {Coughlin}, {Thompson}, {Rowe},
  {Burke}, {Latham}, {Batalha}, {Bryson}, {Christiansen}, {Henze}, {Ofir},
  {Quarles}, {Shporer}, {Van Eylen}, {Van Laerhoven}, {Shah}, {Wolfgang},
  {Chaplin}, {Xie}, {Akeson}, {Argabright}, {Bachtell}, {Barclay}, {Borucki},
  {Caldwell}, {Campbell}, {Catanzarite}, {Cochran}, {Duren}, {Fleming},
  {Fraquelli}, {Girouard}, {Haas}, {He{\l}miniak}, {Howell}, {Huber}, {Larson},
  {Gautier}, {Jenkins}, {Li}, {Lissauer}, {McArthur}, {Miller}, {Morris},
  {Patil-Sabale}, {Plavchan}, {Putnam}, {Quintana}, {Ramirez}, {Silva Aguirre},
  {Seader}, {Smith}, {Steffen}, {Stewart}, {Stober}, {Still}, {Tenenbaum},
  {Troeltzsch}, {Twicken}, \& {Zamudio}}]{mullally2015}
---. 2015, \apjs, 217, 31, \dodoi{10.1088/0067-0049/217/2/31}

\bibitem[{{Narang} {et~al.}(2018){Narang}, {Manoj}, {Furlan}, {Mordasini},
  {Henning}, {Mathew}, {Banyal}, \& {Sivarani}}]{narang2018}
{Narang}, M., {Manoj}, P., {Furlan}, E., {et~al.} 2018, \aj, 156, 221,
  \dodoi{10.3847/1538-3881/aae391}

\bibitem[{{Netopil} {et~al.}(2016){Netopil}, {Paunzen}, {Heiter}, \&
  {Soubiran}}]{netopil2016}
{Netopil}, M., {Paunzen}, E., {Heiter}, U., \& {Soubiran}, C. 2016, \aap, 585,
  A150, \dodoi{10.1051/0004-6361/201526370}

\bibitem[{{Nidever} {et~al.}(2015){Nidever}, {Holtzman}, {Allende Prieto},
  {Beland}, {Bender}, {Bizyaev}, {Burton}, {Desphande}, {Fleming},
  {Garc{\'{\i}}a P{\'e}rez}, {Hearty}, {Majewski}, {M{\'e}sz{\'a}ros}, {Muna},
  {Nguyen}, {Schiavon}, {Shetrone}, {Skrutskie}, {Sobeck}, \&
  {Wilson}}]{nidever2015}
{Nidever}, D.~L., {Holtzman}, J.~A., {Allende Prieto}, C., {et~al.} 2015, \aj,
  150, 173, \dodoi{10.1088/0004-6256/150/6/173}

\bibitem[{{{\"O}berg} {et~al.}(2011){{\"O}berg}, {Murray-Clay}, \&
  {Bergin}}]{oberg2011}
{{\"O}berg}, K.~I., {Murray-Clay}, R., \& {Bergin}, E.~A. 2011, \apjl, 743,
  L16, \dodoi{10.1088/2041-8205/743/1/L16}

\bibitem[{{Owen} \& {Murray-Clay}(2018)}]{owen&murrayclay2018}
{Owen}, J.~E., \& {Murray-Clay}, R. 2018, \mnras, 480, 2206,
  \dodoi{10.1093/mnras/sty1943}

\bibitem[{{Owen} \& {Wu}(2013)}]{owen&wu2013}
{Owen}, J.~E., \& {Wu}, Y. 2013, \apj, 775, 105,
  \dodoi{10.1088/0004-637X/775/2/105}

\bibitem[{Owen \& Wu(2017)}]{owen&wu2017}
Owen, J.~E., \& Wu, Y. 2017, The Astrophysical Journal, 847, 29,
  \dodoi{10.3847/1538-4357/aa890a}

\bibitem[{{Owen} {et~al.}(1994){Owen}, {Siegmund}, {Limmongkol}, \&
  {Hull}}]{owen1994}
{Owen}, R.~E., {Siegmund}, W.~A., {Limmongkol}, S., \& {Hull}, C.~L. 1994, in
  Society of Photo-Optical Instrumentation Engineers (SPIE) Conference Series,
  Vol. 2198, Instrumentation in Astronomy VIII, ed. D.~L. {Crawford} \& E.~R.
  {Craine}, 110--114, \dodoi{10.1117/12.176689}

\bibitem[{pandas~development team(2020)}]{pandas}
pandas~development team, T. 2020, pandas-dev/pandas: Pandas, latest,  Zenodo,
  \dodoi{10.5281/zenodo.3509134}

\bibitem[{Pedregosa {et~al.}(2011)Pedregosa, Varoquaux, Gramfort, Michel,
  Thirion, Grisel, Blondel, Prettenhofer, Weiss, Dubourg, Vanderplas, Passos,
  Cournapeau, Brucher, Perrot, \& Duchesnay}]{scikit-learn}
Pedregosa, F., Varoquaux, G., Gramfort, A., {et~al.} 2011, Journal of Machine
  Learning Research, 12, 2825

\bibitem[{{Pepper} {et~al.}(2003){Pepper}, {Gould}, \& {Depoy}}]{pepper2003}
{Pepper}, J., {Gould}, A., \& {Depoy}, D.~L. 2003, \actaa, 53, 213.
\newblock \doarXiv{astro-ph/0208042}

\bibitem[{{Petigura} {et~al.}(2013){Petigura}, {Howard}, \&
  {Marcy}}]{petigura2013}
{Petigura}, E.~A., {Howard}, A.~W., \& {Marcy}, G.~W. 2013, Proceedings of the
  National Academy of Science, 110, 19273, \dodoi{10.1073/pnas.1319909110}

\bibitem[{{Petigura} {et~al.}(2017){Petigura}, {Sinukoff}, {Lopez},
  {Crossfield}, {Howard}, {Brewer}, {Fulton}, {Isaacson}, {Ciardi}, {Howell},
  {Everett}, {Horch}, {Hirsch}, {Weiss}, \& {Schlieder}}]{petigura2017b}
{Petigura}, E.~A., {Sinukoff}, E., {Lopez}, E.~D., {et~al.} 2017, \aj, 153,
  142, \dodoi{10.3847/1538-3881/aa5ea5}

\bibitem[{{Petigura} {et~al.}(2018){Petigura}, {Marcy}, {Winn}, {Weiss},
  {Fulton}, {Howard}, {Sinukoff}, {Isaacson}, {Morton}, \&
  {Johnson}}]{petigura2018}
{Petigura}, E.~A., {Marcy}, G.~W., {Winn}, J.~N., {et~al.} 2018, \aj, 155, 89,
  \dodoi{10.3847/1538-3881/aaa54c}

\bibitem[{{Price-Whelan} {et~al.}(2020){Price-Whelan}, {Hogg}, {Rix}, {Beaton},
  {Lewis}, {Nidever}, {Almeida}, {Badenes}, {Barba}, {Beers}, {Carlberg}, {De
  Lee}, {Fern{\'a}ndez-Trincado}, {Frinchaboy}, {Garc{\'\i}a-Hern{\'a}ndez},
  {Green}, {Hasselquist}, {Longa-Pe{\~n}a}, {Majewski}, {Nitschelm}, {Sobeck},
  {Stassun}, {Stringfellow}, \& {Troup}}]{pricewhelan2020}
{Price-Whelan}, A.~M., {Hogg}, D.~W., {Rix}, H.-W., {et~al.} 2020, \apj, 895,
  2, \dodoi{10.3847/1538-4357/ab8acc}

\bibitem[{{Rice} \& {Armitage}(2003)}]{rice&armitage2003}
{Rice}, W.~K.~M., \& {Armitage}, P.~J. 2003, \apjl, 598, L55,
  \dodoi{10.1086/380390}

\bibitem[{{Ricker} {et~al.}(2015){Ricker}, {Winn}, {Vanderspek}, {Latham},
  {Bakos}, {Bean}, {Berta-Thompson}, {Brown}, {Buchhave}, {Butler}, {Butler},
  {Chaplin}, {Charbonneau}, {Christensen-Dalsgaard}, {Clampin}, {Deming},
  {Doty}, {De Lee}, {Dressing}, {Dunham}, {Endl}, {Fressin}, {Ge}, {Henning},
  {Holman}, {Howard}, {Ida}, {Jenkins}, {Jernigan}, {Johnson}, {Kaltenegger},
  {Kawai}, {Kjeldsen}, {Laughlin}, {Levine}, {Lin}, {Lissauer}, {MacQueen},
  {Marcy}, {McCullough}, {Morton}, {Narita}, {Paegert}, {Palle}, {Pepe},
  {Pepper}, {Quirrenbach}, {Rinehart}, {Sasselov}, {Sato}, {Seager},
  {Sozzetti}, {Stassun}, {Sullivan}, {Szentgyorgyi}, {Torres}, {Udry}, \&
  {Villasenor}}]{ricker2015}
{Ricker}, G.~R., {Winn}, J.~N., {Vanderspek}, R., {et~al.} 2015, Journal of
  Astronomical Telescopes, Instruments, and Systems, 1, 014003,
  \dodoi{10.1117/1.JATIS.1.1.014003}

\bibitem[{{Rizzuto} {et~al.}(2017){Rizzuto}, {Mann}, {Vanderburg}, {Kraus}, \&
  {Covey}}]{rizzuto2017}
{Rizzuto}, A.~C., {Mann}, A.~W., {Vanderburg}, A., {Kraus}, A.~L., \& {Covey},
  K.~R. 2017, \aj, 154, 224, \dodoi{10.3847/1538-3881/aa9070}

\bibitem[{{Rizzuto} {et~al.}(2018){Rizzuto}, {Vanderburg}, {Mann}, {Kraus},
  {Dressing}, {Ag{\"u}eros}, {Douglas}, \& {Krolikowski}}]{rizzuto2018}
{Rizzuto}, A.~C., {Vanderburg}, A., {Mann}, A.~W., {et~al.} 2018, \aj, 156,
  195, \dodoi{10.3847/1538-3881/aadf37}

\bibitem[{{Santerne} {et~al.}(2016){Santerne}, {Moutou}, {Tsantaki}, {Bouchy},
  {H{\'e}brard}, {Adibekyan}, {Almenara}, {Amard}, {Barros}, {Boisse},
  {Bonomo}, {Bruno}, {Courcol}, {Deleuil}, {Demangeon}, {D{\'\i}az}, {Guillot},
  {Havel}, {Montagnier}, {Rajpurohit}, {Rey}, \& {Santos}}]{santerne2016}
{Santerne}, A., {Moutou}, C., {Tsantaki}, M., {et~al.} 2016, \aap, 587, A64,
  \dodoi{10.1051/0004-6361/201527329}

\bibitem[{{Santos} {et~al.}(2004){Santos}, {Israelian}, \&
  {Mayor}}]{santos2004}
{Santos}, N.~C., {Israelian}, G., \& {Mayor}, M. 2004, \aap, 415, 1153,
  \dodoi{10.1051/0004-6361:20034469}

\bibitem[{{Schlaufman}(2015)}]{schlaufman2015}
{Schlaufman}, K.~C. 2015, \apjl, 799, L26, \dodoi{10.1088/2041-8205/799/2/L26}

\bibitem[{{Schlaufman} \& {Laughlin}(2011)}]{schlaufman&laughlin2011}
{Schlaufman}, K.~C., \& {Laughlin}, G. 2011, \apj, 738, 177,
  \dodoi{10.1088/0004-637X/738/2/177}

\bibitem[{Scott(2010)}]{scott2010}
Scott, D.~W. 2010, WIREs Comput. Stat., 2, 497–502, \dodoi{10.1002/wics.103}

\bibitem[{{Sellwood} \& {Binney}(2002)}]{sellwood2002}
{Sellwood}, J.~A., \& {Binney}, J.~J. 2002, \mnras, 336, 785,
  \dodoi{10.1046/j.1365-8711.2002.05806.x}

\bibitem[{{Serenelli} {et~al.}(2013){Serenelli}, {Bergemann}, {Ruchti}, \&
  {Casagrande}}]{serenelli2013}
{Serenelli}, A.~M., {Bergemann}, M., {Ruchti}, G., \& {Casagrande}, L. 2013,
  \mnras, 429, 3645, \dodoi{10.1093/mnras/sts648}

\bibitem[{{Shetrone} {et~al.}(2015){Shetrone}, {Bizyaev}, {Lawler}, {Allende
  Prieto}, {Johnson}, {Smith}, {Cunha}, {Holtzman}, {Garc{\'{\i}}a P{\'e}rez},
  {M{\'e}sz{\'a}ros}, {Sobeck}, {Zamora}, {Garc{\'{\i}}a-Hern{\'a}ndez},
  {Souto}, {Chojnowski}, {Koesterke}, {Majewski}, \& {Zasowski}}]{shetrone2015}
{Shetrone}, M., {Bizyaev}, D., {Lawler}, J.~E., {et~al.} 2015, \apjs, 221, 24,
  \dodoi{10.1088/0067-0049/221/2/24}

\bibitem[{{Skrutskie} {et~al.}(2006){Skrutskie}, {Cutri}, {Stiening},
  {Weinberg}, {Schneider}, {Carpenter}, {Beichman}, {Capps}, {Chester},
  {Elias}, {Huchra}, {Liebert}, {Lonsdale}, {Monet}, {Price}, {Seitzer},
  {Jarrett}, {Kirkpatrick}, {Gizis}, {Howard}, {Evans}, {Fowler}, {Fullmer},
  {Hurt}, {Light}, {Kopan}, {Marsh}, {McCallon}, {Tam}, {Van Dyk}, \&
  {Wheelock}}]{skrutskie2006}
{Skrutskie}, M.~F., {Cutri}, R.~M., {Stiening}, R., {et~al.} 2006, \aj, 131,
  1163, \dodoi{10.1086/498708}

\bibitem[{{Smith} {et~al.}(2021){Smith}, {Bizyaev}, {Cunha}, {Shetrone},
  {Souto}, {Allende Prieto}, {Masseron}, {M{\'e}sz{\'a}ros}, {J{\"o}nsson},
  {Hasselquist}, {Osorio}, {Garc{\'\i}a-Hern{\'a}ndez}, {Plez}, {Beaton},
  {Holtzman}, {Majewski}, {Stringfellow}, \& {Sobeck}}]{smith2021}
{Smith}, V.~V., {Bizyaev}, D., {Cunha}, K., {et~al.} 2021, \aj, 161, 254,
  \dodoi{10.3847/1538-3881/abefdc}

\bibitem[{{Sousa} {et~al.}(2008){Sousa}, {Santos}, {Mayor}, {Udry},
  {Casagrande}, {Israelian}, {Pepe}, {Queloz}, \& {Monteiro}}]{sousa2008}
{Sousa}, S.~G., {Santos}, N.~C., {Mayor}, M., {et~al.} 2008, \aap, 487, 373,
  \dodoi{10.1051/0004-6361:200809698}

\bibitem[{{Souto} {et~al.}(2018){Souto}, {Cunha}, {Smith}, {Allende Prieto},
  {Garc{\'{\i}}a-Hern{\'a}ndez}, {Pinsonneault}, {Holzer}, {Frinchaboy},
  {Holtzman}, {Johnson}, {J{\"o}nsson}, {Majewski}, {Shetrone}, {Sobeck},
  {Stringfellow}, {Teske}, {Zamora}, {Zasowski}, {Carrera}, {Stassun},
  {Fernandez-Trincado}, {Villanova}, {Minniti}, \& {Santana}}]{souto2018a}
{Souto}, D., {Cunha}, K., {Smith}, V.~V., {et~al.} 2018, \apj, 857, 14,
  \dodoi{10.3847/1538-4357/aab612}

\bibitem[{{Souto} {et~al.}(2019){Souto}, {Allende Prieto}, {Cunha},
  {Pinsonneault}, {Smith}, {Garcia-Dias}, {Bovy}, {Garc{\'\i}a-Hern{\'a}ndez},
  {Holtzman}, {Johnson}, {J{\"o}nsson}, {Majewski}, {Shetrone}, {Sobeck},
  {Zamora}, {Pan}, \& {Nitschelm}}]{souto2019}
{Souto}, D., {Allende Prieto}, C., {Cunha}, K., {et~al.} 2019, \apj, 874, 97,
  \dodoi{10.3847/1538-4357/ab0b43}

\bibitem[{{Taylor}(2005)}]{topcat}
{Taylor}, M.~B. 2005, in Astronomical Society of the Pacific Conference Series,
  Vol. 347, Astronomical Data Analysis Software and Systems XIV, ed.
  P.~{Shopbell}, M.~{Britton}, \& R.~{Ebert}, 29

\bibitem[{{Twicken} {et~al.}(2016){Twicken}, {Jenkins}, {Seader}, {Tenenbaum},
  {Smith}, {Brownston}, {Burke}, {Catanzarite}, {Clarke}, {Cote}, {Girouard},
  {Klaus}, {Li}, {McCauliff}, {Morris}, {Wohler}, {Campbell}, {Kamal Uddin},
  {Zamudio}, {Sabale}, {Bryson}, {Caldwell}, {Christiansen}, {Coughlin},
  {Haas}, {Henze}, {Sand erfer}, \& {Thompson}}]{twicken2016}
{Twicken}, J.~D., {Jenkins}, J.~M., {Seader}, S.~E., {et~al.} 2016, \aj, 152,
  158, \dodoi{10.3847/0004-6256/152/6/158}

\bibitem[{{Udry} {et~al.}(2003){Udry}, {Mayor}, \& {Santos}}]{udry2003}
{Udry}, S., {Mayor}, M., \& {Santos}, N.~C. 2003, \aap, 407, 369,
  \dodoi{10.1051/0004-6361:20030843}

\bibitem[{{Valenti} \& {Fischer}(2005)}]{valenti&fischer2005}
{Valenti}, J.~A., \& {Fischer}, D.~A. 2005, \apjs, 159, 141,
  \dodoi{10.1086/430500}

\bibitem[{{Vanderburg} {et~al.}(2018){Vanderburg}, {Mann}, {Rizzuto},
  {Bieryla}, {Kraus}, {Berlind}, {Calkins}, {Curtis}, {Douglas}, {Esquerdo},
  {Everett}, {Horch}, {Howell}, {Latham}, {Mayo}, {Quinn}, {Scott}, \&
  {Stefanik}}]{vanderburg2018}
{Vanderburg}, A., {Mann}, A.~W., {Rizzuto}, A., {et~al.} 2018, \aj, 156, 46,
  \dodoi{10.3847/1538-3881/aac894}

\bibitem[{Virtanen {et~al.}(2020)Virtanen, Gommers, Oliphant, Haberland, Reddy,
  Cournapeau, Burovski, Peterson, Weckesser, Bright, {van der Walt}, Brett,
  Wilson, Millman, Mayorov, Nelson, Jones, Kern, Larson, Carey, Polat, Feng,
  Moore, {VanderPlas}, Laxalde, Perktold, Cimrman, Henriksen, Quintero, Harris,
  Archibald, Ribeiro, Pedregosa, {van Mulbregt}, \& {SciPy 1.0
  Contributors}}]{scipy}
Virtanen, P., Gommers, R., Oliphant, T.~E., {et~al.} 2020, Nature Methods, 17,
  261, \dodoi{10.1038/s41592-019-0686-2}

\bibitem[{{Wang} \& {Fischer}(2015)}]{wang&fischer2015}
{Wang}, J., \& {Fischer}, D.~A. 2015, \aj, 149, 14,
  \dodoi{10.1088/0004-6256/149/1/14}

\bibitem[{{Wang} \& {Chen}(2019)}]{wang2019}
{Wang}, S., \& {Chen}, X. 2019, \apj, 877, 116,
  \dodoi{10.3847/1538-4357/ab1c61}

\bibitem[{{Wilson} {et~al.}(2012){Wilson}, {Hearty}, {Skrutskie}, {Majewski},
  {Schiavon}, {Eisenstein}, {Gunn}, {Gillespie}, {Weinberg}, {Blank},
  {Henderson}, {Smee}, {Barkhouser}, {Harding}, {Hope}, {Fitzgerald},
  {Stolberg}, {Arns}, {Nelson}, {Brunner}, {Burton}, {Walker}, {Lam},
  {Maseman}, {Barr}, {Leger}, {Carey}, {MacDonald}, {Ebelke}, {Beland},
  {Horne}, {Young}, {Rieke}, {Rieke}, {O'Brien}, {Crane}, {Carr}, {Harrison},
  {Stoll}, {Vernieri}, {Holtzman}, {Nidever}, {Shetrone}, {Allende-Prieto},
  {Johnson}, {Frinchaboy}, {Zasowski}, {Garcia Perez}, {Bizyaev}, \&
  {Zhao}}]{wilson2012}
{Wilson}, J.~C., {Hearty}, F., {Skrutskie}, M.~F., {et~al.} 2012, in American
  Astronomical Society Meeting Abstracts, Vol. 219, American Astronomical
  Society Meeting Abstracts \#219, 428.02

\bibitem[{{Wilson} {et~al.}(2019){Wilson}, {Hearty}, {Skrutskie}, {Majewski},
  {Holtzman}, {Eisenstein}, {Gunn}, {Blank}, {Henderson}, {Smee}, {Nelson},
  {Nidever}, {Arns}, {Barkhouser}, {Barr}, {Beland}, {Bershady}, {Blanton},
  {Brunner}, {Burton}, {Carey}, {Carr}, {Colque}, {Crane}, {Damke}, {Davidson},
  {Dean}, {Di Mille}, {Don}, {Ebelke}, {Evans}, {Fitzgerald}, {Gillespie},
  {Hall}, {Harding}, {Harding}, {Hammond}, {Hancock}, {Harrison}, {Hope},
  {Horne}, {Karakla}, {Lam}, {Leger}, {MacDonald}, {Maseman}, {Matsunari},
  {Melton}, {Mitcheltree}, {O'Brien}, {O'Connell}, {Patten}, {Richardson},
  {Rieke}, {Rieke}, {Roman-Lopes}, {Schiavon}, {Sobeck}, {Stolberg}, {Stoll},
  {Tembe}, {Trujillo}, {Uomoto}, {Vernieri}, {Walker}, {Weinberg}, {Young},
  {Anthony-Brumfield}, {Bizyaev}, {Breslauer}, {De Lee}, {Downey}, {Halverson},
  {Huehnerhoff}, {Klaene}, {Leon}, {Long}, {Mahadevan}, {Malanushenko},
  {Nguyen}, {Owen}, {S{\'a}nchez-Gallego}, {Sayres}, {Shane}, {Shectman},
  {Shetrone}, {Skinner}, {Stauffer}, \& {Zhao}}]{wilson2019}
{Wilson}, J.~C., {Hearty}, F.~R., {Skrutskie}, M.~F., {et~al.} 2019, \pasp,
  131, 055001, \dodoi{10.1088/1538-3873/ab0075}

\bibitem[{{Wilson} {et~al.}(2018){Wilson}, {Teske}, {Majewski}, {Cunha},
  {Smith}, {Souto}, {Bender}, {Mahadevan}, {Troup}, {Allende Prieto},
  {Stassun}, {Skrutskie}, {Almeida}, {Garc{\'{\i}}a-Hern{\'a}ndez}, {Zamora},
  \& {Brinkmann}}]{wilson2018}
{Wilson}, R.~F., {Teske}, J., {Majewski}, S.~R., {et~al.} 2018, \aj, 155, 68,
  \dodoi{10.3847/1538-3881/aa9f27}

\bibitem[{{Wyse}(1995)}]{wyse1995}
{Wyse}, R. F.~G. 1995, in IAU Symposium, Vol. 164, Stellar Populations, ed.
  P.~C. {van der Kruit} \& G.~{Gilmore}, 133

\bibitem[{{Youdin}(2011)}]{youdin2011}
{Youdin}, A.~N. 2011, \apj, 742, 38, \dodoi{10.1088/0004-637X/742/1/38}

\bibitem[{{Zamora} {et~al.}(2015){Zamora}, {Garc{\'\i}a-Hern{\'a}ndez},
  {Allende Prieto}, {Carrera}, {Koesterke}, {Edvardsson}, {Castelli}, {Plez},
  {Bizyaev}, {Cunha}, {Garc{\'\i}a P{\'e}rez}, {Gustafsson}, {Holtzman},
  {Lawler}, {Majewski}, {Manchado}, {M{\'e}sz{\'a}ros}, {Shane}, {Shetrone},
  {Smith}, \& {Zasowski}}]{zamora2015}
{Zamora}, O., {Garc{\'\i}a-Hern{\'a}ndez}, D.~A., {Allende Prieto}, C.,
  {et~al.} 2015, \aj, 149, 181, \dodoi{10.1088/0004-6256/149/6/181}

\bibitem[{{Zasowski} {et~al.}(2013){Zasowski}, {Johnson}, {Frinchaboy},
  {Majewski}, {Nidever}, {Rocha Pinto}, {Girardi}, {Andrews}, {Chojnowski},
  {Cudworth}, {Jackson}, {Munn}, {Skrutskie}, {Beaton}, {Blake}, {Covey},
  {Deshpande}, {Epstein}, {Fabbian}, {Fleming}, {Garcia Hernandez}, {Herrero},
  {Mahadevan}, {M{\'e}sz{\'a}ros}, {Schultheis}, {Sellgren}, {Terrien}, {van
  Saders}, {Allende Prieto}, {Bizyaev}, {Burton}, {Cunha}, {da Costa},
  {Hasselquist}, {Hearty}, {Holtzman}, {Garc{\'{\i}}a P{\'e}rez}, {Maia},
  {O'Connell}, {O'Donnell}, {Pinsonneault}, {Santiago}, {Schiavon}, {Shetrone},
  {Smith}, \& {Wilson}}]{zasowski2013}
{Zasowski}, G., {Johnson}, J.~A., {Frinchaboy}, P.~M., {et~al.} 2013, \aj, 146,
  81, \dodoi{10.1088/0004-6256/146/4/81}

\bibitem[{{Zasowski} {et~al.}(2017){Zasowski}, {Cohen}, {Chojnowski},
  {Santana}, {Oelkers}, {Andrews}, {Beaton}, {Bender}, {Bird}, {Bovy},
  {Carlberg}, {Covey}, {Cunha}, {Dell'Agli}, {Fleming}, {Frinchaboy},
  {Garc{\'\i}a-Hern{\'a}ndez}, {Harding}, {Holtzman}, {Johnson}, {Kollmeier},
  {Majewski}, {M{\'e}sz{\'a}ros}, {Munn}, {Mu{\~n}oz}, {Ness}, {Nidever},
  {Poleski}, {Rom{\'a}n-Z{\'u}{\~n}iga}, {Shetrone}, {Simon}, {Smith},
  {Sobeck}, {Stringfellow}, {Szigeti{\'a}ros}, {Tayar}, \&
  {Troup}}]{zasowski2017}
{Zasowski}, G., {Cohen}, R.~E., {Chojnowski}, S.~D., {et~al.} 2017, \aj, 154,
  198, \dodoi{10.3847/1538-3881/aa8df9}

\bibitem[{{Zhou} {et~al.}(2017){Zhou}, {Bakos}, {Hartman}, {Latham}, {Torres},
  {Bhatti}, {Penev}, {Buchhave}, {Kov{\'a}cs}, {Bieryla}, {Quinn}, {Isaacson},
  {Fulton}, {Falco}, {Csubry}, {Everett}, {Szklenar}, {Esquerdo}, {Berlind},
  {Calkins}, {B{\'e}ky}, {Knox}, {Hinz}, {Horch}, {Hirsch}, {Howell}, {Noyes},
  {Marcy}, {de Val-Borro}, {L{\'a}z{\'a}r}, {Papp}, \& {S{\'a}ri}}]{zhou2017}
{Zhou}, G., {Bakos}, G.~{\'A}., {Hartman}, J.~D., {et~al.} 2017, \aj, 153, 211,
  \dodoi{10.3847/1538-3881/aa674a}

\bibitem[{{Zhu}(2019)}]{zhu2019}
{Zhu}, W. 2019, \apj, 873, 8, \dodoi{10.3847/1538-4357/ab0205}

\bibitem[{{Zinn} {et~al.}(2019{\natexlab{a}}){Zinn}, {Pinsonneault}, {Huber},
  \& {Stello}}]{zinn2019}
{Zinn}, J.~C., {Pinsonneault}, M.~H., {Huber}, D., \& {Stello}, D.
  2019{\natexlab{a}}, \apj, 878, 136, \dodoi{10.3847/1538-4357/ab1f66}

\bibitem[{{Zinn} {et~al.}(2019{\natexlab{b}}){Zinn}, {Pinsonneault}, {Huber},
  {Stello}, {Stassun}, \& {Serenelli}}]{zinn2019b}
{Zinn}, J.~C., {Pinsonneault}, M.~H., {Huber}, D., {et~al.} 2019{\natexlab{b}},
  \apj, 885, 166, \dodoi{10.3847/1538-4357/ab44a9}

\end{thebibliography}
\bibliographystyle{aasjournal}

\begin{longtable*}{cccccccccc}
\caption{Significance testing for the abundances between each planet subsample and \csamp. $^a$Note: For Fe, we use [Fe/H]. }\\ \hline 
[X/Fe]$^a$ & $P$-class & $R_p$-class & $n_{pl}$& $\langle$[X/Fe]$\rangle$ & Norm? & $p_{ks}$& $p_{mw}$& $p_{bf}$ & Sig? \\ \hline \hline
Fe  & All & All & 544 & -0.010$\pm$0.163 &  No  &  0.054  &  0.0081  &  0.16  & No \\
Fe  & All & SE & 212 & -0.032$\pm$0.177 &  No  &  0.47  &  0.16  &  0.57  & No \\
Fe  & All & SN & 260 & 0.003$\pm$0.160 &  No  &  0.025  &  0.0029  &  0.061  & No \\
Fe  & All & SS & 13 & 0.062$\pm$0.099 &  Yes  &  0.015  &  0.0017  &  0.033  & No \\
Fe  & All & JP & 17 & 0.128$\pm$0.186 &  Yes  &  0.011  &  0.0016  &  0.6  & No \\
Fe  & Hot & All & 248 & 0.034$\pm$0.141 &  No  &  0.00083  &  1.3e-05  &  0.014  & Yes \\
Fe  & Hot & SE & 135 & -0.012$\pm$0.140 &  No  &  0.013  &  0.0054  &  0.067  & No \\
Fe  & Hot & SN & 71 & 0.092$\pm$0.136 &  Yes  &  1.5e-05  &  1.5e-07  &  0.0024  & Yes \\
Fe  & Hot & SS & 2 & -- & -- & --  & --  \\
Fe  & Hot & JP & 6 & -- & -- & --  & --  \\
Fe  & Warm & All & 262 & -0.037$\pm$0.163 &  No  &  0.68  &  0.22  &  0.25  & No \\
Fe  & Warm & SE & 72 & -0.134$\pm$0.194 &  Yes  &  0.17  &  0.084  &  0.91  & No \\
Fe  & Warm & SN & 170 & -0.013$\pm$0.142 &  Yes  &  0.13  &  0.057  &  0.057  & No \\
Fe  & Warm & SS & 7 & -- & -- & --  & --  \\
Fe  & Warm & JP & 6 & -- & -- & --  & --  \\
Fe  & Cool & All & 34 & -0.053$\pm$0.248 &  Yes  &  0.42  &  0.37  &  0.51  & No \\
Fe  & Cool & SE & 5 & -- & -- & --  & --  \\
Fe  & Cool & SN & 19 & -0.205$\pm$0.169 &  Yes  &  0.18  &  0.25  &  0.62  & No \\
Fe  & Cool & SS & 4 & -- & -- & --  & --  \\
Fe  & Cool & JP & 5 & -- & -- & --  & --  \\
\hline
C  & All & All & 544 & -0.019$\pm$0.079 &  No  &  0.0059  &  0.0084  &  0.1  & No \\
C  & All & SE & 212 & -0.016$\pm$0.068 &  No  &  0.0069  &  0.029  &  0.078  & No \\
C  & All & SN & 260 & -0.028$\pm$0.091 &  Yes  &  0.0045  &  0.0032  &  0.2  & No \\
C  & All & SS & 13 & -0.008$\pm$0.056 &  Yes  &  0.76  &  0.46  &  0.18  & No \\
C  & All & JP & 17 & 0.020$\pm$0.089 &  Yes  &  0.98  &  0.49  &  0.58  & No \\
C  & Hot & All & 248 & -0.020$\pm$0.081 &  No  &  0.003  &  0.0051  &  0.32  & No \\
C  & Hot & SE & 135 & -0.019$\pm$0.071 &  Yes  &  0.0026  &  0.013  &  0.23  & No \\
C  & Hot & SN & 71 & -0.021$\pm$0.102 &  Yes  &  0.084  &  0.0069  &  0.74  & No \\
C  & Hot & SS & 2 & -- & -- & --  & --  \\
C  & Hot & JP & 6 & -- & -- & --  & --  \\
C  & Warm & All & 262 & -0.022$\pm$0.070 &  Yes  &  0.006  &  0.011  &  0.037  & No \\
C  & Warm & SE & 72 & -0.008$\pm$0.071 &  Yes  &  0.17  &  0.15  &  0.033  & No \\
C  & Warm & SN & 170 & -0.030$\pm$0.074 &  Yes  &  0.002  &  0.004  &  0.12  & No \\
C  & Warm & SS & 7 & -- & -- & --  & --  \\
C  & Warm & JP & 6 & -- & -- & --  & --  \\
C  & Cool & All & 34 & 0.022$\pm$0.079 &  Yes  &  0.89  &  0.36  &  0.14  & No \\
C  & Cool & SE & 5 & -- & -- & --  & --  \\
C  & Cool & SN & 19 & 0.011$\pm$0.100 &  Yes  &  0.84  &  0.38  &  0.46  & No \\
C  & Cool & SS & 4 & -- & -- & --  & --  \\
C  & Cool & JP & 5 & -- & -- & --  & --  \\
\hline
Mg  & All & All & 544 & 0.006$\pm$0.060 &  No  &  0.021  &  0.035  &  0.13  & No \\
Mg  & All & SE & 212 & 0.019$\pm$0.070 &  No  &  0.2  &  0.15  &  0.31  & No \\
Mg  & All & SN & 260 & -0.001$\pm$0.058 &  No  &  0.012  &  0.022  &  0.1  & No \\
Mg  & All & SS & 13 & -0.009$\pm$0.047 &  Yes  &  0.023  &  0.031  &  0.58  & No \\
Mg  & All & JP & 17 & -0.034$\pm$0.039 &  Yes  &  0.022  &  0.018  &  0.22  & No \\
Mg  & Hot & All & 248 & 0.000$\pm$0.053 &  No  &  0.0036  &  0.0064  &  0.04  & No \\
Mg  & Hot & SE & 135 & 0.014$\pm$0.066 &  No  &  0.046  &  0.071  &  0.24  & No \\
Mg  & Hot & SN & 71 & -0.012$\pm$0.047 &  No  &  0.001  &  0.001  &  0.0054  & Yes \\
Mg  & Hot & SS & 2 & -- & -- & --  & --  \\
Mg  & Hot & JP & 6 & -- & -- & --  & --  \\
Mg  & Warm & All & 262 & 0.013$\pm$0.068 &  No  &  0.19  &  0.18  &  0.24  & No \\
Mg  & Warm & SE & 72 & 0.026$\pm$0.074 &  No  &  0.46  &  0.41  &  0.56  & No \\
Mg  & Warm & SN & 170 & 0.005$\pm$0.064 &  No  &  0.081  &  0.1  &  0.2  & No \\
Mg  & Warm & SS & 7 & -- & -- & --  & --  \\
Mg  & Warm & JP & 6 & -- & -- & --  & --  \\
Mg  & Cool & All & 34 & -0.003$\pm$0.073 &  Yes  &  0.44  &  0.12  &  0.79  & No \\
Mg  & Cool & SE & 5 & -- & -- & --  & --  \\
Mg  & Cool & SN & 19 & -0.001$\pm$0.068 &  Yes  &  0.75  &  0.3  &  0.72  & No \\
Mg  & Cool & SS & 4 & -- & -- & --  & --  \\
Mg  & Cool & JP & 5 & -- & -- & --  & --  \\
\hline
Al  & All & All & 540 & 0.067$\pm$0.121 &  No  &  0.034  &  0.28  &  0.027  & No \\
Al  & All & SE & 212 & 0.076$\pm$0.125 &  Yes  &  0.12  &  0.47  &  0.0031  & No \\
Al  & All & SN & 258 & 0.068$\pm$0.124 &  No  &  0.069  &  0.28  &  0.26  & No \\
Al  & All & SS & 13 & -0.004$\pm$0.100 &  Yes  &  0.26  &  0.15  &  0.13  & No \\
Al  & All & JP & 15 & 0.083$\pm$0.132 &  Yes  &  0.28  &  0.38  &  0.12  & No \\
Al  & Hot & All & 247 & 0.064$\pm$0.127 &  No  &  0.089  &  0.27  &  0.16  & No \\
Al  & Hot & SE & 135 & 0.075$\pm$0.133 &  Yes  &  0.32  &  0.49  &  0.03  & No \\
Al  & Hot & SN & 70 & 0.064$\pm$0.152 &  No  &  0.44  &  0.21  &  0.73  & No \\
Al  & Hot & SS & 2 & -- & -- & --  & --  \\
Al  & Hot & JP & 6 & -- & -- & --  & --  \\
Al  & Warm & All & 259 & 0.071$\pm$0.112 &  No  &  0.046  &  0.34  &  0.006  & No \\
Al  & Warm & SE & 72 & 0.076$\pm$0.110 &  No  &  0.052  &  0.41  &  0.0032  & No \\
Al  & Warm & SN & 169 & 0.071$\pm$0.120 &  No  &  0.12  &  0.37  &  0.052  & No \\
Al  & Warm & SS & 7 & -- & -- & --  & --  \\
Al  & Warm & JP & 4 & -- & -- & --  & --  \\
Al  & Cool & All & 34 & 0.056$\pm$0.086 &  Yes  &  0.045  &  0.26  &  0.014  & No \\
Al  & Cool & SE & 5 & -- & -- & --  & --  \\
Al  & Cool & SN & 19 & 0.044$\pm$0.081 &  Yes  &  0.063  &  0.34  &  0.028  & No \\
Al  & Cool & SS & 4 & -- & -- & --  & --  \\
Al  & Cool & JP & 5 & -- & -- & --  & --  \\
\hline
Si  & All & All & 544 & 0.002$\pm$0.058 &  No  &  0.17  &  0.47  &  0.00062  & Yes \\
Si  & All & SE & 212 & 0.012$\pm$0.056 &  No  &  0.22  &  0.32  &  0.0047  & No \\
Si  & All & SN & 260 & -0.000$\pm$0.057 &  No  &  0.24  &  0.37  &  0.0029  & No \\
Si  & All & SS & 13 & 0.004$\pm$0.061 &  Yes  &  0.26  &  0.2  &  0.22  & No \\
Si  & All & JP & 17 & 0.020$\pm$0.059 &  Yes  &  0.84  &  0.39  &  0.19  & No \\
Si  & Hot & All & 248 & 0.001$\pm$0.056 &  Yes  &  0.14  &  0.36  &  0.00088  & Yes \\
Si  & Hot & SE & 135 & 0.008$\pm$0.053 &  Yes  &  0.18  &  0.44  &  0.0055  & No \\
Si  & Hot & SN & 71 & -0.002$\pm$0.065 &  Yes  &  0.49  &  0.25  &  0.03  & No \\
Si  & Hot & SS & 2 & -- & -- & --  & --  \\
Si  & Hot & JP & 6 & -- & -- & --  & --  \\
Si  & Warm & All & 262 & 0.003$\pm$0.059 &  No  &  0.28  &  0.48  &  0.0029  & No \\
Si  & Warm & SE & 72 & 0.015$\pm$0.059 &  Yes  &  0.34  &  0.25  &  0.081  & No \\
Si  & Warm & SN & 170 & 0.001$\pm$0.056 &  No  &  0.17  &  0.39  &  0.0051  & No \\
Si  & Warm & SS & 7 & -- & -- & --  & --  \\
Si  & Warm & JP & 6 & -- & -- & --  & --  \\
Si  & Cool & All & 34 & 0.008$\pm$0.061 &  Yes  &  0.45  &  0.27  &  0.17  & No \\
Si  & Cool & SE & 5 & -- & -- & --  & --  \\
Si  & Cool & SN & 19 & 0.004$\pm$0.053 &  Yes  &  0.63  &  0.24  &  0.29  & No \\
Si  & Cool & SS & 4 & -- & -- & --  & --  \\
Si  & Cool & JP & 5 & -- & -- & --  & --  \\
\hline
S  & All & All & 542 & 0.013$\pm$0.098 &  No  &  0.31  &  0.071  &  0.14  & No \\
S  & All & SE & 211 & 0.008$\pm$0.089 &  No  &  0.097  &  0.029  &  0.32  & No \\
S  & All & SN & 259 & 0.026$\pm$0.104 &  No  &  0.7  &  0.23  &  0.24  & No \\
S  & All & SS & 13 & 0.009$\pm$0.098 &  Yes  &  0.81  &  0.31  &  0.17  & No \\
S  & All & JP & 17 & 0.068$\pm$0.132 &  Yes  &  0.74  &  0.24  &  0.38  & No \\
S  & Hot & All & 247 & 0.001$\pm$0.096 &  No  &  0.028  &  0.0061  &  0.26  & No \\
S  & Hot & SE & 134 & 0.008$\pm$0.088 &  No  &  0.094  &  0.023  &  0.4  & No \\
S  & Hot & SN & 71 & -0.003$\pm$0.144 &  Yes  &  0.17  &  0.022  &  0.87  & No \\
S  & Hot & SS & 2 & -- & -- & --  & --  \\
S  & Hot & JP & 6 & -- & -- & --  & --  \\
S  & Warm & All & 261 & 0.025$\pm$0.098 &  No  &  0.66  &  0.32  &  0.14  & No \\
S  & Warm & SE & 72 & 0.008$\pm$0.108 &  No  &  0.29  &  0.11  &  0.49  & No \\
S  & Warm & SN & 169 & 0.031$\pm$0.098 &  No  &  0.77  &  0.45  &  0.12  & No \\
S  & Warm & SS & 7 & -- & -- & --  & --  \\
S  & Warm & JP & 6 & -- & -- & --  & --  \\
S  & Cool & All & 34 & 0.040$\pm$0.107 &  Yes  &  0.6  &  0.48  &  0.032  & No \\
S  & Cool & SE & 5 & -- & -- & --  & --  \\
S  & Cool & SN & 19 & 0.056$\pm$0.130 &  Yes  &  0.74  &  0.38  &  0.21  & No \\
S  & Cool & SS & 4 & -- & -- & --  & --  \\
S  & Cool & JP & 5 & -- & -- & --  & --  \\
\hline
K  & All & All & 540 & 0.014$\pm$0.077 &  No  &  0.00018  &  0.00018  &  0.27  & Yes \\
K  & All & SE & 209 & 0.019$\pm$0.076 &  No  &  0.0021  &  0.0022  &  0.023  & No \\
K  & All & SN & 260 & 0.012$\pm$0.079 &  No  &  0.00045  &  0.00023  &  0.85  & Yes \\
K  & All & SS & 13 & 0.022$\pm$0.090 &  Yes  &  0.1  &  0.067  &  0.35  & No \\
K  & All & JP & 17 & 0.005$\pm$0.077 &  No  &  0.067  &  0.021  &  0.92  & No \\
K  & Hot & All & 246 & 0.012$\pm$0.069 &  No  &  6.7e-05  &  7.8e-05  &  0.27  & Yes \\
K  & Hot & SE & 134 & 0.021$\pm$0.068 &  No  &  0.00079  &  0.0013  &  0.032  & Yes \\
K  & Hot & SN & 71 & 0.000$\pm$0.086 &  No  &  0.00058  &  0.00037  &  0.5  & Yes \\
K  & Hot & SS & 2 & -- & -- & --  & --  \\
K  & Hot & JP & 6 & -- & -- & --  & --  \\
K  & Warm & All & 260 & 0.013$\pm$0.080 &  No  &  0.0015  &  0.001  &  0.085  & No \\
K  & Warm & SE & 70 & 0.010$\pm$0.080 &  Yes  &  0.018  &  0.021  &  0.13  & No \\
K  & Warm & SN & 170 & 0.013$\pm$0.077 &  No  &  0.0031  &  0.001  &  0.16  & Yes \\
K  & Warm & SS & 7 & -- & -- & --  & --  \\
K  & Warm & JP & 6 & -- & -- & --  & --  \\
K  & Cool & All & 34 & 0.027$\pm$0.097 &  No  &  0.23  &  0.064  &  0.35  & No \\
K  & Cool & SE & 5 & -- & -- & --  & --  \\
K  & Cool & SN & 19 & 0.022$\pm$0.093 &  No  &  0.21  &  0.078  &  0.2  & No \\
K  & Cool & SS & 4 & -- & -- & --  & --  \\
K  & Cool & JP & 5 & -- & -- & --  & --  \\
\hline
Ca  & All & All & 544 & 0.008$\pm$0.046 &  No  &  0.59  &  0.37  &  0.14  & No \\
Ca  & All & SE & 212 & 0.012$\pm$0.047 &  No  &  0.71  &  0.45  &  0.033  & No \\
Ca  & All & SN & 260 & 0.009$\pm$0.049 &  No  &  0.64  &  0.34  &  0.45  & No \\
Ca  & All & SS & 13 & 0.003$\pm$0.043 &  Yes  &  0.31  &  0.2  &  0.43  & No \\
Ca  & All & JP & 17 & 0.003$\pm$0.046 &  Yes  &  0.94  &  0.38  &  0.49  & No \\
Ca  & Hot & All & 248 & 0.005$\pm$0.041 &  No  &  0.36  &  0.3  &  0.027  & No \\
Ca  & Hot & SE & 135 & 0.011$\pm$0.042 &  No  &  0.67  &  0.4  &  0.0094  & No \\
Ca  & Hot & SN & 71 & 0.001$\pm$0.046 &  No  &  0.49  &  0.25  &  0.48  & No \\
Ca  & Hot & SS & 2 & -- & -- & --  & --  \\
Ca  & Hot & JP & 6 & -- & -- & --  & --  \\
Ca  & Warm & All & 262 & 0.010$\pm$0.050 &  No  &  0.89  &  0.43  &  0.4  & No \\
Ca  & Warm & SE & 72 & 0.015$\pm$0.053 &  Yes  &  0.96  &  0.34  &  0.5  & No \\
Ca  & Warm & SN & 170 & 0.010$\pm$0.049 &  No  &  0.87  &  0.4  &  0.43  & No \\
Ca  & Warm & SS & 7 & -- & -- & --  & --  \\
Ca  & Warm & JP & 6 & -- & -- & --  & --  \\
Ca  & Cool & All & 34 & 0.013$\pm$0.065 &  No  &  1.0  &  0.44  &  0.86  & No \\
Ca  & Cool & SE & 5 & -- & -- & --  & --  \\
Ca  & Cool & SN & 19 & 0.017$\pm$0.060 &  Yes  &  0.94  &  0.49  &  0.94  & No \\
Ca  & Cool & SS & 4 & -- & -- & --  & --  \\
Ca  & Cool & JP & 5 & -- & -- & --  & --  \\
\hline
Mn  & All & All & 544 & -0.003$\pm$0.074 &  No  &  0.58  &  0.23  &  0.04  & No \\
Mn  & All & SE & 212 & -0.008$\pm$0.074 &  No  &  0.78  &  0.36  &  0.16  & No \\
Mn  & All & SN & 260 & 0.002$\pm$0.074 &  No  &  0.42  &  0.12  &  0.12  & No \\
Mn  & All & SS & 13 & 0.032$\pm$0.040 &  Yes  &  0.041  &  0.013  &  0.2  & No \\
Mn  & All & JP & 17 & 0.018$\pm$0.069 &  Yes  &  0.061  &  0.016  &  0.54  & No \\
Mn  & Hot & All & 248 & 0.006$\pm$0.065 &  No  &  0.075  &  0.037  &  0.033  & No \\
Mn  & Hot & SE & 135 & 0.002$\pm$0.061 &  No  &  0.28  &  0.27  &  0.12  & No \\
Mn  & Hot & SN & 71 & 0.028$\pm$0.073 &  Yes  &  0.055  &  0.0034  &  0.16  & No \\
Mn  & Hot & SS & 2 & -- & -- & --  & --  \\
Mn  & Hot & JP & 6 & -- & -- & --  & --  \\
Mn  & Warm & All & 262 & -0.007$\pm$0.073 &  No  &  0.8  &  0.5  &  0.13  & No \\
Mn  & Warm & SE & 72 & -0.022$\pm$0.072 &  Yes  &  0.12  &  0.069  &  0.56  & No \\
Mn  & Warm & SN & 170 & -0.005$\pm$0.075 &  No  &  0.8  &  0.31  &  0.15  & No \\
Mn  & Warm & SS & 7 & -- & -- & --  & --  \\
Mn  & Warm & JP & 6 & -- & -- & --  & --  \\
Mn  & Cool & All & 34 & -0.036$\pm$0.084 &  Yes  &  0.5  &  0.14  &  0.7  & No \\
Mn  & Cool & SE & 5 & -- & -- & --  & --  \\
Mn  & Cool & SN & 19 & -0.043$\pm$0.092 &  Yes  &  0.58  &  0.23  &  0.96  & No \\
Mn  & Cool & SS & 4 & -- & -- & --  & --  \\
Mn  & Cool & JP & 5 & -- & -- & --  & --  \\
\hline
Ni  & All & All & 544 & 0.019$\pm$0.041 &  No  &  0.56  &  0.38  &  0.23  & No \\
Ni  & All & SE & 212 & 0.018$\pm$0.040 &  No  &  0.25  &  0.2  &  0.13  & No \\
Ni  & All & SN & 260 & 0.021$\pm$0.048 &  No  &  0.94  &  0.45  &  0.7  & No \\
Ni  & All & SS & 13 & 0.035$\pm$0.045 &  Yes  &  0.9  &  0.27  &  0.34  & No \\
Ni  & All & JP & 17 & 0.047$\pm$0.027 &  Yes  &  0.12  &  0.059  &  1.0  & No \\
Ni  & Hot & All & 248 & 0.021$\pm$0.038 &  No  &  0.54  &  0.38  &  0.1  & No \\
Ni  & Hot & SE & 135 & 0.018$\pm$0.039 &  No  &  0.36  &  0.21  &  0.099  & No \\
Ni  & Hot & SN & 71 & 0.029$\pm$0.045 &  No  &  0.38  &  0.16  &  0.74  & No \\
Ni  & Hot & SS & 2 & -- & -- & --  & --  \\
Ni  & Hot & JP & 6 & -- & -- & --  & --  \\
Ni  & Warm & All & 262 & 0.019$\pm$0.046 &  Yes  &  0.8  &  0.41  &  0.48  & No \\
Ni  & Warm & SE & 72 & 0.018$\pm$0.042 &  Yes  &  0.6  &  0.35  &  0.42  & No \\
Ni  & Warm & SN & 170 & 0.017$\pm$0.053 &  Yes  &  0.87  &  0.38  &  0.76  & No \\
Ni  & Warm & SS & 7 & -- & -- & --  & --  \\
Ni  & Warm & JP & 6 & -- & -- & --  & --  \\
Ni  & Cool & All & 34 & 0.008$\pm$0.057 &  Yes  &  0.66  &  0.31  &  0.68  & No \\
Ni  & Cool & SE & 5 & -- & -- & --  & --  \\
Ni  & Cool & SN & 19 & 0.009$\pm$0.045 &  Yes  &  0.94  &  0.5  &  0.37  & No \\
Ni  & Cool & SS & 4 & -- & -- & --  & --  \\
Ni  & Cool & JP & 5 & -- & -- & --  & --  \\
\hline
\label{tab:tests}
\end{longtable*}

\newpage

\begin{longtable*}{cccccccccc}
\caption{Best fit parameters for the planet occurrence rate distributions from Equation \ref{eq:xpfit} }. \\ 
\hline  \hline 

$R_p$-Class & $P$-class& [X/H] & $\log C$ & $\alpha$ & $\beta$\\ \hline
SE & hot & C & $ -1.67^{+0.11}_{-0.11}$  & $ +1.46^{+0.15}_{-0.15}$  & $ +1.16^{+0.47}_{-0.43}$ \\
SE & hot & Mg & $ -1.86^{+0.11}_{-0.11}$  & $ +1.47^{+0.16}_{-0.15}$  & $ +0.59^{+0.53}_{-0.51}$ \\
SE & hot & Al & $ -1.86^{+0.10}_{-0.11}$  & $ +1.47^{+0.15}_{-0.15}$  & $ +0.51^{+0.39}_{-0.37}$ \\
SE & hot & Si & $ -1.81^{+0.11}_{-0.12}$  & $ +1.44^{+0.16}_{-0.15}$  & $ +0.83^{+0.52}_{-0.50}$ \\
SE & hot & S & $ -1.76^{+0.10}_{-0.11}$  & $ +1.36^{+0.15}_{-0.15}$  & $ +0.77^{+0.46}_{-0.46}$ \\
SE & hot & K & $ -1.82^{+0.10}_{-0.11}$  & $ +1.42^{+0.15}_{-0.15}$  & $ +0.41^{+0.44}_{-0.42}$ \\
SE & hot & Ca & $ -1.81^{+0.11}_{-0.11}$  & $ +1.40^{+0.16}_{-0.16}$  & $ +0.95^{+0.46}_{-0.43}$ \\
SE & hot & Mn & $ -1.78^{+0.10}_{-0.11}$  & $ +1.39^{+0.15}_{-0.15}$  & $ +0.58^{+0.33}_{-0.32}$ \\
SE & hot & Fe & $ -1.82^{+0.11}_{-0.11}$  & $ +1.43^{+0.16}_{-0.15}$  & $ +0.72^{+0.46}_{-0.43}$ \\
SE & hot & Ni & $ -1.82^{+0.10}_{-0.11}$  & $ +1.41^{+0.15}_{-0.15}$  & $ +0.57^{+0.45}_{-0.39}$ \\
SE & warm & C & $ +0.07^{+0.32}_{-0.31}$  & $ -0.50^{+0.23}_{-0.24}$  & $ -0.25^{+0.52}_{-0.54}$ \\
SE & warm & Mg & $ -0.12^{+0.32}_{-0.31}$  & $ -0.47^{+0.23}_{-0.24}$  & $ -0.76^{+0.58}_{-0.59}$ \\
SE & warm & Al & $ -0.05^{+0.29}_{-0.29}$  & $ -0.44^{+0.22}_{-0.22}$  & $ -0.34^{+0.39}_{-0.41}$ \\
SE & warm & Si & $ +0.02^{+0.32}_{-0.31}$  & $ -0.55^{+0.23}_{-0.24}$  & $ -0.44^{+0.46}_{-0.48}$ \\
SE & warm & S & $ +0.04^{+0.31}_{-0.30}$  & $ -0.55^{+0.23}_{-0.24}$  & $ -0.51^{+0.47}_{-0.46}$ \\
SE & warm & K & $ -0.13^{+0.31}_{-0.31}$  & $ -0.42^{+0.23}_{-0.24}$  & $ -0.66^{+0.49}_{-0.52}$ \\
SE & warm & Ca & $ -0.10^{+0.32}_{-0.31}$  & $ -0.47^{+0.23}_{-0.24}$  & $ -0.36^{+0.51}_{-0.53}$ \\
SE & warm & Mn & $ -0.08^{+0.31}_{-0.31}$  & $ -0.51^{+0.23}_{-0.24}$  & $ -0.66^{+0.37}_{-0.37}$ \\
SE & warm & Fe & $ -0.05^{+0.31}_{-0.31}$  & $ -0.51^{+0.23}_{-0.24}$  & $ -0.62^{+0.45}_{-0.46}$ \\
SE & warm & Ni & $ +0.01^{+0.31}_{-0.30}$  & $ -0.55^{+0.23}_{-0.24}$  & $ -0.69^{+0.41}_{-0.42}$ \\
 \hline 
SN & hot & C & $ -2.60^{+0.21}_{-0.23}$  & $ +2.35^{+0.28}_{-0.27}$  & $ +2.33^{+0.67}_{-0.61}$ \\
SN & hot & Mg & $ -2.83^{+0.21}_{-0.23}$  & $ +2.40^{+0.28}_{-0.26}$  & $ +2.86^{+0.76}_{-0.69}$ \\
SN & hot & Al & $ -2.83^{+0.21}_{-0.22}$  & $ +2.35^{+0.28}_{-0.26}$  & $ +1.06^{+0.46}_{-0.42}$ \\
SN & hot & Si & $ -2.66^{+0.20}_{-0.22}$  & $ +2.25^{+0.27}_{-0.26}$  & $ +2.02^{+0.63}_{-0.59}$ \\
SN & hot & S & $ -2.61^{+0.19}_{-0.20}$  & $ +2.19^{+0.26}_{-0.25}$  & $ +1.38^{+0.53}_{-0.52}$ \\
SN & hot & K & $ -2.65^{+0.19}_{-0.20}$  & $ +2.14^{+0.26}_{-0.24}$  & $ +1.53^{+0.63}_{-0.56}$ \\
SN & hot & Ca & $ -2.95^{+0.24}_{-0.26}$  & $ +2.52^{+0.31}_{-0.30}$  & $ +2.61^{+0.80}_{-0.72}$ \\
SN & hot & Mn & $ -2.82^{+0.21}_{-0.23}$  & $ +2.43^{+0.28}_{-0.27}$  & $ +1.76^{+0.52}_{-0.49}$ \\
SN & hot & Fe & $ -2.81^{+0.22}_{-0.23}$  & $ +2.39^{+0.29}_{-0.27}$  & $ +2.34^{+0.68}_{-0.61}$ \\
SN & hot & Ni & $ -2.88^{+0.22}_{-0.24}$  & $ +2.44^{+0.29}_{-0.28}$  & $ +2.00^{+0.58}_{-0.56}$ \\
SN & warm & C & $ -0.75^{+0.19}_{-0.20}$  & $ +0.44^{+0.13}_{-0.13}$  & $ +0.63^{+0.43}_{-0.46}$ \\
SN & warm & Mg & $ -0.90^{+0.18}_{-0.19}$  & $ +0.46^{+0.12}_{-0.12}$  & $ +0.63^{+0.47}_{-0.51}$ \\
SN & warm & Al & $ -0.97^{+0.19}_{-0.19}$  & $ +0.49^{+0.13}_{-0.13}$  & $ +0.27^{+0.35}_{-0.32}$ \\
SN & warm & Si & $ -0.88^{+0.19}_{-0.19}$  & $ +0.44^{+0.13}_{-0.13}$  & $ +0.42^{+0.45}_{-0.39}$ \\
SN & warm & S & $ -0.82^{+0.19}_{-0.19}$  & $ +0.40^{+0.13}_{-0.13}$  & $ +0.43^{+0.42}_{-0.41}$ \\
SN & warm & K & $ -0.92^{+0.18}_{-0.18}$  & $ +0.47^{+0.12}_{-0.13}$  & $ -0.06^{+0.45}_{-0.42}$ \\
SN & warm & Ca & $ -0.84^{+0.19}_{-0.19}$  & $ +0.42^{+0.13}_{-0.13}$  & $ +0.59^{+0.49}_{-0.45}$ \\
SN & warm & Mn & $ -0.91^{+0.19}_{-0.19}$  & $ +0.46^{+0.13}_{-0.13}$  & $ +0.24^{+0.32}_{-0.30}$ \\
SN & warm & Fe & $ -0.89^{+0.19}_{-0.19}$  & $ +0.45^{+0.13}_{-0.13}$  & $ +0.47^{+0.41}_{-0.40}$ \\
SN & warm & Ni & $ -0.91^{+0.19}_{-0.19}$  & $ +0.46^{+0.13}_{-0.13}$  & $ +0.43^{+0.34}_{-0.37}$ \\
 \hline 
SS & hot &C & -- & -- & -- \\
SS & hot &Mg & -- & -- & -- \\
SS & hot &Al & -- & -- & -- \\
SS & hot &Si & -- & -- & -- \\
SS & hot &S & -- & -- & -- \\
SS & hot &K & -- & -- & -- \\
SS & hot &Ca & -- & -- & -- \\
SS & hot &Mn & -- & -- & -- \\
SS & hot &Fe & -- & -- & -- \\
SS & hot &Ni & -- & -- & -- \\
SS & warm & C & $ -4.84^{+0.96}_{-1.11}$  & $ +2.29^{+0.62}_{-0.55}$  & $ +3.43^{+1.33}_{-1.17}$ \\
SS & warm & Mg & $ -3.31^{+0.63}_{-0.68}$  & $ +1.38^{+0.41}_{-0.39}$  & $ +2.26^{+1.18}_{-1.06}$ \\
SS & warm & Al & $ -3.35^{+0.63}_{-0.68}$  & $ +1.38^{+0.41}_{-0.39}$  & $ +1.02^{+0.69}_{-0.65}$ \\
SS & warm & Si & $ -3.26^{+0.63}_{-0.68}$  & $ +1.38^{+0.41}_{-0.39}$  & $ +2.22^{+0.99}_{-0.91}$ \\
SS & warm & S & $ -3.49^{+0.68}_{-0.73}$  & $ +1.45^{+0.44}_{-0.42}$  & $ +2.37^{+1.02}_{-0.93}$ \\
SS & warm & K & $ -3.18^{+0.60}_{-0.64}$  & $ +1.31^{+0.39}_{-0.38}$  & $ +1.05^{+0.84}_{-0.81}$ \\
SS & warm & Ca & $ -3.53^{+0.69}_{-0.75}$  & $ +1.52^{+0.45}_{-0.42}$  & $ +1.90^{+1.12}_{-1.02}$ \\
SS & warm & Mn & $ -3.61^{+0.71}_{-0.77}$  & $ +1.55^{+0.45}_{-0.43}$  & $ +1.10^{+0.75}_{-0.69}$ \\
SS & warm & Fe & $ -3.18^{+0.63}_{-0.68}$  & $ +1.32^{+0.41}_{-0.40}$  & $ +2.14^{+1.05}_{-0.95}$ \\
SS & warm & Ni & $ -3.64^{+0.70}_{-0.77}$  & $ +1.55^{+0.45}_{-0.43}$  & $ +1.50^{+0.93}_{-0.85}$ \\
 \hline 
JP & hot & C & $ -3.68^{+0.64}_{-0.80}$  & $ +1.47^{+0.88}_{-0.81}$  & $ +7.69^{+3.62}_{-2.84}$ \\
JP & hot & Mg & $ -3.84^{+0.64}_{-0.80}$  & $ +1.47^{+0.89}_{-0.81}$  & $ +7.31^{+3.93}_{-3.14}$ \\
JP & hot & Al & $ -3.56^{+0.58}_{-0.73}$  & $ +0.53^{+0.77}_{-0.80}$  & $ +4.58^{+2.28}_{-1.85}$ \\
JP & hot & Si & $ -4.62^{+0.94}_{-1.23}$  & $ +1.74^{+1.11}_{-0.96}$  & $ +10.32^{+6.00}_{-4.45}$ \\
JP & hot & S & $ -3.60^{+0.63}_{-0.81}$  & $ +0.83^{+0.94}_{-0.93}$  & $ +5.47^{+3.27}_{-2.58}$ \\
JP & hot & K & $ -3.75^{+0.60}_{-0.75}$  & $ +1.47^{+0.87}_{-0.79}$  & $ +4.21^{+2.64}_{-2.14}$ \\
JP & hot & Ca & $ -3.88^{+0.90}_{-1.28}$  & $ +0.16^{+1.73}_{-1.95}$  & $ +5.56^{+6.91}_{-4.72}$ \\
JP & hot & Mn & $ -3.91^{+0.73}_{-0.95}$  & $ +0.98^{+0.94}_{-0.90}$  & $ +6.70^{+4.01}_{-3.00}$ \\
JP & hot & Fe & $ -3.77^{+0.84}_{-1.20}$  & $ +0.16^{+1.71}_{-1.91}$  & $ +3.82^{+5.26}_{-3.78}$ \\
JP & hot & Ni & $ -4.67^{+0.96}_{-1.26}$  & $ +1.35^{+1.01}_{-0.93}$  & $ +10.26^{+5.69}_{-4.33}$ \\
JP & warm & C & $ -2.71^{+1.16}_{-1.25}$  & $ +0.65^{+0.79}_{-0.81}$  & $ +0.65^{+1.59}_{-1.51}$ \\
JP & warm & Mg & $ -2.92^{+1.16}_{-1.26}$  & $ +0.65^{+0.79}_{-0.80}$  & $ -0.62^{+1.78}_{-1.78}$ \\
JP & warm & Al & $ -1.55^{+1.46}_{-1.40}$  & $ -0.42^{+1.00}_{-1.16}$  & $ +0.30^{+1.45}_{-1.40}$ \\
JP & warm & Si & $ -2.32^{+0.92}_{-0.97}$  & $ +0.46^{+0.63}_{-0.65}$  & $ +2.23^{+1.55}_{-1.37}$ \\
JP & warm & S & $ -2.84^{+1.16}_{-1.24}$  & $ +0.65^{+0.79}_{-0.81}$  & $ +0.68^{+1.52}_{-1.44}$ \\
JP & warm & K & $ -2.03^{+1.03}_{-1.05}$  & $ +0.12^{+0.71}_{-0.76}$  & $ -0.83^{+1.29}_{-1.31}$ \\
JP & warm & Ca & $ -1.95^{+1.03}_{-1.05}$  & $ +0.12^{+0.71}_{-0.76}$  & $ +0.64^{+1.55}_{-1.41}$ \\
JP & warm & Mn & $ -1.95^{+1.04}_{-1.06}$  & $ +0.12^{+0.72}_{-0.77}$  & $ +0.75^{+1.14}_{-1.03}$ \\
JP & warm & Fe & $ -1.95^{+1.03}_{-1.05}$  & $ +0.12^{+0.71}_{-0.76}$  & $ +0.86^{+1.49}_{-1.34}$ \\
JP & warm & Ni & $ -1.97^{+1.04}_{-1.06}$  & $ +0.12^{+0.72}_{-0.77}$  & $ +0.66^{+1.33}_{-1.20}$ \\
 \hline 
  \label{tab:occrate_fits}
\end{longtable*}

\end{document}